\documentclass[12pt,a4paper,twoside,openright]{book} 

\usepackage{amssymb, amsthm, paralist, amsmath, mathrsfs} 

\usepackage{tikz} 
\usetikzlibrary{calc}

\usepackage{graphicx} 

\usepackage{hyperref} 

\usepackage{setspace} 
\usepackage{msor_thesis} 

\usepackage{mathptmx} 

\usepackage[utf8]{inputenc}

\usepackage{mleftright}

\usepackage{booktabs}

\makeatletter
\renewcommand{\@chapapp}{}
\newenvironment{chapquote}[2][2em]
{\setlength{\@tempdima}{#1}%
	\def\chapquote@author{#2}%
	\parshape 1 \@tempdima \dimexpr\textwidth-2\@tempdima\relax%
	\itshape}
{\par\normalfont\hfill--\ \chapquote@author\hspace*{\@tempdima}\par\bigskip}
\makeatother

\newtheorem{theorem}{Theorem}[chapter]

\newtheorem{mydef}{Definition}[chapter]

\newtheorem{corollary}{Corollary}[chapter]

\begin{document}
\frontmatter 

\author{Del Rajan}
\title{Complex Spacetimes and the Newman-Janis Trick}
\subject{Mathematics}


\abstract{In this thesis, we explore the subject of complex spacetimes, in which the mathematical theory of complex manifolds gets modified for application to General Relativity.  We will also explore the mysterious Newman-Janis trick, which is an elementary and quite short method to obtain the Kerr black hole from the Schwarzschild black hole through the use of complex variables.  This exposition will cover variations of the Newman-Janis trick, partial explanations, as well as original contributions.}


\ack{I want to thank my supervisor Professor Matt Visser for many things, but three things in particular.  First, I want to thank him for taking me on board as his research student and providing me with an opportunity, when it was not a trivial decision.  I am forever grateful for that. 
	
I also want to thank Matt for his amazing support as a supervisor for this research project.  This includes his time spent on this project, as well as teaching me on other current issues of theoretical physics and shaping my understanding of the Universe.  I couldn't have asked for a better mentor. 

Last but not least, I feel absolutely lucky that I had a supervisor who has the personal characteristics of being very kind and being very supportive.
	
I want to thank my fellow graduate students in the School of Mathematics \& Statistics.  Your company and our time spent together is a highlight of 2015 for me.

I also want to thank Baktash and Padideh Fazelzadeh, Gayathara De Silva, Ashwyn Sathanantham and Chirag Ahuja for 2014.  Without you, this would not be possible.}


\mscthesisonly 

\maketitle

\thispagestyle{empty}
\null\vspace{\stretch {1}}
\begin{flushright}
	Dedicated to my family.
\end{flushright}
\vspace{\stretch{2}}\null

\tableofcontents

\mainmatter 


\chapter{Introduction}

What is space, time and the quantum?  This is the fundamental question of theoretical physics today, exactly 100 years after Albert Einstein completed his theory of space and time, i.e. General Relativity. This question captures the essence of the technical issues of quantum gravity as well as the quantum measurement problem.  

Space and time as described by Einstein is depicted in a beautiful differential geometric framework, while quantum theory resides primarily in complex vector spaces.  A hope of trying to answer the above question would be to explore the interaction between differential geometry and complex vector spaces through the lens of physics.

In this thesis, we are going to discuss such a view, highlighting established work on complex spacetimes and exploring a mysterious procedure known as the Newman-Janis trick.   

To elaborate on what the trick does, it is important to recall that a solution of Einstein's field equations was found within a year after the theory was completed.  This was known as the Schwarzschild spacetime, and it described a black hole.  Another solution called the Kerr spacetime, that described a rotating black hole, was found almost 50 years later, whose derivation involved incredible algebraic complexity and hundreds of pages of non-trivial calculations.  These two spacetimes are considered to be the most important solutions of the theory of General Relativity.

The Newman-Janis trick is a short elementary procedure where one is able to obtain the latter solution from the former through the introduction of complex variables. However, no one fully understands why this trick works.

Before we move into the main chapters of this thesis, we will provide a short review of General Relativity in \S 2.  Specifically, we will elaborate on describing the Schwarzschild and Kerr solutions.  In addition to this, we will introduce the null tetrad formalism where the central geometric variables will be a specific type of tetrad, instead of the metric.

In chapters \S 2 and \S 3, the thesis will move into describing the theory of complex manifolds and its relationship to spacetime geometry.  Complex manifold theory in the hands of the mathematicians, has evolved into a vast rich subject but mainly, requires the use of a Riemannian signature metric.  We shall then review the work of Flaherty \cite{flaherty1976hermitian}, in which the mathematical theory gets modified for manifolds with Lorentzian signature metrics.  This provides us with a framework in which to explore research problems regarding complex variables in General Relativity.

The second part of this thesis starts in \S 4, where the Newman-Janis trick is introduced, including variations of the trick.  This will follow the original argument from the paper that introduced the trick exactly 50 years ago \cite{newman1965note}. 

Since then, a number of partial explanations and analysis have been provided in regards to the trick.  This includes explanations by Roy Kerr and Ezra Newman themselves and the exposition of these insights will be highlighted in \S 5.

Finally, we aim to provide an original contribution to this research problem by writing the trick in a different way.  This enables us to show an equivalence between different approaches and a possible framework to study the physics of the situation.


\chapter{Review of General Relativity}

\begin{chapquote}{Albert Einstein \textit{}}
	``The state of mind which  enables a man to do work of this kind... is akin to that of the religious worshiper or the lover; the daily effort comes from no deliberate intention or program, but straight from the heart.''
\end{chapquote}

General Relativity is our current theory for describing space and time and was formulated in its final form by Albert Einstein in 1915 \cite{isaacson2007einstein, einstein1915field}.  In addition to this, the theory advances our understanding of gravitation from the Newtonian perspective of forces to the modern day understanding of curved spacetimes, described by Lorentzian metric manifolds satisfying the Einstein field equations.

For a detailed background to the theory of General Relativity, refer to Wald \cite{wald2010general} or Carroll \cite{carroll2004spacetime} and for a detailed mathematical description of Lorentzian manifolds, refer to \cite{beem1996global, chrusciel2010mathematical}.

Before outlining the key equations, we should note some typical conventions are used,  such as that the geometric units are used by default.  Hence, the speed of light is given $c \equiv 1$, and Newton's constant of gravitation is given by $G_{N} \equiv 1$.  Another important point to note is the use of the Einstein summation convention where one omits the summation symbol whenever a pair of contravariant and covariant indices appears in one term.  We usually let the indices range over the four spacetime dimensions unless otherwise stated.

\section{Lorentzian geometry}

A metric is said to have a Lorentzian signature for $(+---)$ or $(-+++)$, and a Euclidean signature for $(++++)$.  In particular, we shall use the Lorentzian signature $(+---)$ throughout the thesis except when otherwise stated.  

Spacetime is described by a Lorentzian metric manifold and more specifically by the four dimensional metric tensor, $g_{ab}$, and its invariant quantity $ds^{2} = g_{ab}\:dx^{a}\:dx^{b}$.   

The metric tensor plays the crucial role of determining the geometry of the manifold and the important geometric quantities are built from this tensor and its derivatives.

The connexion/Christoffel symbol,  which is not a tensor, is given by 

\begin{equation}
\Gamma_{\;\;bc}^{\:a} = \frac{1}{2}\:g^{ad}\:(g_{db,c} + g_{dc,b} - g_{bc,d})
\end{equation}

and defines the notion of parallelism in a manifold.  This connexion can be defined in terms of the covariant derivative of a tensor

\begin{equation}
\nabla_{b} \: V^{a} = \partial_{b}\:V^{a} + \Gamma_{\;\;cb}^{\:a}\:V^{c}.
\end{equation}

This is a generalization of taking a derivative in curved spaces. Notice the deviation from flat space is represented by the connexion. 

The Riemann curvature tensor is a quantity which measures the extent to which the covariant derivative fails to commute, and in that sense, the data about the curvature is located in the components of this tensor.  The explicit formula for this tensor is given by

\begin{equation}
R_{\;\;\;bcd}^{\;a} = \partial_{c}\:\Gamma_{\;\;bd}^{\:a}\: - \partial_{d}\:\Gamma_{\;\;bc}^{\:a}\: + \Gamma_{\;\;ec}^{\:a}\:\Gamma_{\;\;bd}^{\:e}\: - \Gamma_{\;\;ed}^{\:a}\:\Gamma_{\;\;bc}^{\:e}.
\end{equation}  

From this, one can see that since the Riemann tensor is made up of the metric and its derivatives, the geometry of curvature is ultimately contained in the metric.

The Ricci tensor and the associated Ricci scalar can be built out of the Riemann tensor as 

\begin{align}
\begin{split}
R_{ab} = R_{\;\;\;acb}^{c}, \\
R = g^{ab}\:R_{ab},
\end{split}
\end{align}

and the Einstein tensor is given by

\begin{equation}
G_{ab} \equiv R_{ab} - \frac{1}{2} \:R\:g_{ab}.
\end{equation}

The celebrated Einstein field equations is beautifully written as 

\begin{equation} \label{EinsteinEquations}
G_{ab} = 8\:\pi\: T_{ab},
\end{equation}

where $T_{ab}$ is the stress-energy tensor, that describes the matter or field which is creating the curvature in spacetime.  

Vacuum spacetimes are solutions where $T_{ab} = 0$ in (\ref{EinsteinEquations}).  This can be shown to be equivalent to the statement that $R_{ab} = 0$ and is known as a Ricci-flat solution.

\section{Vacuum spacetimes}

In this section, we introduce the two most studied solutions in all of General Relativity.  They also happen to be vacuum spacetimes and in addition to that form the basis of our study of the Newman-Janis trick.

\subsubsection*{Schwarzschild spacetime}

The Schwarzschild solution was published in 1916 by Karl Schwarzschild \cite{schwarzschild1916gravitationsfeld} and was obtained within one year after the completion of General Relativity.  

The gravitational fields that are important to us in every day life such as the ones from the Sun or the Earth are described by slowly rotating, nearly spherically symmetric objects.  These can be best described by the exact spherically symmetric solution of Einstein's equations (\ref{EinsteinEquations}), namely the Schwarzschild solution.  

The Schwarzschild metric in coordinates $(t, r, \theta, \phi)$ is given by

\begin{equation}
ds^{2}= (1-\frac{2\:m}{r})\:{dt}^{2}-\frac{1}{1-\frac{2\:m}{r}}dr^{2}-{r}^2({d\theta}^{2}+{{\sin}^{2}\theta}\:{d\phi}^2).
\end{equation}

The parameter $m$ measures the amount of mass inside the radius $r$ and in the region $r \leq 2m$, the metric describes a black hole region. Observers can enter it, but can never leave the region.  The exact surface between this region and the outside is known as the event horizon of a black hole.

Notice that a true singularity exists for $r=0$ and a coordinate singularity exists $r=2m$ as can be shown if one were to put this metric in another coordinate system.

We can express this vacuum solution in the advanced Eddington-Finkelstein coordinates by performing a coordinate transformation, 

\begin{align}
\begin{split}
u &= t - r - 2m\:\text{ln}\:\Bigl(\frac{r}{2m}-1\Bigl), \\
r' &= r, \\
\theta' &= \theta, \\
\phi' &= \phi,
\end{split}
\end{align}

and dropping the primes, to obtain

\begin{equation}\label{SchinEF}
ds^{2}= \Bigl(1-\frac{2\:m}{r}\Bigl)\:{du}^{2}+2\:du\:dr-{r}^2({d\theta}^{2}+{{\sin}^{2}\theta}\:{d\phi}^2).
\end{equation}

If one sets the mass parameter $m$ to zero, we obtain the flat metric given by 

\begin{equation}
ds^{2}= {du}^{2}+2\:du\:dr-{r}^2({d\theta}^{2}+{{\sin}^{2}\theta}\:{d\phi}^2).
\end{equation}

\subsubsection*{Kerr spacetime}

The Kerr solution was discovered by Roy Kerr almost 50 years after the completion of General Relativity \cite{kerr1963gravitational}.  The derivation involved an enormous amount of algebraic complexity and hundreds of pages of non-trivial calculations.  For a detailed account of the construction of the solution by Kerr, refer to \cite{wiltshire2009kerr}.    

The Kerr metric is a mathematical description of rotating black holes and is the rotating generalization of the Schwarzschild metric.  The physical parameters involved extend from considering only mass to now including the parameter $a$ which is the angular momentum per unit mass.

The advanced Eddington-Finkelstein form of the Kerr spacetime is given by coordinates $(u, r, \theta, \phi)$ and the metric is expressed as

	\begin{multline}
	ds^{2}= (1-\frac{2\:m\:r}{r^{2}+a^{2}{\cos}^{2}\theta})\:{du}^{2}+2\:du\:dr+\frac{4\:m\:r\:a\sin^{2}\theta}{r^{2}+a^{2}{\cos}^{2}\theta}du\:d\phi \\
	-2\:a\:\sin^{2}\theta\:d\phi\:dr 
	-(({r^{2}+a^{2}{\cos}^{2}\theta})\:a^{2}\sin^{2}\theta + 2\:m\:r\:a^{2}\sin^{2}\theta\\
	+({r^{2}+a^{2}{\cos}^{2}\theta})^{2})\frac{\sin^{2}\theta}{({r^{2}+a^{2}{\cos}^{2}\theta})}d\phi^{2} 
	-({r^{2}+a^{2}{\cos}^{2}\theta})\:{d\theta}^{2}.
	\end{multline}

Notice that if one sets $a=0$, one obtains the Schwarzschild geometry and if one sets $m=0$, then the flat space metric is obtained.

A true singularity exists for Kerr where the singularity takes the shape of a ring given by 

\begin{equation}
r = 0; \qquad \theta = \frac{\pi}{2}.
\end{equation}

The Kerr spacetime has a large number of differences compared to the Schwarzschild spacetime as there are different surfaces associated to it such the outer and inner event horizons as well as surfaces known as ergospheres.  Detailed analysis on these specific structures and other characteristics can be found in \cite{wiltshire2009kerr, chandrasekhar1998mathematical}.

\section{Null Tetrads}

Null tetrads and their associated Newman-Penrose field equations \cite{o2003introduction, penrose1988spinors} is another framework for expressing the theory of General Relativity.  For the purposes of this thesis, we will only consider null tetrads themselves and not delve into this new Newman-Penrose framework.

In a spacetime endowed with a physically meaningful extra property known as a spinor structure \cite{geroch1968spinor}, one can define a local null tetrad.  Hence, at each point on the manifold, there are four null vectors ${l^{a}, n^{a}, m^{a}, \overline{m}^{a}}$ with specific properties.

The vectors $l^{a}$ and  $n^{a}$ are real and satisfy $l^{a}\:n_{a} = 1$.  The other two vectors, $m^{a}, \overline{m}^{a}$ are complex null vectors and have the property that they are complex conjugates of each other and satisfy the condition, $m^{a}\:\overline{m}_{a} = -1$.

The null tetrad can become the central variable of General Relativity and its relationship to the metric tensor can be expressed as  

\begin{equation} \label{nulltetradmetric}
g_{ab} = l_{a}\:n_{b} + n_{a}\:l_{b} - m_{a}\:\overline{m}_{b} - \overline{m}_{a}\:m_{b},
\end{equation}

\begin{equation} 
g^{ab} = l^{a}\:n^{b} + n^{a}\:l^{b} - m^{a}\:\overline{m}^{b} - \overline{m}^{a}\:m^{b}.
\end{equation} 
 
But finding a null tetrad for a metric can sometimes be difficult (and is never unique) and hence it is a science, but as well an art.

An important set of transformations to consider are the proper Lorentz transformations  on the null tetrads.  The first involves a null rotation about $l^{a}$ and is given by 

\begin{align}
\begin{split}
l^{a} &\rightarrow \hat{l}^{a} = l^{a}, \\
n^{a} &\rightarrow \hat{n}^{a} = n^{a} + a\:\overline{m}^{a} + \overline{a}\:m^{a} + a\:\overline{a}\:l^{a}, \\
m^{a} &\rightarrow \hat{m}^{a} = m^{a} + a\:l^{a}, \\
\overline{m}^{a} &\rightarrow \hat{\overline{m}}^{a} = \overline{m}^{a} + \overline{a}\:l^{a},
\end{split}
\end{align}

where $a$ is an arbitrary complex function.  The second Lorentz transformation represents a boost in the $l^{a}$ - $n^{a}$ plane and a rotation in the $m^{a}$ - $\overline{m}^{a}$ plane.  This is expressed by

\begin{align} 
\begin{split}
l^{a} \; &\longrightarrow \; \hat{l}^{a} = A^{-1}\:l^{a}, \\
n^{a} \; &\longrightarrow \; \hat{n}^{a} = A\:n^{a}, \\
m^{a} \; &\longrightarrow \; \hat{m}^{a} =e^{i\:\phi}\:m^{a}, \\
\overline{m}^{a} \; &\longrightarrow \; \hat{\overline{m}}^{a} =e^{-i\:\phi}\: \overline{m}^{a},
\end{split}
\end{align}

where $A$ and $\phi$ are arbitrary real functions.  The final proper Lorentz transformation is a null rotation about $n^{a}$, given by 

\begin{align}
\begin{split}
{l}^{a} &\rightarrow \hat{l}^{a} = l^{a} + b\:\overline{m}^{a} + \overline{b}\:m^{a} + b\:\overline{b}\:n^{a},  \\
n^{a} &\rightarrow \hat{n}^{a} = n^{a}, \\
m^{a} &\rightarrow \hat{m}^{a} = m^{a} + b\:n^{a}, \\
\overline{m}^{a} &\rightarrow \hat{\overline{m}}^{a} = \overline{m}^{a} + \overline{b}\:n^{a},
\end{split}
\end{align}

where $b$ is an arbitrary complex function.

\section{Discussion}

In this chapter, we outlined the key points of the theory of General Relativity.  Spacetime can be described by a Lorentzian manifold satisfying the Einstein equations (\ref{EinsteinEquations}).  The first vacuum solution considered was the Schwarzschild solution and was found within one year after General Relativity was finalized.  The Kerr solution took almost an astounding 50 years to find.  The comparison between these solutions represent the opposite ends of a spectrum when it comes to analytically solving the Einstein equations.


\chapter{Complex Manifold Theory}

\begin{chapquote}{Michael Atiyah \textit{}}
	``Algebra is the offer made by the devil to the mathematician... All you need to do, is give me your soul: give up geometry.''
\end{chapquote}

To begin the investigation of complexified spacetimes and their relevance to the Newman-Janis trick, this chapter will provide the necessary mathematical background of complex manifold theory. 

Complex manifolds represent the synthesis of complex variables with the field of differential geometry.  These mathematical constructions have found applications in various areas of physics including proposed theories of quantum gravity such as supersymmetric string theory as well as twistor theory.  A particularly nice general summary is presented by Penrose in \cite{penrose2006road}. Within the context of General Relativity, complex manifolds are introduced in a variety of ways, of which the one that is explored in this thesis is largely due to Flaherty \cite{flaherty1976hermitian}.  On a historical note, Einstein himself introduced a complex valued metric tensor in an attempt to include General Relativity into a unified field theory \cite{einstein1945generalization}.     

Standard references for the subject of complex manifolds include \cite{huybrechts2006complex, griffiths2014principles, kodaira2012complex}, but we will be closely following the material presented by Flaherty in \cite{flaherty1976hermitian}.

\section{Complex Linear Algebra}

We start this chapter by considering vector spaces, before moving to the case of manifolds.

\subsubsection*{Complexification \& Complex Structure}

Let $V$ be a real finite-dimensional vector space which we denote ($V$,\: $\mathbb{R}$). 

\begin{mydef}
	The complexification of ($V$,\:$\mathbb{R}$) is the complex vector space $V^{\mathbb{C}}$, also denoted ($V^{\mathbb{C}}$, $\mathbb{C}$), where:
	
	(i)  \hspace{1cm} $Z = X+iY \in V^{\mathbb{C}}  \hspace{0.1cm}\longleftrightarrow \hspace{0.1cm} X, Y \in V$;
	
	(ii) $(X_{1} + iY_{1}) + (X_{2} + iY_{2}) \equiv (X_{1} + X_{2})+i(Y_{1} + Y_{2})$ 
	for all $X_{1}, X_{2}, Y_{1}, Y_{2} \in V$;
	
	(iii)  ($\alpha +i\beta$)($X + iY$)$\equiv$ ($\alpha X - \beta Y$) + i($\beta X + \alpha Y$) for all $X, Y \in V$ and $\alpha, \beta \in \mathbb{R}$.
	
\end{mydef}

A crucial point is that $V^{\mathbb{C}}$ satisfies the axioms of a complex vector space and  that complex conjugation in $V^{\mathbb{C}}$ is defined by $\overline{Z} = \overline{X+iY} \equiv X-iY$. 

\begin{mydef}
	A complex structure on a finite-dimensional real vector space $V$ is an endomorphism $J$ such that $J(J(X)) = -X$ for all $X\in V$.  We will denote a vector space with a complex structure by ($V,\: \mathbb{R};\:J$).
\end{mydef} 

\begin{theorem}
	($V,\:\mathbb{R};\:J$) is even-dimensional.  (See e.g. Flaherty \cite{flaherty1976hermitian}.)
\end{theorem}

From the above theorem, we can see that the dimensionality of a vector space with a complex structure, which we denote by $n$, can be expressed as $n=2m$ for a particular $m$.  This allows one to create a notation where we can let unbarred indices range from index value $1$ to index value $m$ and let barred indices range from $m+1$ to $m+m=n$. Thus producing expressions of the form $A_{\alpha}B^{\overline{\alpha}} = \sum\limits_{\alpha=1}^{m}A_{\alpha}B^{m+\alpha}$.  

The following theorem will show that a complex structure on a real vector space ($V,\:\mathbb{R};\:J$),  has a relationship to the complexification of that vector space $V^{\mathbb{C}}$, in particular to the subspaces of the complexified vector space.  Given ($V,\:\mathbb{R};\:J$), we can define the following subspaces of $V^{\mathbb{C}}$:

\begin{equation}
W(J)= \{Z|Z = X-iJ(X) \text{ and }  X \in V\}; 
\end{equation}

\begin{equation}
\overline{W}(J)= \{Z|Z = X+iJ(X) \text{ and }  X \in V\}. 
\end{equation}

The subspaces $W(J)$ and $\overline{W}(J)$ are said to be complex conjugates of each other, since given an element $Z\in W(J)$, one can see that the complex conjugate of $Z$ is in $\overline{W}(J)$, for all elements in $V^{\mathbb{C}}$.

\begin{theorem}
	Given ($V,\:\mathbb{R};\:J$), then $V^{\mathbb{C}} = W(J) \bigoplus \overline{W}(J)$.  Conversely, given ($V,\:\mathbb{R}$), where $U$ and $\overline{U}$ are subspaces of ($V^{\mathbb{C}}, \mathbb{C}$) which are complex conjugates to each other and $V^{\mathbb{C}} = U \bigoplus \overline{U}$, then there is exists a complex structure $J$ for $V$ such that $W(J) = U$ and $\overline{W}(J)=\overline{U}$.  (See e.g. Flaherty \cite{flaherty1976hermitian}.)
\end{theorem}

One can extend the complex structure, which was initially defined to be a structure on a real vector space $V$, to the endomorphism $J:V^{\mathbb{C}}\to V^{\mathbb{C}}$.  This has the implication that elements of the set $W(J)$ satisfy the equation $J(Z)=iZ$.  These vectors are referred to as type $(1,0)$.  Consequently, $\overline{W}(J)$ is then the subspace of $V^{\mathbb{C}}$ consisting of vectors $\overline{Z}$ where $J(\overline{Z})=-i\overline{Z}$ and these vectors are referred to as type $(0,1)$.

The operator $-J$ is also a complex structure on $V$ and is referred to as the complex structure conjugate to $J$.

\subsubsection*{Complex Structure on the Dual Space}

\begin{mydef}
	The dual space $V^{*}$ of ($V,\:\mathbb{R};\:J$) consists of all linear functions $\omega: V \to \mathbb{R}$.
\end{mydef}

If we are given an $X \in V$, the notation that we use for the value of the real number in the above  mapping is $\omega(X)$ or $<X, \omega>$.

The complexification of $V^{*}$ is denoted by $V^{* \mathbb{C}}$ and this is the complex vector space consisting of all linear functions $\omega +i\theta: V \to \mathbb{C}$.  An explicit expanded expression for this map is given by $(\omega +i\theta)(X) = \omega(X) + i\theta(X)$ for all $X \in V$ and for all $\omega +i\theta \in V^{* \mathbb{C}}$.  

A unique complex structure, $J^{*}$, can be constructed for the dual space $V^{*}$, given that there is a complex structure on the vector space $V$.  Given a $\omega \in V^{*}$, we define $J^{*}(\omega) \in V^{*}$ by 

\begin{equation}
J^{*}(\omega)(X)=\omega(J(X)) \text{ for all } X \in V.
\end{equation}

From this, we can deduce that $J^{*}$ is a complex structure for ${V^{*}}$.  To see this note $(J^{*})^{2}(\omega)(X)=J^{*}(\omega)(J(X)) = \omega(J^{2}(X))=\omega(-X)=(-\omega)(X)$.  Therefore, $(J^{*})^{2} (\omega)= -\omega$ for all $\omega \in V^{*}$.

The operator $J^{*}$ can also be extended to map the complexified dual space $V^{*\mathbb{C}}$ to itself.  This allows us to define type $(1,0)$ elements as those linear functions $\omega$ for which $J^{*}(\omega)=i\omega$.  Similarly type $(0,1)$ forms are elements, $\overline{\omega}$ in $V^{*\mathbb{C}}$ for which  $J^{*}(\overline{\omega})=-i \overline{\omega}$.   As for the case of the complexified vector space, we can get a direct sum decomposition expression of the form $V^{*\mathbb{C}} = W(J^{*}) \bigoplus \overline{W}(J^{*})$.

\subsubsection*{Coordinate representation}

The next step in our introduction to complex structures on vector spaces is to introduce these expressions in terms of coordinates.  At this stage, these are merely coordinates on the vector space, not coordinates on any underlying manifold.  In particular, we'll set up bases and this enables one to then perform calculations.  

Let \{$e_{1}, ... , e_{n}$\} be a basis for $V$, and let the corresponding basis for $V^{*}$ be labelled as \{$e^{*1}, ... , e^{*n}$\}, such that $e^{*a}(e_{b}) = \delta_{a}^{b}$.  An arbitrary element of $V$ can then be represented as $X = x^{a}e_{a}$.  In addition to this, an element of $V^{*}$ can now be written with respect to this basis as $\omega = \omega_{a}\;e^{*a}$.  

The complex structure acting on an element of $V$, with respect to this basis, is written as $J(e_{a})=J_{a}^{\:\:b}\:e_{b}$.  By using the definition of a complex structure, we see that:

\begin{equation}
-e_{a}=J^{2}(e_{a})=J_{a}^{\:\:b}\;J_{b}^{\:\:c}\;e_{c} \quad \longrightarrow \quad J_{a}^{\:\:b}\:J_{b}^{\:\:c}=-\delta_{a}^{c} 
\end{equation}

To illustate an example, suppose $J(X)=Y$, where $Y=y^{b}\;e_{b}$, then we see in coordinates that $y^{b}=J_{a}^{\:\:b}\;x^{a}$. 

Since our vector space is of dimensionality $n= 2m$, it follows that $J_{a}^{\:\:b}$ is a $2m\times 2m$ matrix.  Furthermore, it can be shown that $J_{a}^{\:\:b}$ has $m$ eigenvalues $+i$ with eigenvectors of type $(1,0)$ vectors and $m$ eigenvalues $-i$ with eigenvectors of type $(0,1)$ vectors.

A useful expression for calculations is that given an arbitrary basis \{$e_{1}, ... , e_{n}$\}, the most general complex structure on a vector space is given by the expression

\begin{equation}
J_{a}^{\;d} = S_{a}^{\:\:b}\;(J_{0})_{b}^{\:\:c}\;(S^{-1})_c^{\:\:d}\text{,}
\end{equation}  

where 

\begin{equation}
(J_{0})_{b}^{\:\:c} = 
\begin{pmatrix}
0 & I_{m} \\
-I_{m} & 0
\end{pmatrix}\text{,}
\end{equation}

and $I_{m}$ is a $m \times m$ identity matrix. In addition to this, it is crucial that $S_{a}^{\:\:b}$ has to have the property of being a non-singular matrix.

We will now consider how one goes about setting up a basis for $V^{*\mathbb{C}}$ by using a complex structure $J$.  The construction would involve starting with same basis \{$e^{*1}, ... , e^{*n}$\} for $V^{*}$, and use this to construct a basis $\lambda^{a}$ for $V^{*\mathbb{C}}$, using $J$ as follows:

\begin{equation}
\lambda^{a}(X) = e^{*a}(J(X))+ie^{*a}(X).
\end{equation}

We can then choose the \{$e^{*1}, ... , e^{*n}$\} basis in such a way that the set \{$\lambda^{1}, ... , \lambda^{m}$\} is $\mathbb{C}$-linearly independent.  These form the basis for the subspace $W(J^{*})$. The elements \{$\overline{\lambda^{1}}, ... , \overline{\lambda^{m}}$\} form the basis for the subspace $\overline{W}(J^{*})$.  

Writing this basis as $\lambda^{\alpha}$, and then splitting it up into real and imaginary parts 
$\lambda^{\alpha} = \mu^{\alpha}+i\mu^{\overline{\alpha}}$, (recall that unbarred indices range and sum over $1,...,m$ and barred indices range and sum over $m+1, ... , m+m=n$), we can find that $\{\mu^{\alpha}, \mu^{\overline{\alpha}}\}$ is a basis for $V^{*}$. One can then construct the dual basis to $V^{*}$ which provides a basis for $V$ denoted by 
  \{$E_{\alpha}, E_{\overline{\alpha}}$\}.  
  
  The action of the complex structure $J$ on such a basis is $J(E_{\alpha})=E_{\overline{\alpha}}$, while $J(E_{\overline{\alpha}})=-E_{\alpha}$. 
  
  One can use this coordinate based construction to prove the following theorem.

\begin{theorem}
	A complex structure $J$ determines an orientation of V.  (See e.g. Flaherty \cite{flaherty1976hermitian}.)
\end{theorem}

In this section, we concentrated our study on complex structures acting on vector spaces. This will naturally connect to later aspects of this chapter when we consider complex structures on vector spaces at points of a type of manifold called a complex manifold.

\section{Complex Manifolds}

The mathematical construction of a complex manifold involves starting with the basic constituents of a manifold. 

\subsubsection*{Complex Structure on Manifolds}
 
One starts off with a manifold, denoted by $M$, which by definition has the property that each point $p \in M$ has a neighborhood $U$ homeomorphic to $\mathbb{R}^{n}$ for some value $n$.  The coordinate charts of a manifold denoted by $x:U\subseteq M \rightarrow \mathbb{R}$ provide a system of local coordinates for points of the manifold.  

In this thesis, we will only consider manifolds of real dimension $n=2m$, i.e. even-dimensional.  In other studies of complex manifolds one usually starts with defining a manifold where the chart is covered by open sets homeomorphic to $\mathbb{C}^{m}$.  But since $\mathbb{R}^{2m}$ is homeomorphic to $\mathbb{C}^{m}$, we will stick to this particular construction of a complex manifold.    

Given that our manifolds must be even-dimensional, an arbitrary point $p \in U$ has a coordinate representation by the chart $x$ as $x(p)=(x^{1}, ... , x^{n})$.  We refer to this as the real coordinates. To get what we call the complex coordintes $(z^{1}, ... , z^{m})$, we use the formula

\begin{equation}
z^{\alpha}\equiv x^{\alpha} + ix^{\overline{\alpha}}.
\end{equation}

This allows us to create complex coordinates to points of any even-dimensional manifold.  The complex coordinates and real coordinates are in one-to-one relationship by the relations

\begin{align}
x^{\alpha}= \frac{1}{2} (z^{\alpha}+\overline{z^{\alpha}}); \\
{x^{\overline{\alpha}}}= \frac{1}{2i} (z^{\alpha}-\overline{z^{\alpha}}). 
\end{align}

To construct the essential component of a complex manifold, one has to consider the transition functions of the atlas of the manifold.  

Suppose $p \in U\cap U'$ where $x:U\cap U'\rightarrow W\subseteq \mathbb{R}^{n}$  and $x':U\cap U'\rightarrow W'\subseteq \mathbb{R}^{n}$.  Hence, $p$ has two sets of real coordinates, $x^{a}$ and $x^{a'}$, and the transition functions are given by $x^{a'}=x^{a'}(x^{b})$ and $x^{a}=x^{a}(x^{b'})$.  Given that we can interchange between real and complex coordinates of $p$, one can also rewrite these transition functions as $z^{\alpha '}=z^{\alpha '}(z^{\beta}, \overline{z^{\beta}})$ and $z^{\alpha}=z^{\alpha}(z^{\beta '}, \overline{z^{\beta '}})$.

\begin{mydef}
	A structure on a manifold is constructed by restricting the allowed sets $U, U', ...$ to those for which the associated transition functions belong to some specific pseudogroup of transformations. (See e.g. Flaherty\:\cite{flaherty1976hermitian}.)
\end{mydef}

A common example of a structure on a manifold that is found in General Relativity is one with a differentiable structure.  Here the transition functions are required to be at least $C^{2}$ functions.  

\begin{mydef}
	A manifold has a complex structure if it can be covered by sets $U, U', ...$ such that in the intersections of these sets, the transition functions $z^{\alpha '}=z^{\alpha '}(z^{\beta})$ are \textbf{holomorphic} functions. 
\end{mydef}
	
\begin{mydef}
	A complex manifold is a manifold which admits a complex structure.
\end{mydef}	

In other words, a complex manifold is a manifold that consists of an atlas of open sets $U, U', ...$ such that if $z^{\alpha},z^{\beta}$ are the complex coordinates associated with $U, U'$ where $U\cap U'$ is non-empty, then

\begin{equation}
z^{\beta '}=z^{\beta '}(z^{\alpha}), \hspace{1cm} \text{det}\;(\partial  z^{\beta '} / \partial z^{\alpha}) \neq 0, 
\end{equation}

where $z^{\beta '}(z^{\alpha})$ are holomorphic functions of $z^{\alpha}$.

Equivalently, using real coordinates for the transition functions, one can write them as  $x^{\beta '} = x^{\beta '}(x^{\alpha}, x^{\overline{\alpha}})$ and $x^{\overline{\beta '}} = x^{\overline{\beta '}}(x^{\alpha}, x^{\overline{\alpha}})$.  If these functions satisfy the Cauchy-Riemann conditions, 

\begin{equation}
\frac{\partial x^{\beta '}}{\partial x^{\alpha}} = \frac{\partial x^{\overline{\beta} '}}{\partial x^{\overline{\alpha}}}\text{,}   \hspace{1cm}\text{and} \hspace{1cm} \frac{\partial x^{\beta '}}{\partial x^{\overline{\alpha}}} = -\frac{\partial x^{\overline{\beta} '}}{\partial x^{{\alpha}}}\text{,}
\end{equation}

then the manifold is a complex manifold.

\subsubsection*{Conditions to admit a complex structure}

There are a number of conditions that must be met for a manifold to admit a complex structure.  One of them is that the manifold has to be of even dimensionality.  Another condition involves a property called orientability. 

\begin{mydef}
	A manifold is orientable if and only if it can be covered by open sets such that if whenever $x^{a'} = x^{a'}(x^{b})$ are the transition functions for intersecting sets, then det$(\partial x^{a'}/\partial{x^{b}}) > 0$.
\end{mydef}

\begin{theorem}
	A complex manifold is orientable.  (See e.g. Flaherty \cite{flaherty1976hermitian}.)
\end{theorem}

Even-dimensionality and orientability are not sufficient conditions for a manifold to be a complex manifold.  Figuring out whether a manifold has a complex structure is a non-trivial problem \cite{hwangcomplex}.  In the following, a few examples of complex manifolds are provided including spaces that may be familiar to us.

\subsubsection*{Examples of Complex Manifolds}

The trivial example of a complex manifold is Euclidean space $\mathbb{R}^{2m}$.  It only needs a single chart to cover it since it is homeomorphic to $\mathbb{R}^{2m}$.  The real coordintates would be $x^{1}, ..., x^{m}, x^{m+1}, ... , x^{2m}$ everywhere, and the corresponding complex coordinates are $z^{1} = x^{1}+ix^{m+1}, ... , z^{m}=x^{m}+ix^{2m}$.  This coordinate system provides a complex structure for the space in consideration.

A particularly important class of complex manifolds that occurs in areas of pure mathematics such as algebraic geometry \cite{griffiths2014principles}, and in areas of physics such as in geometric quantum mechanics \cite{brody2001geometric, bengtsson2006geometry}, are complex projective spaces. An intuitive notion of these spaces is that a point in the complex projective space is a line through $\mathbb{C}^{m+1}$.  Basically one starts off with the punctured space $\mathbb{C}^{m+1}/0$ and using an equivalence relation, we identify 

\begin{equation}
(z^{1}, ... , z^{m+1}) \approx (\lambda z^{1}, ... , \lambda z^{m+1})
\end{equation}

for any non-zero complex scalar $\lambda$. To obtain the complex structure for this space, one can form an atlas with charts given by 

\begin{equation}
U_{j}=\{z^{\alpha}: \alpha = 1, ... , m+1 | z_{j}  \neq 0\}, \hspace{1cm} w_{(j)}^{\: \alpha} \equiv \frac{z^{\alpha}}{z^{j}},
\end{equation} 

for some fixed $j$. The atlas covers the entire complex projective space and each individual chart is homeomorphic to $\mathbb{R}^{2m}$.  To check whether it has a complex structure, we look at the transition functions given by 

\begin{equation}
w_{(i)}^{\: \alpha} \equiv \frac{z^{\alpha}}{z^{i}} = \frac{{z^{\alpha}}/{z^{j}}}{{z^{i}}/{z^{j}}} = \frac{w_{(j)}^{\: \alpha}}{w_{(j)}^{\: i}},
\end{equation}

and indeed, these are holomorphic functions.

\subsubsection*{Conjugate and Product Manifolds}

To define the analytic continuation of functions on a complex manifold, requires the introduction of various mathematical structures.  These are the conjugate complex structure, the conjugate manifold and finally the case of the product manifold.

\begin{mydef}
	Given a manifold, $M$, with complex structure defined by the set of coordinate patches $\{(U, z^{\alpha}), (U', z^{\alpha '}), ...\}$, the conjugate complex structure is defined as the complex structure $\{(U,\overline{z^{\alpha}}), (U', \overline{z^{\alpha '}}), ...\}$.
\end{mydef}

\begin{theorem}
	The conjugate complex structure is a complex structure for $M$.  (See e.g. Flaherty \cite{flaherty1976hermitian}.)
\end{theorem}

We have seen a complex structure for $\mathbb{R}^{2m}$, given by complex coordinates  $z^{1} = x^{1}+ix^{m+1}, ... , z^{m}=x^{m}+ix^{2m}$.  The conjugate complex structure is put on $\mathbb{R}^{2m}$ by specifying the coordinates $z^{1'} = x^{1}-ix^{m+1}, ... , z^{m'}=x^{m}-ix^{2m}$.  To really see that these complex structures are inequivalent despite covering the same manifold, one can look at the coordinate transformations between them, and show that they are non-holomorphic.

\begin{mydef}
	Given a complex manifold $M$, the conjugate manifold $\overline{M}$ is that complex manifold for which there exists a homeomorphism $*:M\rightarrow \overline{M}$ such that if $(U, z^{\alpha})$ is a chart of $M$ and $*(U)=V$, then $(V, z^{\overline{\alpha}})$ is a chart of $\overline{M}$ with $z^{\overline{\alpha}}(*(p)) = \overline{z^{\alpha}(p)}$ for all $p \in U$. 
\end{mydef}

The homeomorphism in coordinate patch $V$ was labelled $z^{\overline{\alpha}}$ to be consistent with our previous notation, and one can label those coordinates to range from $z^{m+1}$ to $z^{2m}$.

\begin{mydef}
	Given a manifold $M$ with complex structure $\{(U, z^{\alpha}), (U', z^{\beta '}), ...\}$, and its conjugate $\overline{M}$ with complex structure $\{(V, z^{\overline{\alpha}}), (V', z^{\overline{\beta '}}), ...\}$, the product manifold denotes the manifold $M \times \overline{M}$ with complex structure consisting of a maximal atlas that includes $\{(U \times V, z^{\alpha}, z^{\overline{\alpha}}), (U' \times V',z^{\beta '}, z^{\overline{\beta '}}), ...\}$.  In addition to this, in the intersection of the charts, we have the following holomorphic transition functions $z^{\beta '} = f^{\beta '}(z^{\alpha})$ and $z^{\overline{\beta '}} = \overline{f^{\beta '}}(z^{\overline{\alpha}})$.
\end{mydef} 

The product manifold $M \times \overline{M}$ has twice the dimensionality of $M$.  In particular, $M$ can be shown to be a submanifold of $M \times \overline{M}$ with the condition that $z^{\overline{\alpha}}=\overline{z^{\alpha}}$.

In the following, we proceed to define a number of different mathematical objects on complex manifolds.  

\section{Functions on a Complex Manifold}

The notion of a product manifold will prove to be useful when considering properties of functions on a complex manifold.

If we are given a real analytic function in real coordinates on a complex manifold, it can be shown that the same function in terms of complex coordinates will not, in general, be holomorphic.  What is surprising though, is that one can construct a holomorphic function given an underlying real analytic function, as long as we define the respective holomorphic function on a product manifold.

To be more explicit, our discussion of functions $f:M\rightarrow \mathbb{R}$ on complex manifolds starts with considering an arbitrary coordinate patch of that manifold, say $(U, x)$.  One can then proceed to consider the associated coordinate functions $(f\circ x^{-1})\:(z^{\alpha}, \overline{z^{\alpha}})$ and $(f\circ x^{-1})\:(x^{\alpha}, x^{\overline{\alpha}})$ which we denote $f \:(z^{\alpha}, \overline{z^{\alpha}})$ and $f(x^{\alpha}, x^{\overline{\alpha}})$ respectively.  As we have stated above, $f \:(z^{\alpha}, \overline{z^{\alpha}})$ will not necessarily be a complex analytic function of $f(x^{\alpha}, x^{\overline{\alpha}})$  but there does exist a relevant relationship which the following theorem reveals.

\begin{theorem}
	Every real analytic function $f(x^{\alpha}, x^{\overline{\alpha}})$ on $M$ gives rise to a complex analytic function $F(z^{\alpha}, z^{\overline{\alpha}})$ on $M \times \overline{M}$, which we refer to as the analytic continuation of $f(x^{\alpha}, x^{\overline{\alpha}})$.  (See e.g. Flaherty \cite{flaherty1976hermitian}.)
\end{theorem}

The operation of taking a partial derivative of complex-valued functions on a complex manifold also requires the construction of the product manifold.

\begin{mydef}
	Given a real analytic function $f(x^{\alpha}, x^{\overline{\alpha}})$ on $M$ and the corresponding complex coordinate function $f(z^{\alpha}, \overline{z^{\alpha}})$ we define
	
	\begin{equation}
	\frac{\partial f(z^{\alpha}, \overline{z^{\alpha}}) }{\partial z^{\beta}} \equiv \frac{\partial F(z^{\alpha}, z^{\overline{\alpha}})}{\partial z^{\beta} } \bigg|_{z^{\overline{\alpha}}=\overline{z^{\alpha}}}
	\end{equation}
	
	and 
	
	\begin{equation}
	\frac{\partial f(z^{\alpha}, \overline{z^{\alpha}}) }{\partial \overline{z^{\beta}}} \equiv \frac{\partial F(z^{\alpha}, z^{\overline{\alpha}})}{\partial z^{\overline{\beta}} } \bigg|_{z^{\overline{\alpha}}=\overline{z^{\alpha}}}
	\end{equation}
	
	where $F$ is the analytic continuation of $f$.

\end{mydef}

\section{Vectors on a Complex Manifold}

The foregoing discussion on complex linear algebra will prove to be useful for the consideration of vectors on complex manifolds.  For a general background on vectors and vector spaces at points on a manifold, see Wald \cite{wald2010general}. 

Before we consider the case of a complex manifold, consider two significant vector spaces associated to a point $p$ on a differentiable manifold $M$.  This refers to the tangent space $M_{p}$ and the cotangent space $M_{p}^{\: *}$.  Furthermore, one can use the procedure of complexification to construct the complexified tangent space $M_{p}^{\: \mathbb{C}}$ from the tangent space, and complexified cotangent space $M_{p}^{\: \mathbb{C}}$ from the cotangent space.  This also implies that $M_{p}$ and $M_{p}^{\: *}$ are subspaces of $M_{p}^{\: \mathbb{C}}$ and $M_{p}^{\: * \mathbb{C}}$ respectively.  

We will now focus on these four vector spaces and study the effects of introducing complex structures on them.  As in the previous section, we assume that our manifold is even-dimensional and in addition to this, we advance our analysis by studying the more restricted case of a complex manifold.

\subsubsection*{Tangent spaces}

The tangent space $M_{p}$ of an even-dimensional differentiable manifold consists of a basis of $n$ tangent vectors $\partial / \partial x^{a}$, defined by its operation on coordinate functions $f(x^{a})$ as $(\partial / \partial x^{a})(f) = \partial f(x^{b}) / \partial x^{a}$. One can write an arbitrary vector (tangent vector) as 

\begin{equation}
X = \zeta^{\alpha} (\partial / \partial x^{\alpha}) + \zeta^{\overline{\alpha}} (\partial / \partial x^{\overline{\alpha}}),
\end{equation} 

which has basis $\{\partial / \partial x^{\alpha}, \partial / \partial x^{\overline{\alpha}} \}$.  We refer to this basis as the real basis.  Furthermore, the components $\{\zeta^{\alpha}, \zeta^{\overline{\alpha}}  \}$ (which are real numbers) are called the ``real components'' with respect to the real basis. The reason for the name will become clear later, when we consider a different type of basis for the case when we look at the complexified tangent space of a complex manifold.  

If our manifold underlying the tangent space is a complex manifold, one can introduce a particular complex structure for the tangent space as shown by the following definition.

\begin{mydef}\label{canonicalcomplexstructure}
	The canonical complex structure for the tangent space $M_{p}$ at a point $p$ of a complex manifold $M$ is the complex structure $J$ given by the operation
	
	\begin{equation}
	J(\partial / \partial x^{\alpha})= \partial / \partial x^{\overline{\alpha}}  \qquad  	J(\partial / \partial x^{\overline{\alpha}})= - \partial / \partial x^{\alpha}, 
	\end{equation}
	
	where \{$x^{\alpha}, x^{\overline{\alpha}}$\}  are the real coordinates of a coordinate patch $(U,x^{\alpha}, x^{\overline{\alpha}} )$ of a complex structure containing $p$.  
	
\end{mydef}

Due to the complex structure of the manifold, the canonical complex structure of the tangent space acting on the basis vectors has the same expression regardless of what chart of the complex structure of the manifold one uses, provided it contains the point $p$.  To be more explicit, if $(U', x^{\alpha '}, x^{\overline{\alpha '}})$ is another chart of the complex structure of the manifold, then

\begin{equation}
J(\partial / \partial x^{\alpha '})= \partial / \partial x^{\overline{\alpha '}},  \qquad  	J(\partial / \partial x^{\overline{\alpha '}})= - \partial / \partial x^{\alpha '}.
\end{equation}

\subsubsection*{Cotangent spaces}

Our attention turns to the cotangent space $M_{p}^{\:\:*}$ of an even-dimensional differentiable manifold.  It has a basis which is dual to $\{\partial / \partial x^{\alpha}, \partial / \partial x^{\overline{\alpha}} \}$, denoted by $\{dx^{\alpha}, dx^{\overline{\alpha}} \}$ and this is also called the real basis, but now with respect to the cotangent space.  An arbitrary cotangent vector can be represented in terms of the real basis as 

\begin{equation}
\omega = \omega_{\alpha}\: dx^{\alpha} + \omega_{\overline{\alpha}}\: dx^{\overline{\alpha}},
\end{equation} 

where the components with respect to the real basis $\{dx^{\alpha}, dx^{\overline{\alpha}} \}$ are referred to as real components.

In the more restricted case of a complex manifold, the canonical complex structure $J$ on $M_{p}$ induces a complex structure $J^{*}$ on $M_{p}^{\:\:*}$.  The operation of such a structure on the real basis of $M_{p}^{\:\:*}$ is given by

\begin{equation}
J^{*}(dx^{\alpha}) = - dx^{\overline{\alpha}}; \qquad J^{*}(dx^{\overline{\alpha}}) = dx^{\alpha};
\end{equation}  

and this expression is maintained regardless of whether one uses another chart for the complex structure.

So far we have been referring to the bases above as real --- the introduction of a complex basis is tied into the following discussion on the complexified tangent space.

\subsubsection*{Complexified Tangent Spaces}

The complexified tangent space $M_{p}^{\: \mathbb{C}}$ at a point on a even-dimensional differentiable manifold contains the set of complexified tangent vectors at that point.  In other words, if  $X \in M_{p}$ is given by real components $\{\zeta^{\alpha}, \zeta^{\overline{\alpha}} \}$ and the real component of a vector $Y \in M_{p}$ is $\{\eta^{\alpha}, \eta^{\overline{\alpha}} \}$, the resulting complexified tangent vector in $M_{p}^{\: \mathbb{C}}$ has the expression

\begin{equation}
X+iY = (\zeta^{\alpha}+i \eta^{\alpha}) \: \partial / \partial x^{\alpha} + (\zeta^{\overline{\alpha}}+i \eta^{\overline{\alpha}}) \: \partial / \partial x^{\overline{\alpha}},
\end{equation}  

where the real components are complex numbers, $\zeta^{\alpha}+i \eta^{\alpha}$ and $\zeta^{\overline{\alpha}}+i \eta^{\overline{\alpha}}$.

If we focus on the complex coordinates of a chart, $z^{\alpha} = x^{\alpha} +ix^{\overline{\alpha}}$ and $\overline{z^{\alpha}} = x^{\alpha} -ix^{\overline{\alpha}}$, one can derive a new basis for the complexified tangent space which we refer to as the complex basis.

\begin{theorem}
	Given the basis $\{\partial / \partial x^{\alpha}, \partial / \partial x^{\overline{\alpha}} \}$ for $M_{p}^{\:\mathbb{C}}$, one can construct a basis called the complex basis, for this vector space given by expression
	
	\begin{equation}
	\partial / \partial z^{\alpha} = \frac{1}{2}(\partial / \partial x^{\alpha} - i \: \partial / \partial x^{\overline{\alpha}}), 
	\end{equation}
	
	\begin{equation}
	\partial / \partial \overline{z^{\alpha}} = \frac{1}{2}(\partial / \partial x^{\alpha} + i \: \partial / \partial x^{\overline{\alpha}}). 
	\end{equation}
	
	(See e.g. Flaherty \cite{flaherty1976hermitian}.)	
\end{theorem}

An arbitrary complexified tangent vector that is represented using the complex basis will have components that we refer to as complex components.

Limiting ourselves now to the case of a complex manifold, we saw in the section on complex linear algebra, one was able to extend the complex structure from a vector space $V$ to $V^{\mathbb{C}}$.  Therefore, it is also possible to extend the canonical complex structure to act on all of the complexified tangent spaces $M_{p}^{\:\:\mathbb{C}}$.  This extension will reveal further properties of a complex manifold as expressed by the following theorem.

\begin{theorem}\label{Extendedcanonicalcomplexstructure}
	In a complex manifold, the action of the canonical complex structure on the complex basis of $M_{p}^{\:\:\mathbb{C}}$ on any chart of the complex structure of the manifold is given by
	
	\begin{equation}
	J(\partial / \partial z^{\alpha}) = i \: (\partial / \partial z^{\alpha}), \qquad J(\partial / \partial \overline{z^{\alpha}}) = -i \: (\partial / \partial \overline{z^{\alpha}}).
	\end{equation}
	
	Conversely, if $J$ has this form with respect to both of two intersecting charts, then the complex coordinate transformations are holomorphic (provided it is real analytic).  (See e.g. Flaherty \cite{flaherty1976hermitian}.)

\end{theorem}

\subsubsection*{Complexified Cotangent spaces}

Finally, for the case of the complexified cotangent space $M_{p}^{*\mathbb{C}}$, an arbitrary element of the space can be represented using the real basis $\{dx^{\alpha}, dx^{\overline{\alpha}} \} $.  If we perform the complex coordinate transformation $z^{\alpha} = x^{\alpha} +ix^{\overline{\alpha}}$ and $\overline{z^{\alpha}} = x^{\alpha} -ix^{\overline{\alpha}}$, one can introduce a new type of basis called the complex basis  $\{dz^{\alpha}, dz^{\overline{\alpha}} \}$.  

  The relationship between the complex basis and real basis of the complexified cotangent space is given by

\begin{equation}
dz^{\alpha} = dx^{\alpha} + i dx^{\overline{\alpha}}, 
\end{equation}  

\begin{equation}
dz^{\overline{\alpha}} = dx^{\alpha} - i dx^{\overline{\alpha}}.
\end{equation}

Therefore, an arbitrary complexified cotangent vector can be expressed in terms of a real basis or a complex basis.  Once again, if one is representing a complexified cotangent vector with respect to the complex basis, the components are referred to as complex components.

\section{Tensors on a Complex Manifold}

Tensors play a crucial role in General Relativity, particularly in representing meaningful physical quantities.  Our mathematical excursion into complex manifolds, has finally brought us into the subject of tensors and tensor fields.  We provide a basic introduction as well as some explicit examples of the interaction of tensors on a complex manifold.  In particular, we shall see that having a complex structure on a manifold makes some transformation laws particularly elegant for objects known as complex tensors.

\subsubsection*{Complex Tensors}

Generalizing from vectors in vector spaces, one can construct multi-linear mappings from the products of the complexified tangent space and complexified cotangent space to the complex numbers  $T: M_{p}^{\:\mathbb{*C}} \times M_{p}^{\:\mathbb{*C}} \times ... \times M_{p}^{\:\mathbb{C}} \times M_{p}^{\:\mathbb{C}} \rightarrow \mathbb{C}$. These are referred to as complex tensors.  Our notation will involve lower case letters for the real components of a tensor and upper case for complex components of a tensor.

An example of a complex tensor which can be represented in terms of the real bases is the complex tensor $T: M_{p}^{\:\mathbb{*C}} \times M_{p}^{\:\mathbb{*C}} \times  M_{p}^{\:\mathbb{C}} \rightarrow \mathbb{C}$,

\begin{align}
\begin{split}
T = t_{\alpha \beta}^{\;\;\;\; \gamma} \; dx^{\alpha} \otimes \; dx^{\beta} \otimes \;  \frac{\partial}{\partial x^{\gamma}} + t_{\overline{\alpha} \beta}^{\;\;\;\; \gamma} \; dx^{\overline{\alpha}} \otimes \; dx^{\beta} \otimes \;  \frac{\partial}{\partial x^{\gamma}} \\
+\; t_{\alpha \overline{\beta}}^{\;\;\;\; \gamma} \; dx^{\alpha} \otimes \; dx^{\overline{\beta}} \otimes \;  \frac{\partial}{\partial x^{\gamma}} +
t_{\alpha \beta}^{\;\;\;\; \overline{\gamma}} \; dx^{\alpha} \otimes \; dx^{\beta} \otimes \;  \frac{\partial}{\partial x^{\overline{\gamma}}}  \\
+ \; t_{\overline{\alpha} \overline{\beta}}^{\;\;\;\; \gamma} \; dx^{\overline{\alpha}} \otimes \; dx^{\overline{\beta}} \otimes \;  \frac{\partial}{\partial x^{\gamma}} +
t_{\overline{\alpha} \beta}^{\;\;\;\; \overline{\gamma}} \; dx^{\overline{\alpha}} \otimes \; dx^{\beta} \otimes \;  \frac{\partial}{\partial x^{\overline{\gamma}}} \\
\; + t_{\alpha \overline{\beta}}^{\;\;\;\; \overline{\gamma}} \; dx^{\alpha} \otimes \; dx^{\overline{\beta}} \otimes \;  \frac{\partial}{\partial x^{\overline{\gamma}}} +
t_{\overline{\alpha} \overline{\beta}}^{\;\;\;\; \overline{\gamma}} \; dx^{\overline{\alpha}} \otimes \; dx^{\overline{\beta}} \otimes \;  \frac{\partial}{\partial x^{\overline{\gamma}}} .
\end{split}
\end{align}

One can also express a complex tensor in terms of the complex basis and the tensor above in complex basis would be explicitly written as

\begin{align}
\begin{split}
T = T_{\alpha \beta}^{\;\;\;\; \gamma} \; dz^{\alpha} \otimes \; dz^{\beta} \otimes \;  \frac{\partial}{\partial z^{\gamma}} + T_{\overline{\alpha} \beta}^{\;\;\;\; \gamma} \; d\overline{z^{\alpha}} \otimes \; dz^{\beta} \otimes \;  \frac{\partial}{\partial z^{\gamma}}  \\
+ \; T_{\alpha \overline{\beta}}^{\;\;\;\; \gamma} \; dz^{\alpha} \otimes \; d\overline{z^{\beta}} \otimes \;  \frac{\partial}{\partial z^{\gamma}} +
T_{\alpha \beta}^{\;\;\;\; \overline{\gamma}} \; dz^{\alpha} \otimes \; dz^{\beta} \otimes \;  \frac{\partial}{\partial \overline{z^{\gamma}}}    \\
+ \; T_{\overline{\alpha} \overline{\beta}}^{\;\;\;\; \gamma} \; d \overline{z^{\alpha}} \otimes \; d \overline{z^{\beta}} \otimes \;  \frac{\partial}{\partial z^{\gamma}} +
T_{\overline{\alpha} \beta}^{\;\;\;\; \overline{\gamma}} \; d \overline{z^{\alpha}} \otimes \; dz^{\beta} \otimes \;  \frac{\partial}{\partial \overline{z^{\gamma}}}  \\
+ \; T_{\alpha \overline{\beta}}^{\;\;\;\; \overline{\gamma}} \; dz^{\alpha} \otimes \; d\overline{z^{\beta}} \otimes \;  \frac{\partial}{\partial \overline{z^{\gamma}}} +
T_{\overline{\alpha} \overline{\beta}}^{\;\;\;\; \overline{\gamma}} \; d \overline{z^{\alpha}} \otimes \; d \overline{z^{\beta}} \otimes \;  \frac{\partial}{\partial \overline{z^{\gamma}}}.
\end{split}
\end{align}

Given that $z^{\alpha} = x^{\alpha} + i\:x^{\overline{\alpha}}$, and taking into account the conjugate $\overline{z^{\alpha}}$, the following formula provides a convenient transformation relationship between the real $(t_{\;\;\;\;c\:d...}^{a\:b...})$ and complex components $(T_{\;\;\;\;c\:d...}^{a\:b...})$ of an arbitrary tensor, 

\begin{equation}\label{tensortransformation}
T_{\;\;\;\;c\:d...}^{a\:b...} = \frac{\partial z^{a}}{\partial x^{e}}\;\frac{\partial z^{b}}{\partial x^{f}}\; ... \; \frac{\partial x^{g}}{\partial z^{c}}\; \frac{\partial x^{h}}{\partial z^{d}}\; ... \; t_{\;\;\;\;g\:h...}^{e\:f...}.
\end{equation}

Note that the Latin letters range from $1,\: ...\:, \:m$ for the index $\alpha$ and range from $m+1,\:...\:,\:m+m=2m$ for $\overline{\alpha}$.   

A calculation that highlights the effectiveness of formula (\ref{tensortransformation}) is shown as follows.  Suppose we have a vector $V$ which has real components $\{\zeta^{\alpha}, \: \zeta^{\overline{\alpha}}\}$ and complex components $\{V^{\alpha}, \: V^{\overline{\alpha}}\}$.  Applying the transformation formula to this vector, one can proceed to find the expression

\begin{equation}
V^{\alpha} = \frac{\partial z^{\alpha}}{\partial x^{\alpha}}\: \zeta^{\alpha} \;+\; \frac{\partial z^{\alpha}}{\partial x^{\overline{\alpha}}}\: \zeta^{\overline{\alpha}} \;=\;\zeta^{\alpha} \;+\; i\:\zeta^{\overline{\alpha}}.
\end{equation}

Similarly, it can be found that $V^{\overline{\alpha}} = \;\zeta^{\alpha} - i\:\zeta^{\overline{\alpha}}$.
   
\subsubsection*{Real Tensors} 

Computationally, real tensors are tensors whose real components are real numbers.  An interesting question we would like to answer is that given a tensor is represented using a complex basis, how can we identify whether or not it is a real tensor?

To start our analysis, consider the simple case of an element of the complexified tangent space $M_{p}^{\;\mathbb{C}}$, 

\begin{equation}
V=\zeta^{\alpha} \frac{\partial}{\partial x^{\alpha}} + \zeta^{\overline{\alpha}} \frac{\partial}{\partial x^{\overline{\alpha}}}.
\end{equation}      

The components are complex numbers, but if we restrict them to be real numbers, i.e. $\zeta^{\alpha} = \overline{\zeta^{\alpha}}$ and $\zeta^{\overline{\alpha}} = \overline{\zeta^{\overline{\alpha}}}$, then elements of such set are part of the subspace, $M_{p}$.  Such elements are called the real vectors, since their real components are real.

If we now represent $V$ in terms of the complex basis, with complex components $\{V^{\alpha}, \:V^{\overline{\alpha}}\}$, and we enforce the condition that $V=\overline{V}$, it produces a real vector.  Explicitly, one finds that for the vector to be a real vector, the complex components have to satisfy $V^{\alpha}=\overline{V^{\overline{\alpha}}}$ and $V^{\overline{\alpha}}=\overline{V^{\alpha}}$. 

We now have two ways to identify whether a vector in the complexified tangent space is a real vector.  Both of these definitions can be shown to be equivalent, and even in the extended case of a real tensor.

It is important to note that a real tensor expressed in a complex basis has to have complex components that satisfy the condition, $\overline{T_{\;\;\;\;c\:d\:...}^{a\:b ...}} = T_{\;\;\;\;\overline{c\:d}\:...}^{\overline{a\:b} ...}$.

\subsubsection*{Type of Tensor}

The section on complex linear algebra refers to certain vectors as type $(1,0)$ and type $(0,1)$ vectors, when acted upon by a complex structure.  We will show that on a complex manifold using a complex basis, the type of a particular vector will have very simple expressions. 

On a complex manifold, given the canonical complex structure $J$, one can see that the vector $V=\zeta^{\alpha} \frac{\partial}{\partial x^{\alpha}} - i\: \zeta^{{\alpha}} \frac{\partial}{\partial x^{\overline{\alpha}}}$ satisfies $J(V)=i\:V$, thereby making it a type $(1,0)$ vector.  Thus we have the expression that $\zeta^{\overline{\alpha}} = -i\:\zeta^{\alpha}$ which eventually leads us to the fact that the complex components of the vector, $\{V^{\alpha}, V^{\overline{\alpha}}\}$, require that $V^{\overline{\alpha}}=0$.  Thus type $(1,0)$ vectors in a complex basis are all of the form  $V=V^{\alpha}\:\partial /\partial z^{\alpha}$.  A type $(0,1)$ vector with respect to the canonical complex structure is expressed as $V=V^{\overline{\alpha}}\:\partial /\partial \overline{z^{\alpha}}$.

In addition to this, simpler expressions in terms of the complex basis also exist for type $(1,0)$ forms and type $(0,1)$ forms.  These are respectively $\omega = \omega_{\alpha} \:dz^{\alpha}$ and $\omega = \omega_{\overline{\alpha}} \:d \overline{z^{\alpha}}$.

The generalization to tensors exist in the sense that we can speak of contravariant type $(p, q)$ and covariant type $(r, s)$ tensors.  An example is given by a tensor that is of covariant type $(0,2)$ and it is expressed as $T\:=\:T_{\overline{\alpha \beta}} \; d \overline{z^{\alpha}} \otimes \; d \overline{z^{\beta}}$.

\subsubsection*{Complex Tensors on Complex Manifolds}

The highlight of this section is the following transformation law which utilizes the structure of a complex manifold, in particular the fact that the Cauchy-Riemann equations hold in the intersection of two charts.  This transformation law of tensors with respect to their complex components has the formula

\begin{equation}
T_{\;\;\;\;\eta '\:...\:\overline{\zeta} '...}^{\alpha ' ...\:\overline{\beta '}...} = 
\frac{\partial z^{\alpha '}}{\partial z^{\mu}}\;...\:
\overline{\Big(\frac{\partial z^{\beta '}}{\partial z^{\phi}}\Big)}\;...\:
\frac{\partial z^{\omega}}{\partial z^{\eta '}}\;...\:
\overline{\Big(\frac{\partial z^{\theta}}{\partial z^{\zeta '}}\Big)}\;...\: T_{\;\;\;\;\omega \:...\:\overline{\theta} ...}^{\mu  ...\:\overline{\phi }...}
\end{equation}

An application of this transformation law can be applied to a vector which has the expression $V^{\alpha} = V^{\alpha}\: \partial /\partial z^{\alpha} + V^{\overline{\alpha}}\: \partial /\partial \overline{z^{\alpha}}$ according to a chart of the manifold and the expression $V^{\alpha '} = V^{\alpha '}\: \partial /\partial z^{\alpha '} + V^{\overline{\alpha '}}\: \partial /\partial \overline{z^{\alpha '}}$ of another intersecting chart.  Using the transformation law, one can find that in the intersection of the two charts, we have the transformations for complex components as

\begin{equation}
V^{\alpha '} = \frac{\partial z^{\alpha '}}{\partial z^{\beta}} V^{\beta}, \qquad V^{\overline{\alpha '}} = \overline{\Big(\frac{\partial z^{\alpha '}}{\partial z^{\beta}}\Big)} V^{\overline{\beta}}.
\end{equation}

If the underlying manifold did not satisfy the requirement of a complex manifold, then the transformation law would not have such a simple expression.

\subsubsection*{Tensor Fields and Analytic Continuation}

For a general background on tensor bundles and the rigorous construction of tensor fields, refer to Lee \cite{lee2006riemannian}. A crucial point is that for the case of tensor fields, the components of a tensor now become functions of points on the manifold.  

We will now discuss the concept of analytic continuation of tensor fields on a complex manifold which creates a new tensor field residing on the product manifold.

The general procedure for analytic continuation of tensor fields is that we replace the complex component functions with their analytic continuation.  In addition to this, the basis vectors $\partial /\partial \overline{z^{\alpha}}$ and $d\overline{z^{\alpha}}$ are replaced by $\partial /\partial z^{\overline{\alpha}}$ and $d{z^{\overline{\alpha}}}$.

As an example consider a vector field on a complex manifold, $M$, expressed as

\begin{equation}
V = V^{\alpha} (z^{\beta}, \overline{z^{\beta}}) \; \partial /\partial {z^{\alpha}} + V^{\overline{\alpha}} (z^{\beta}, \overline{z^{\beta}}) \; \partial /\partial \overline{{z^{\alpha}}}.
\end{equation}

The analytic continuation of such a vector field is a new vector field that is defined on the product manifold given by

\begin{equation}
Y = Y^{\alpha} (z^{\beta}, {z^{\overline{\beta}}}) \; \partial /\partial {z^{\alpha}} + Y^{\overline{\alpha}} (z^{\beta}, {z^{\overline{\beta}}}) \; \partial /\partial {{z^{\overline{\alpha}}}},
\end{equation}

where the functions $Y^{\alpha}$ and  $Y^{\overline{\alpha}}$ are respectively the analytic continuations of $V^{\alpha}$ and $V^{\overline{\alpha}}$.

\section{Almost Complex Manifolds}

	Almost complex manifolds are objects in the mathematical universe that are studied for their own elegant properties.  But within the context of this thesis, we concentrate on their applicability and their computational properties in answering a significant question about complex manifolds. That is, how can we tell if a manifold admits a complex structure?  
	
	To start exploring this question, one needs to first set up a few definitions regarding these new types of manifolds.

\begin{mydef}
	An almost complex structure on $M$ is a real differentiable tensor field $J$ of rank $(1,1)$ with the property 
	\begin{equation}
	J(J(V))=-V
	\end{equation}
	for any differentiable vector field $V$.
\end{mydef}

In other words the real components of this tensor $j_{a}^{\;\;b}$ satisfies $j_{a}^{\;\;s}\:j_{s}^{\;\;b}= -\delta_{a}^{\;\;b}$. 

\begin{mydef}
	A manifold which admits an almost complex structure is called an almost complex manifold.
\end{mydef} 

\begin{theorem}
	A manifold $M$ with an almost complex structure $J$ is even-dimensional and orientable. (See e.g. Flaherty \cite{flaherty1976hermitian}.)
\end{theorem}  
  
The proof of the theorem requires an equivalent definition of an almost complex structure on a manifold, which is that of a differentiable field of linear maps $J_{p}\: :\: M_{p} \rightarrow M_{p}$ on each tangent space $M_{p}$ such that $J_{p}(J_{p}(\eta_{p}))=-\eta_{p}$ for all $\eta_{p} \in M_{p}$. Thus the tensor field $J$ can be used to construct a complex structure on vector spaces $M_{p}$ for each $p \in M$.

An example of an almost complex manifold is $\mathbb{R}^{4}$ where one can cover the manifold with standard coordinates $\{x^{1}, x^{2}, x^{3}, x^{4}\}$, and the almost complex structure is represented using the tensor

\begin{equation}
j_{a}^{\;\;b}\;=\;
  \begin{bmatrix}
  	0 & 1 & 0    & 0 \\
  	-1 & 0 & 0  & 0 \\
  	0 & 0 & 0  & 1 \\
  	0 & 0 & -1  & 0
  \end{bmatrix}.
  \end{equation}

Concentrating our focus back towards complex manifolds, we shall see that the relationship between complex manifolds and almost complex manifolds is that the former is a subset of the latter.

\begin{theorem}
	A complex manifold admits an almost complex structure.  (See e.g. Flaherty \cite{flaherty1976hermitian}.)
\end{theorem}

In other words, a complex manifold is an almost complex manifold.  

A sketch of the proof is that a complex manifold admits the canonical complex structure on $M_{p}$ for all $p \in M$ (See Definition \ref*{canonicalcomplexstructure}). This canonical complex structure fits the requirement of what is required to be an almost complex structure.  Thereby, allowing a complex manifold to be expressed as an almost complex manifold.

Equivalently one can use the extended canonical complex structure (See Theorem \ref{Extendedcanonicalcomplexstructure}) and conclude that it also fits the requirement of an almost complex structure.  The complex components of that tensor are given by

\begin{equation}
J_{a}^{\;\;b}\;=\;
\begin{bmatrix}
i \delta_{\alpha}^{\;\;\beta}    & 0 \\
 0  & -i \delta_{\overline{\alpha}}^{\;\;\overline{\beta}} 
\end{bmatrix},
\end{equation}

regardless of the choice of chart of the complex structure.    

Through the lens of almost complex manifolds, a canonical complex structure of a complex manifold is known as an integrable almost complex structure.  One can say that such an almost complex structure is integrable or was induced by an underlying complex structure.  

Every complex manifold admits an almost complex structure but not every almost complex structure is induced.  Therefore only a subset of almost complex manifolds are complex manifolds.  

Let us return to the main question of how to determine if a manifold has a complex structure?  We have seen that one can set up an almost complex structure and we would like to check if it's an integrable structure.  If it is, then we have the canonical complex structure for a complex manifold and thereby, allowing us to put complex coordinates on the manifold and treat is as a complex manifold. 

To see an algorithmic process for determining the integrability of an almost complex structure i.e. if an almost complex manifold is a complex manifold, requires the following definition.

\begin{mydef}
	The Nijenhuis tensor $N$ of a manifold \cite{nijenhuis} with an almost complex structure $J$ is the tensor whose real components are given by 
	\begin{equation}
	n_{a b}^{\;\;\;\;c} = j_{a}^{\;\;s}\;(j_{b \;\; ,\:s}^{\;\;c} - j_{s \;\; ,\:b}^{\;\;c}) - j_{b}^{\;\;s}\;(j_{a \;\; ,\:s}^{\;\;c} - j_{s \;\; ,\:a}^{\;\;c}),
	\end{equation}
	
	where $j_{a}^{\;\; b}$ are the real components of $J$ and the commas represent partial derivatives with respect to coordinates.
	
\end{mydef}

We will find that the vanishing of the Nijenhuis tensor allows one to determine if the almost complex manifold admits a complex structure.  The following theorems are known as the integrability theorems.

\begin{theorem}
	In order for an almost complex structure $J$ to be integrable it is necessary that $n_{a b}^{\;\;\;\;c} = 0$. (See e.g. Flaherty \cite{flaherty1976hermitian}.)
\end{theorem}

The above theorem is not particularly useful since we would like the converse statement, thereby producing an algorithmic procedure for determining an integrable almost complex structure.

\begin{theorem}
	If (and only if) $J$ is a real analytic almost complex structure where $n_{a b}^{\;\;\;\;c} = 0$, then $J$ is integrable. (See e.g. Flaherty \cite{flaherty1976hermitian}.)
\end{theorem}

This theorem provides one with a simple yet powerful algorithm to determine if the underlying manifold is a complex manifold.  But it still rests on the assumption of real analyticity of $J$, though the following theorem weakens that condition.

\begin{theorem}
	\textbf{(Newlander -- Nirenberg)}  If $M$ is a $2m$-dimensional manifold of differentiability class $C^{2m+1}$ which admits an almost complex structure $J$ of class $C^{2m}$, then $J$ is induced by a complex structure on $M$ if and only $n_{a b}^{\;\;\;\;c} = 0$.
\end{theorem}

From the discussion in this section, it can be said that an integrable almost complex manifold is completely equivalent to a complex manifold. 

The question of whether a manifold admits a complex structure can be somewhat answered for the cases where one is able to construct almost complex structures for a manifold.  

An important point to make is that one can construct different almost complex structures for a manifold, where one is integrable and the other is not.  For example it is an open problem whether the six-sphere admits a complex structure \cite{bryant2014s}.  The currently known almost complex structure is not integrable but there could exist another almost complex structure that is.

\section{Hermitian Manifolds}

This central idea of this section involves introducing a specific type of complex manifold called a Hermitian manifold.  The definition of such an object requires the mathematical necessity of a Riemannian metric, as opposed to a Lorentzian metric that we typically encounter in General Relativity.  Hence this theory will be studied from a mathematical perspective as per the theme of this chapter, but the necessarily modifications for relativity will be discussed in the next chapter.

\subsubsection*{Hermitian Structures on Vector Spaces}

We start with the notion of Hermitian structures on real vector spaces as this will be applicable for the later case of tangent space of a manifold.

\begin{mydef}
	A Hermitian structure on a real vector space $V$ with a complex structure $J$ is a map $H: V \times V \rightarrow \mathbb{C}$ with the properties
	
	(i) \; $H(\alpha \: X_{1} + \beta \: X_{2},\; Y) = \alpha \: H(X_{1}, Y) + \beta \: H(X_{2}, Y)$, 
	
	(ii)\; $\overline{H(X, Y)} = H(Y, X)$, 
	
	(iii) \; $H(J(X),\: Y) = i\; H(X, Y)$,   
	
	for all $\alpha, \: \beta \: \in \mathbb{R}$ and $X_{1}, X_{2}, X, Y \in V$.
	
\end{mydef}

The Hermitian structure can be decomposed in terms of its real and imaginary parts, $H(X, Y) = F(X, Y) +i \: G(X,Y)$.  It can be shown that the imaginary part has antisymmetric properties, $G(X, Y) = -\:G(Y, X)$ and this leads to the following definition.  

\begin{mydef}
	The K\"{a}hler form, $K$,  of a Hermitian structure $H$ is the two-form given by $K=-\frac{1}{2} G(X, Y)$.
\end{mydef}

\subsubsection*{Almost Hermitian Manifolds}

We shall see that in the following the construction of an almost Hermitian manifold has a close relationship to a Hermitian structure on vector spaces.

\begin{mydef}
	Suppose we have a manifold $M$ with a Riemannian metric $g$ and an almost complex structure $J$.  Then $M$ is called an \textbf{almost Hermitian manifold} (manifold which admits an almost Hermitian structure) if and only if $g(JX, JY) = g(X, Y)$ for any vectors $X$ and $Y$. 
\end{mydef}

In this context, $g$ is called the Hermitian metric.  In terms of components the equation $g(JX, JY) = g(X, Y)$ can be expressed as $J_{a}^{\;\;m}\:J_{b}^{\;\;n}\:g_{m\:n} = g_{a\:b}$ and moreover, one can construct the almost Hermitian structure tensor, given by $H_{a\:b} = g_{a\:b} - i\: J_{a\:b}$.

As we shall see in the next subsection when one moves from an underlying almost complex manifold to an underlying complex manifold, the Hermitian structure tensor will have the property of a Hermitian matrix.

\begin{theorem}
	The tangent space over a point of an almost Hermitian manifold admits a vector space Hermitian structure, $H_{a\:b} = g_{a\:b} - i\: J_{a\:b}$. (See e.g. Flaherty \cite{flaherty1976hermitian}.)
\end{theorem}

Also important to note is that the K\"{a}hler form, $K$ is given by $K = \frac{i}{2}\: \: J_{a\:b} \; dx^{a}  \wedge dx^{b}$.

\subsubsection*{Hermitian Manifolds}

The central objects of this section, and an important subset of almost Hermitian manifolds, are the Hermitian manifolds.

\begin{mydef}
	An Hermitian manifold (manifold which admits an Hermitian structure) is an almost Hermitian manifold for which the almost complex structure tensor is integrable.  In this case the tensor $H_{a\:b}$ is called the Hermitian structure tensor.
\end{mydef}

The implications of this can be expressed in the following theorems.

\begin{theorem}
	A complex manifold with complex structure $J$ is Hermitian if and only if it admits an almost Hermitian structure i.e. the metric tensor that satisfies the condition $g(JX, JY)=g(X,Y)$. (See e.g. Flaherty \cite{flaherty1976hermitian}.)
\end{theorem}

\begin{theorem}
	Every complex manifold admits a Hermitian structure. (See e.g. Morrow \cite{morrow1971complex}.)
\end{theorem}

In the following, we shall see that in the case of a Hermitian manifold, the coordinate expressions for common objects are much simpler than the general case of a complex manifold, and that the differential geometry can be seen to be quite distinctive from standard Riemannian geometry. 

For example, if $z^{\alpha}$ and $\overline{z^{\beta}}$ are complex coordinates, the metric for a Hermitian manifold can be written as 

\begin{equation}
ds^{2} = 2 \; g_{\alpha \: \overline{\beta}}\; dz^{\alpha}\:\overline{dz^{\beta}}.
\end{equation}

Delving into the details for the derivation of such an expression, we find the following theorem.

\begin{theorem}
	If $z^{\alpha}$ and $\overline{z^{\beta}}$ are complex coordinates on an Hermitian manifold then, 
	
	(i)  $\; g_{\alpha \; \beta} = g_{\overline{\alpha \; \beta}} = g^{\alpha \; \beta} = g^{\overline{\alpha \; \beta}}  = 0   $, 
	
	(ii)  $\; J_{\alpha \; \beta} = J_{\overline{\alpha \; \beta}} = J^{\alpha \; \beta} = J^{\overline{\alpha \; \beta}}  = 0   $,
	
	(iii) $\; J_{\alpha \; \overline{\beta}} = i \; g_{\alpha \; \overline{\beta}}$, 
	
	(iv) $\; J_{\overline{\alpha} \; {\beta}} = -i \; g_{\overline{\alpha} \; {\beta}}$,
	
	(v) $J^{\alpha \; \overline{\beta}} = i\; g^{\alpha \; \overline{\beta}}$,
	
	(vi) $J^{\overline{\alpha} \; {\beta}} = -i\; g^{\overline{\alpha} \; {\beta}}$.
	
	(See e.g. Flaherty \cite{flaherty1976hermitian}.)
	
\end{theorem}

Turning our attention now to the curvature of Hermitian manifolds, one would like to construct an appropriate affine connection.  This construction, which is not identical to the Riemannian connection, is motivated by the fact that the Hermitian connection acting on the Hermitian metric produces a vanishing result.  We start by defining the notion of covariant differentiation on a Hermitian manifold.  

\begin{mydef}
	On a Hermitian manifold with metric $g_{\alpha \; \overline{\beta}}$, the Hermitian covariant derivative $\mathscr{D}$ is defined by
	
	$\mathscr{D}_{\lambda} \; T_{\;\;\; \eta ... \overline{\psi} ... }^{\alpha ... \overline{\beta} ... } = (g^{\overline{\mu} \: \alpha} ... g_{\overline{\nu} \:\eta}...) \; \partial_{\lambda} \;
	(T_{\;\;\; \sigma ... \overline{\psi} ... }^{\rho ... \overline{\beta} ... } \; g_{\rho\:\overline{\mu}} ... g^{\sigma\: \overline{\nu}})$, 
	
	and
	
		$\mathscr{D}_{\overline{\lambda}} \; T_{\;\;\; \eta ... \overline{\psi} ... }^{\alpha ... \overline{\beta} ... } = (g^{\overline{\beta} \: \mu} ... g_{{\nu} \:\overline{\psi}}...) \; \partial_{\overline{\lambda}} \;
		(T_{\;\;\; \eta ... \overline{\sigma} ... }^{\alpha ... \overline{\rho} ... } \; g_{\overline{\rho}\:{\mu}} ... g^{\overline{\sigma}\: {\nu}})$.
	
\end{mydef} 

Just as we have in some sense captured the idea of a Hermitian covariant derivatives, we will also be able to identify what we can regard as the ``Hermitian Christoffel symbols" as presented in the following theorem.  

\begin{theorem}
	
	\begin{equation}
	\mathscr{D}_{\lambda} \; T_{\;\;\; \eta ... \overline{\psi} ... }^{\alpha ... \overline{\beta} ... } = \partial_{\lambda} \: T_{\;\;\; \eta ... \overline{\psi} ... }^{\alpha ... \overline{\beta} ... } 
	+ \Theta_{\lambda \; \rho}^{\alpha} \; 
	T_{\;\;\; \eta ... \overline{\psi} ... }^{\rho ... \overline{\beta} ... } +\; ... \; - 
	\Theta_{\lambda \: \eta}^{\sigma}\; 
	T_{\;\;\; \sigma ... \overline{\psi} ... }^{\alpha ... \overline{\beta} ... } \; - ... ,
	\end{equation}
	
	and 
	
	\begin{equation}
	\mathscr{D}_{\overline{\lambda}} \; T_{\;\;\; \eta ... \overline{\psi} ... }^{\alpha ... \overline{\beta} ... } = \partial_{\overline{\lambda}} \: T_{\;\;\; \eta ... \overline{\psi} ... }^{\alpha ... \overline{\beta} ... }
	+ \Theta_{\overline{\lambda \; \rho}}^{\overline{\beta}} \; 
	T_{\;\;\; \eta ... \overline{\psi} ... }^{\alpha ... \overline{\rho} ... } +\; ... \; - 
	\Theta_{\overline{\lambda \: \psi}}^{\overline{\sigma}}\; 
	T_{\;\;\; \eta ... \overline{\sigma} ... }^{\alpha ... \overline{\beta} ... } \; - ... ,
	\end{equation}
	
	where $\; \Theta_{\lambda \: \rho}^{\alpha} = g^{\alpha \overline{\mu}}\:\partial_{\lambda}\:g_{\rho\:\overline{\mu}}\;$, and 
	$\; \Theta_{\overline{\lambda \: \rho}}^{\overline{\beta}} = g^{\overline{\beta} {\mu}}\:\partial_{\overline{\lambda}}\:g_{\overline{\rho}\:{\mu}} = \overline{\Theta_{{\lambda \: \rho}}^{{\beta}}} \;$. 
	
	(See e.g. Flaherty \cite{flaherty1976hermitian}.)
	
\end{theorem}

These Hermitian Christoffel symbols, $\Theta_{b\:c}^{a}$ are not symmetric with respect to their covariant components, hence the covariant derivative $\mathscr{D}$ is not in general torison-free.  We will see in certain cases that there exists a relationship to their respective Riemannian counterparts, namely the Christoffel symbols $\Gamma_{b\:c}^{a}$ and the Riemannian covariant derivative operator $\nabla_{l}$.

The construction of the Hermitian connection is also largely motivated by the following theorem.

\begin{theorem}
	\begin{equation}
	\mathscr{D}_{\alpha} \: g_{\beta\: \overline{\eta}} = 
	\mathscr{D}_{\overline{\alpha}} \: g_{\beta\: \overline{\eta}} = 0.
	\end{equation}
	(See e.g. Flaherty \cite{flaherty1976hermitian}.)
\end{theorem}

\section{K\"{a}hler Manifolds}

We move onto our final type of manifold that we study in this chapter which is known as a K\"{a}hler manifold.  An important example of such a manifold is the complex projective space, $\mathbb{C}P^{n}$.  But not all complex manifolds are K\"{a}hler and they are only a subset of the set of Hermitian manifolds, thus they are a very restrictive case.  The beauty of these mathematical objects is their astounding simplicity in calculating differential geometric quantities and their applications to different types of proposed theories for quantum gravity \cite{penrose2006road}.

\subsubsection*{Almost K\"{a}hler manifolds and K\"{a}hler manifolds}

We start by defining an almost K\"{a}hler manifold and then use that definition in our construction of a K\"{a}hler manifold.

\begin{mydef}
	An almost K\"{a}hler manifold (manifold which admits an almost K\"{a}hler structure) is an almost Hermitian manifold for which the almost complex structure tensor $J_{a}^{\;\;b}$ satisfies
	
	\begin{equation}
	d(J_{ab} \: dx^{a} \wedge dx^{b}) = 0,
	\end{equation}
	
	where ``d" denotes the exterior derivative.
\end{mydef}

Using the K\"{a}hler form, $K$, of the almost Hermitian manifold, one can rewrite the above definition of an almost K\"{a}hler manifold as an almost Hermitian manifold for which $d(K)=0$.

\begin{mydef}
	A K\"{a}hler manfiold (manifold which admits a K\"{a}hler structure) is an almost K\"{a}hler manifold whose almost complex structure tensor $J$ is integrable.
\end{mydef} 

We call the metric in this context, a K\"{a}hler metric.

\begin{theorem}
	A K\"{a}hler manifold is an Hermitian manifold. (See e.g. Flaherty \cite{flaherty1976hermitian}.) 
\end{theorem}

\subsubsection*{Relationship to Hermitian Geometry}

The differential geometric aspects of Hermitian manifolds can also play an important role in determining if a manifold is a K\"{a}hler manifold.  This can be highlighted in the following theorems.

\begin{theorem}
	A Hermitian manifold is K\"{a}hler if and only if 
	\begin{equation}
	\Theta_{\beta\:\eta} ^{\alpha} = \Theta_{\eta\:\beta}^{\alpha} \qquad \text{and} \qquad \Theta_{\overline{\beta\:\eta}} ^{\overline{\alpha}} = \Theta_{\overline{\eta\:\beta}}^{\overline{\alpha}}.  
	\end{equation}
	(See e.g. Flaherty \cite{flaherty1976hermitian}.)
\end{theorem}

\begin{corollary}
	The Hermitian connection $\Theta$ coincides with the Riemannian connection $\Gamma$ if and only if the metric is K\"{a}hler.
\end{corollary}

\begin{theorem}
	In a K\"{a}hler manifold, we have 
	\begin{equation}
	\Theta_{\alpha\:\beta} ^{\alpha} = g^{-1}\:\partial_{\beta}\: g \qquad \text{and} \qquad \Theta_{\overline{\alpha\:\beta}} ^{\overline{\alpha}} = g^{-1}\:\partial_{\overline{\beta}}\:g, 
	\end{equation}
	where $g\: = \: det(g_{\alpha \overline{\beta}})$. (See e.g. Flaherty \cite{flaherty1976hermitian}.)
\end{theorem}

\begin{theorem}
	For an Hermitian manifold, $g_{a\: b} - i\: J_{a\: b}$ is K\"{a}hler if and only if $\nabla_{c}\:J_{a\: b} = 0$, where $\nabla_{c}$ is the Riemannian covariant derivative. (See e.g. Flaherty \cite{flaherty1976hermitian}.)
\end{theorem}

\subsubsection*{Curvature of K\"{a}hler manifolds}

Another power of endowing the K\"{a}hlerian structure to a manifold lies in its simple formulae of geometric quantities.  These are reflected in the following theorems and explicitly show how the Riemann tensor and Ricci tensor are expressed in terms of just scalar functions.

\begin{theorem} \label{kahlerscalar}
	On a K\"{a}hler manifold, $g_{\alpha\:\overline{\beta}}$ is locally expressible as $g_{\alpha\:\overline{\beta}} = \partial_{\alpha}\: \partial_{\overline{\beta}} K$, where $K$ is a real scalar function. (See e.g. Flaherty \cite{flaherty1976hermitian}.)
\end{theorem}

	The following theorem expresses an explicit formula for the Riemann curvature tensor for a K\"{a}hler manifold.

\begin{theorem}
	On a K\"{a}hler manifold, all components are zero except 
	
	\begin{equation}
	R_{\;\;\beta\:\overline{\mu}\psi}^{\alpha} = \partial_{\overline{\mu}}\;\Theta_{\psi\:\beta}^{\alpha}, \quad 
	R_{\;\;\overline{\beta\:{\mu}\psi}}^{\overline{\alpha}} = -\partial_{{\psi}}\;\Theta_{\overline{\mu\:\beta}}^{\overline{\alpha}},\quad 
	\end{equation}
	
	\begin{equation}
		R_{\;\;\beta\:{\mu}\:\overline{\psi}}^{\alpha} = -\partial_{\overline{\psi}}\;\Theta_{{\mu\:\beta}}^{{\alpha}}, \quad
		R_{\;\;\overline{\beta}\:{\mu}\overline{\psi}}^{\overline{\alpha}} = \partial_{\overline{\mu}}\;\Theta_{\overline{\psi\:\beta}}^{\overline{\alpha}}
	\end{equation}

	and the ones calculated taking into account symmetry.  (See e.g. Flaherty \cite{flaherty1976hermitian}.)
\end{theorem}

\begin{theorem}
	On a K\"{a}hler manifold, 
	\begin{equation}
	R_{\alpha \: \beta \: c \: d} = R_{\overline{\alpha \: \beta} \: c \: d} = 0, \qquad 
	R_{a \: b \: \psi \: \mu} = R_{a \: b \: \overline{\psi \: \mu}} = 0,
	\end{equation}
	
	\begin{equation}
	R_{\overline{\alpha} \:\beta \: \overline{\psi} \:\mu} = g_{\overline{\alpha} \eta}\: \partial_{\overline{\psi}} \;\Theta_{\mu \beta}^{\:\eta}, \qquad
	R_{\alpha \:\overline{\beta} \: {\psi} \:\overline{\mu}} = 
	g_{{\alpha} \overline{\eta}}\: \partial_{{\psi}} \;\Theta_{\overline{\mu \: \beta}}^{\:\overline{\eta}}.
	\end{equation}
	
	(See e.g. Flaherty \cite{flaherty1976hermitian}.)
	
\end{theorem}

\begin{theorem}
	On a K\"{a}hler manifold, 
	
	\begin{equation}
	R_{\overline{\alpha} \beta \overline{\psi} \mu} = 
	\partial_{\overline{\alpha}}\: \partial_{\beta} \: \partial_{\overline{\psi}} \: \partial_{\mu} K \: - \: 
	g^{\overline{\eta} \sigma} (\partial_{\overline{\eta}} \: \partial_{\beta}\: \partial_{\mu} K) \:(\partial_{\sigma}\: \partial_{\overline{\alpha}} \:\partial_{\overline{\psi}}\: K), 
	\end{equation}
	
	where $g_{\alpha \overline{\beta}} = \partial_{\alpha}\: \partial_{\overline{\beta}} \: K$, as shown in Theorem \ref{kahlerscalar}. (See e.g. Flaherty \cite{flaherty1976hermitian}.)
	 
\end{theorem}

Finally we turn to the Ricci tensor which has an extremely simple representation with respect to a K\"{a}hler manifold.

\begin{theorem}
	On a K\"{a}hler manifold, 
	\begin{equation}
	R_{\alpha \beta} = R_{\overline{\alpha \beta}} = 0, \qquad R_{\alpha \overline{\beta}} = \partial_{\alpha} \: \partial_{\overline{\beta}} (\text{ln} \: g),
	\end{equation}
	
	where $g = \text{det}\:(g_{\alpha \overline{\beta}})$.
\end{theorem}

\section{Discussion}

In this chapter, we have outlined the basics and key aspects of complex manifold theory.  

We have seen that K\"{a}hler manifolds are subsets of Hermitian manifolds, which are in turn subsets of complex manifolds.  And these are subsets of almost complex manifolds.  

We have also seen that the integrability conditions are a powerful calculational tool to understand the underlying properties of a manifold from an almost complex structure.  This is particularly remarkable given the simple condition of a vanishing Nijenhuis tensor.

Lastly, an important aspect of this chapter was highlighting the elegant geometric formulae of K\"{a}hler manifolds.  

As this thesis moves to the next chapter, an important question one can ask is: how are these results applicable to General Relativity?  The condition of a Riemannian metric in defining Hermitian and K\"{a}hler manifolds do provide an obstacle in the path to answer this question, since in relativity one considers primarily Lorentzian manifolds.

\chapter{Complex Spacetimes}

\begin{chapquote}{Roger Penrose, \textit{The Complex Geometry of the Natural World}}
	``Moreover, the macroscopic geometry of relativity has many special features about it that are suggestive of a hidden complex-manifold origin, and of certain deep underlying physical connections between the normal spatio-temporal relations between things and the complex linear superposition of quantum mechanics.''
\end{chapquote}

In the previous section, a brief summary was presented on the differential geometry of complex manifolds.  Much of the theoretical underpinning of that piece of mathematics assumes a Riemannian signature metric.  In fact the definition of Hermitian and K\"{a}hler manifolds demands the metric signature to be Riemannian.

In marrying General Relativity with complex variables, this presents an obstacle owing to the fact that spacetime is modelled with a Lorentzian signature metric, as opposed to a Riemannian one.

Consequently, we outline two approaches to overcome this obstacle, the first being that one can modify the relevant metrics of General Relativity to become Euclidean signature.  From this, the conversion to a complex manifold can be done with some straightforward calculations, without changing the underlying theory.  In some sense, this has been motivated by themes in Euclidean gravity and similar approaches \cite{gibbons1993euclidean, ortin2004gravity}. 

The second approach is to modify complex manifold theory to fit into the framework of Lorentzian signature manifolds.  The coverage of the latter subject, in this thesis is largely due to Flaherty \cite{flaherty1976hermitian}, but there exist other approaches to combining Lorentzian signature metrics with complex manifolds \cite{duggal1986cr, robinson2002holomorphic}.

\section{Modifying Lorentzian signature metrics}

In this section, we present some results that I obtained (in collaboration with my supervisor) regarding modifying the Lorentzian signature metrics from General Relativity to fit the requirements of a complex Hermitian manifold.  The examples will feature the Schwarzschild metric and general static spherically symmetric metrics.   In this section, we employ the Lorentzian signature $(-+++)$.

\subsubsection*{Modifying Schwarzschild spacetimes}

We now show how the Lorentzian Schwarzschild metric can be modified and rewritten as a complex Hermitian metric.

First, one starts by writing the Schwarzschild metric in the Kruskal-Szekers form

\begin{equation}
ds^{2} = \frac{32m^{3}}{r} e^{-r/2m} (-dT^{2} + dX^{2}) +r^{2} d\Omega^{2}; \qquad r = 2m\Bigl(1+W\Bigl(\frac{X^{2}-T^{2}}{e}\Bigl)\Bigl),
\end{equation} 

where $d\Omega^{2}=d\theta^{2} + \text{sin}^{2}\theta \:d\phi^{2}$ and $W(x)$ is the Lambert W-function.

From this form of the metric, one can perform a Wick rotation, which transforms the Lorentzian metric into a Euclidean metric by a coordinate transformation of the form $T \rightarrow iT$.  This procedure has been used extensively for the flat space case in quantum field theory \cite{zee2010quantum}, as  well as being extensively investigated within the context of curved spacetime \cite{mattvisserwick}.  The resulting metric now takes the Euclidean form

\begin{equation}
ds^{2} = \frac{32m^{3}}{r} e^{-r/2m} (dT^{2} + dX^{2}) +r^{2} d\Omega^{2}; \qquad r = 2m\Bigl(1+W\Bigl(\frac{X^{2}+T^{2}}{e}\Bigl)\Bigl),
\end{equation}

which is also Ricci-flat (i.e. Ricci tensor is zero) with four real dimensions.  To introduce complex coordinates onto this modified metric, we define $z = X+iT$, and thereby set

\begin{equation}
ds^{2} = \frac{32m^{3}}{r} e^{-r/2m} \; dz \: d\overline{z} +r^{2} d\Omega^{2}; \qquad r = 2m\Bigl(1+W\Bigl(\frac{z \overline{z}}{e}\Bigl)\Bigl).
\end{equation}

Introducing coordinate $w$ as a complex stereographic projection \cite{o2003introduction} on the unit sphere, results in 

\begin{equation}\label{stereographic}
d\Omega^{2} = d\theta^{2} + \text{sin}^{2}\theta \:d\phi^{2} = \frac{dw \: d\overline{w}}{(1 + \frac{1}{4} w \:\overline{w})^{2}}.
\end{equation}

Substituting this all into one expression, it follows that the metric can be recast as a complex Hermitian metric which is

\begin{equation}
ds^{2} = \frac{16\:m^{2}\:W(z \: \overline{z}/e)}{z \: \overline{z}(1 + W(z \: \overline{z}/e))} \;\: dz \: d\overline{z} + \frac{4\: m^{2} (1+W(z \: \overline{z}/e))^{2}}{(1 + \frac{1}{4} w \: \overline{w})^{2}} \;\: dw \: d\overline{w}.
\end{equation}

This concludes a calculation in which we have shown that a Euclidean Schwarzschild metric can be written as a complex Hermitian metric.

\subsubsection*{Modifying static spherically symmetric spacetimes: \; Route 1}

We now show that any static spherically symmetric Euclideanized spacetime can be converted to a 2-complex dimensional manifold with a Hermitian metric.  A number of different approaches will be discussed.

One can start the derivation from the general form of a static spherically symmetric metric, given by

\begin{equation} \label{staticspherical}
ds^{2} = -e^{-2 \phi(r)}\: (1 - 2m(r)/r)\: dt^{2} + \frac{dr^{2}}{1-2m(r)/r} + r^{2}\: d\Omega^{2}.
\end{equation}

Utilizing the tortoise coordinate, 

\begin{equation}
r_{*} = \int_{0}^{r} \frac{dr}{e^{\phi(r)}(1 - 2m(r)/r)}; \qquad dr_{*} = \frac{dr}{e^{\phi(r)}(1 - 2m(r)/r)},
\end{equation}
 
the metric can be written as 

\begin{equation} \label{staticspherical2}
ds^{2} = -e^{-2 \phi(r)}\: (1 - 2m(r)/r)\: [dt^{2} - dr_{*}^{2}] + r^{2}\: d\Omega^{2}; \qquad r = r(r_{*}).
\end{equation}
 
From this form, a number of different approaches can be pursued for complexification.  

The first one is as follows.  We use the Wick rotation $t \rightarrow it$ to obtain a Euclidean metric

\begin{equation}
ds^{2} = e^{-2 \phi(r)}\: (1 - 2m(r)/r)\: [dt^{2} + dr_{*}^{2}] + r^{2}\: d\Omega^{2}; \qquad r = r(r_{*}).
\end{equation}

Introducing complex coordinates $z=r_{*}+ it$, one can rewrite this as 

\begin{equation}
ds^{2} = e^{-2 \phi(r)}\: (1 - 2m(r)/r)\: dz \: d\overline{z} + r^{2}\: d\Omega^{2}; \qquad r = r\Bigl(\frac{1}{2}[z + \overline{z}]\Bigl).
\end{equation}

Using the complex stereographic coordinate $w$ from (\ref{stereographic}), results in the final metric has the expression, 

\begin{equation}
ds^{2} = F(z + \overline{z}) \: dz \: d\overline{z} + H(z + \overline{z}) \frac{dw \: d\overline{w}}{(1+\frac{1}{4}w \: \overline{w})^{2}},  
\end{equation}

where the functions $F$ and $H$ are arbitrary real functions of the indicated argument.  

This derived metric is a complex Hermitian metric, and is adequate for any Euclideanized static spherically symmetric spacetime.

\subsubsection*{Modifying static spherically symmetric spacetimes: \; Route 2}

A second distinct route to proceed from the metric (\ref{staticspherical2}) to start the complexification process, is to introduce coordinates,

\begin{equation}
X = e^{r_{*}} \: \text{cosh} \; t; \qquad T = e^{r_{*}} \: \text{sinh} \; t,
\end{equation}

so that, 

\begin{equation}
dX = e^{r_{*}} [dr_{*} \: \text{cosh}\: t + \text{sinh} \: t dt]; \qquad dT = e^{r_{*}} [dr_{*} \: \text{sinh}\: t + \text{cosh} \: t dt].
\end{equation}

Due to the fact that $-dT^{2} + dX^{2} = -e^{2r_{*}}(dt^{2}-dr_{*}^{2})$, one can write the metric as 

\begin{equation}
ds^{2} = -e^{-2 \phi(r)}\: (1 - 2m(r)/r)\: e^{-2r_{*}}[dT^{2} - dX^{2}] + r^{2}\: d\Omega^{2}, 
\end{equation}

where $r = r(r_{*}) = r(\text{ln}(X^{2}-T^{2}))$.

From this one performs a Wick rotation $T \rightarrow iT$ thus producing the Euclidean metric,

\begin{equation}
ds^{2} = e^{-2 \phi(r)}\: (1 - 2m(r)/r)\: e^{-2r_{*}}[dT^{2} + dX^{2}] + r^{2}\: d\Omega^{2},
\end{equation}

with $r = r(r_{*}) = r(\text{ln}(X^{2}-T^{2}))$.  One can now introduce complex coordinates $z= X+iT$ so that \

\begin{equation}
ds^{2} = e^{-2 \phi(r)}\: (1 - 2m(r)/r)\: e^{-2r_{*}}\: dz \: d\overline{z} + r^{2}\: d\Omega^{2}; \quad r = r(r_{*})=r(\text{ln}(z\overline{z})). 
\end{equation}

Using the complex stereographic coordinate (\ref{stereographic}), and putting all of this together in one expression, results in a complex Hermitian metric

\begin{equation}
ds^{2}= F(z\:\overline{z})\: dz \: d\overline{z} + H(z \: \overline{z}) \frac{dw \: d\overline{w}}{(1+\frac{1}{4}w\: \overline{w})^{2}}, 
\end{equation}

for arbitrary real functions $F$ and $H$ of the indicated argument. 

\subsubsection*{Modifying static spherically symmetric spacetimes: \; Route 3}

For a third approach, going back to our original static spherically symmetric spacetime (\ref{staticspherical}), one can proceed as follows.  We start by regrouping the terms in (\ref{staticspherical}) as 

\begin{equation} \label{regroup}
ds^{2} = (-e^{-2 \phi(r)}\: (1 - 2m(r)/r)\: dt^{2} + r^{2} \: \text{sin}^{2} \theta \: d\phi^{2}) + 
\Bigl(\frac{dr^{2}}{1-2m(r)/r}+r^{2}\: d\theta^{2}\Bigl).
\end{equation}

Now let $r \rightarrow r \tilde{(r)}$ and accordingly, the metric takes the form

\begin{equation}
ds^{2} = (-e^{-2 \phi(r)}\: (1 - 2m(r)/r)\: dt^{2} + r^{2} \: \text{sin}^{2} \theta \: d\phi^{2}) + 
\Bigl(\frac{(dr/d\tilde{r})^{2}\: d\tilde{r}^{2}}{1-2m(r)/r}+(r^{2}/\tilde{r}^{2}) \: \tilde{r}^{2}\: d\theta^{2}\Bigl).
\end{equation}

Now choose 

\begin{equation}
\frac{(dr / d\tilde{r})^{2}}{1-2m(r)/r} = (r^{2}/ \tilde{r}^{2}), 
\end{equation}

so that

\begin{equation}
\frac{dr}{r \sqrt{1-2m(r)/r}} = \frac{d\tilde{r}}{\tilde{r}}.
\end{equation}

Integrating, we see
\begin{equation}
\text{ln} \; \tilde{r} = \int \frac{dr}{r \sqrt{1-2m(r)/r}}, 
\end{equation}

and so
\begin{equation}
\tilde{r} = \text{exp} \Biggl[ \int \frac{dr}{r \sqrt{1-2m(r)/r}}\Biggl].
\end{equation} 

Therefore, it follows that 

\begin{equation}
ds^{2} = (-e^{-2 \phi(r)}\: (1 - 2m(r)/r)\: dt^{2} + r^{2} \: \text{sin}^{2} \theta \: d\phi^{2}) + \frac{r^{2}}{\tilde{r}^{2}} (d\tilde{r}^{2} + \tilde{r}^{2} d\theta^{2}).
\end{equation}

By setting $w=\tilde{r}e^{i\theta}$, thus $r=r(w\:\overline{w})$, and we then have

\begin{equation}
ds^{2} = (-e^{-2 \phi(r)}\: (1 - 2m(r)/r)\: dt^{2} - \frac{r(w \overline{w})^{2}(w-\overline{w})^{2}}{4w \overline{w}} d\phi^{2}) + \frac{r(w\:\overline{w})^{2}}{w\: \overline{w}} dw \: d\overline{w}.
\end{equation}

To proceed forward and get the dimensionality in a correct form, let $m_{\infty} = m(r \rightarrow \infty)$ and write $z =t + im_{\infty}\phi$, to obtain a real valued (but \emph{not} Hermitian) metric on a complex manifold, 

\begin{align}
ds^{2} = \Bigl(-e^{-2 \phi(r)}\: (1 - 2m(r)/r)\: \frac{(dz + d\overline{z})^{2}}{4} + \\ \nonumber
 \frac{r(w \overline{w})^{2}(w-\overline{w})^{2}}{4w \overline{w}} \frac{(dz - d\overline{z})^{2}}{4m_{\infty}^{2}}\Bigl) 
 +  \frac{r(w\:\overline{w})^{2}}{w\: \overline{w}} dw \: d\overline{w}.
\end{align}

\subsubsection*{Modifying static spherically symmetric spacetimes: \; Route 4}

The last complexification route for static spherically symmetric spacetimes presented will involve starting with the metric (\ref{staticspherical}) and regrouping the terms to obtain the expression (\ref{regroup}).  Using this and defining $w=re^{i\theta}$, we obtain

\begin{align}
ds^{2} = \Bigl(-e^{-2 \phi(r)}\: (1 - 2m(r)/r)\: dt^{2} - \frac{(w-\overline{w})^{2}}{4} d\phi^{2}) \\ \nonumber
+ \Bigl( \frac{(\overline{w}\:dw + w d\overline{w})^{2}}{4r(r-2m(r))} + 2w\overline{w}\:dw \: d\overline{w} - \frac{w\:d\overline{w}^{2}}{4\overline{w}} - \frac{\overline{w}\:dw^{2}}{4w}\Bigl).
\end{align}

Now let $m_{H}$ be defined by $m(r_{H}) = r_{H}$ with $m_{H} = 2r_{H}$, and set $z = t + im_{H}\phi$. Consequently, the metric is expressed as

\begin{align}
ds^{2} = \Bigl(-e^{-2 \phi(r)}\: (1 - 2m(r)/r)\: \frac{(dz + d\overline{z})^{2}}{4} + \frac{(w-\overline{w})^{2}}{4} \frac{(dz - d\overline{z})^{2}}{4m_{H}^{2}}\Bigl) \\ \nonumber
+ \Bigl( \frac{(\overline{w}\:dw + w d\overline{w})^{2}}{4r(r-2m(r))} + 2w\overline{w}\:dw \: d\overline{w} - \frac{w\:d\overline{w}^{2}}{4\overline{w}} - \frac{\overline{w}\:dw^{2}}{4w}\Bigl).
\end{align}

and has the form,

\begin{align} \label{complexmanifoldform}
\begin{split}
ds^{2} = \Big(g_{zz}(w, \overline{w})dz^{2} + {g}_{\overline{z}\overline{z}}(w, \overline{w})d\overline{z}^{2} + 2g_{z\overline{z}}(w, \overline{w}) dz\: d\overline{z}\Big) \\
+ \Big(g_{ww}(w, \overline{w})dw^{2} + {g}_{\overline{w}\overline{w}}(w, \overline{w})d\overline{w}^{2} + 2g_{w\overline{w}}(w, \overline{w}) dw\: d\overline{w}\Big).
\end{split}
\end{align}

\section{Modifying Complex Manifold Theory} 

We will see that standard complex manifold theory does not lend itself well to fitting within the framework of Lorentzian signature metrics.  Therefore, the aim of this subsection is to provide an overview of one possible direction, largely due to Flaherty \cite{flaherty1976hermitian}, in which this merger can be successfully implemented.

The underlying assumption of this section is that the manifold in question is equipped with a globally normed null tetrad field.  This extra mathematical structure is reasonable, given that our intention is physically motivated \cite{geroch1968spinor}.  Nonetheless, the mathematics in question will be local, in the sense that we will be focusing only on one coordinate patch at a time. 

\subsubsection*{An almost complex structure on spacetime}

Going back to the General Relativity chapter, we recall that a spacetime metric can be represented by a null tetrad through the expression

\begin{equation} 
g_{ab} = l_{a}\:n_{b} + n_{a}\:l_{b} - m_{a}\:\overline{m}_{b} - \overline{m}_{a}\:m_{b}.
\end{equation} 

Without modifications to either the Lorentzian signature of a metric, nor the definitions of complex manifold theory, one can still construct an almost complex structure for a spacetime.  However, we will find that it is of limited use.

\begin{theorem}
	Given a spacetime with a normed null tetrad, an almost complex structure for the spacetime is given (locally) by 
	\begin{equation}
	J_{a}^{\;b} = -l_{a}\:l^{b}+n_{a}\:n^{b}-i\:m_{a}\:\overline{m}^{b} + i\:\overline{m}_{a}\:m^{b}.
	\end{equation}
	(See Flaherty \cite{flaherty1976hermitian}.)
\end{theorem}

There a number of important properties to observe, such that $J_{a}^{\;s}\:J_{s}^{\;b} =- \delta_{a}^{\;b}$, and that the almost complex structure is real.  

In addition to this, the integrability conditions can be satisfied through the vanishing of the Nijenhuis tensor, since in one coordinate patch the tetrads can be chosen to be real analytic.

Nonetheless, there exists a crucial drawback to this almost complex structure, which is that it is not Hermitian through the standard definition of complex manifold theory. Specifically, one can see that 

\begin{equation*}
J_{a}^{\;m}\:J_{b}^{\;n}\;g_{mn} = -l_{a}\:n_{b} - n_{a}\:l_{b} - m_{a}\:\overline{m}_{b} - \overline{m}_{a}\:m_{b} \; \neq \; g_{ab}.
\end{equation*} 

The question is this: Are there any almost Hermitian structures that can be constructed on a Lorentzian spacetime?  The answer is no, as established by the following theorem.

\begin{theorem}
	A manifold with a metric of Lorentzian signature cannot admit an almost Hermitian structure. (See e.g. Flaherty \cite{flaherty1976hermitian}.)
\end{theorem}

This is reasonably well known result but we present a sketch of the proof provided by Flaherty \cite{flaherty1976hermitian}.  

Suppose we have an almost Hermitian structure $J_{a}^{\;b}$, and we choose coordinates such that the metric take the form $\eta_{ab}$ at that point.  Then by the Hermitian property, we will have $J_{a}^{\;m}\:J_{b}^{\;n}\:\eta_{mn} = \eta_{ab}$.  Hence this allows $J_{a}^{\;b}$ to define a Lorentz transformation whose square $J_{a}^{\;s}\:J_{s}^{\;b}$ is equal to minus the identity.  

Since this is a Lorentz transformation, it would imply that $J_{a}^{\;b}$ is non-singular and hence has four linearly independent eigenvectors with non-zero eigenvalues.  

Furthermore, given the property that $J^{2}=-1$, we can show that the eigenvalues are $\{+i, +i, -i, -i\}$ and that the eigenvectors are null vectors due to the antisymmetry of $J_{ab}$. In addition to this, since $J_{a}^{\;b}$ is real, the eigenvectors come in complex conjugate pairs.  

To summarize, we have four linearly independent complex null vectors in complex conjugate pairs and this is only possible if the metric signature is $(++++)$, $(++--)$, or $(----)$.  Hence we have a contradiction, and the stated result follows.

\subsubsection*{Modified almost complex structure}

One way to overcome this obstacle is to modify specific definitions in complex manifold theory to proceed forward.  To be more precise, we now allow the possibility of a complex-valued almost complex structure which is called the modified almost complex structure.  Similarly other structures which were real, can be termed ``modified" if they are allowed to be complex-valued.  

\begin{mydef}
Given a spacetime and a normed null tetrad, the modified almost Hermitian structure for the spacetime is given by the structure tensor
\begin{equation} \label{modifiedalmostHermitian}
J_{a}^{\;b} = i\:l_{a}\:n^{b} - i\:n_{a}\:l^{b} - i\:m_{a}\:\overline{m}^{b} + i\:\overline{m}_{a}\:m^{b}.
\end{equation}	
\end{mydef}

Despite this modification of now becoming a complex-valued structure (in fact $i(l_{a}\:n^{b}-n_{a}\:l^{b})$ is pure imaginary), the modified almost Hermitian structure satisfies all the requirements one would like to treat the tensor as an almost Hermitian structure. This can be highlighted below with the following theorem.

\begin{theorem}
	For the $J_{a}^{\;b}$ given by (\ref{modifiedalmostHermitian}), we have:
	\begin{equation*}
	J_{a}^{\;m}\: J_{m}^{\;b} = -\delta_{a}^{\;b},  
	\end{equation*}
	\begin{equation*}
	J_{a}^{\;m}\: J_{b}^{\;n}\:g_{mn} = g_{ab},  
	\end{equation*}
	\begin{align*}
	J_{a}^{\;b}\: l_{b} &= i\:l_{a}, \\
	J_{a}^{\;b}\: m_{b} &= i\:m_{a},\\
	J_{a}^{\;b}\: n_{b} &= -i\:n_{a},\\
	 J_{a}^{\;b}\: \overline{m}_{b} &= -i\:\overline{m}_{a}.
	\end{align*}
	(See e.g. Flaherty \cite{flaherty1976hermitian}.)
	\end{theorem}

The eigenvalues of this modified almost Hermitian structure can be seen to be $\{+i, +i, -i,-i\}$ and serves as an important role in the integrability of the structure.

\begin{mydef}
	The modified almost Hermitian structure (\ref{modifiedalmostHermitian}) is said to be integrable, if coordinates (generally complex) exist such that $J_{a}^{\;b}$ has the components
	\begin{equation*}
	J_{a}^{\;b}\;=\;
	\begin{bmatrix}
	+i & 0 & 0    & 0 \\
	0 & +i & 0  & 0 \\
	0 & 0 & -i  & 0 \\
	0 & 0 & 0  & -i
	\end{bmatrix}.
	\end{equation*} 
\end{mydef}

\begin{theorem}
	The modified almost Hermitian structure $J_{a}^{\;b}$ (\ref{modifiedalmostHermitian}) is integrable if and only if the Nijenhuis tensor formed from $J_{a}^{\;b}$ vanishes. (See Flaherty \cite{flaherty1976hermitian}.)
\end{theorem}

The next theorem involves a basic understanding of Petrov classification; a basic introduction on this classification scheme can be found in \cite{stephani2003exact}.  What is of relevance to us is that both the Schwarzschild and the Kerr spacetime are Type D.  

\begin{theorem}
	Among the vacuum spactimes, the modified almost Hermitian structure $J_{a}^{\;b}$ (\ref{modifiedalmostHermitian}) is integrable if and only if the spacetime is of Type D. (See e.g. Flaherty \cite{flaherty1976hermitian}.)
\end{theorem}

\subsubsection*{Analyzing the modified almost Hermitian structure}

In this subsection, we present some analysis and remarks that I developed (in collaboration with my supervisor) regarding Flaherty's approach to modifications of the almost Hermitian structure.

We start by analyzing how an almost complex structure behaves under a range of different semi-Riemannian signatures.  The purpose of this exercise is to better understand the derivation of Flaherty's modified almost Hermitian structure, and to improve it in some suitable manner. 

\textbf{(1+1) dimensions.} 

One can start by looking at Lorentzian signature metrics of $(1+1)$ dimensions.  Suppose we have an almost Hermitian structure then, 

\begin{equation*}
	J_{ab}\;=\;
	\begin{bmatrix}
0 & -s \\
	+s & 0 
	\end{bmatrix}_{ab}.
\end{equation*}

At any point, we can choose coordinates such that 

\begin{equation*}
	g_{ab}=\; \eta_{ab}\: =
	\begin{bmatrix}
	-1 & 0 \\
	0 & +1 
	\end{bmatrix}.
\end{equation*}

It follows that

\begin{equation*}
J_{\;\;b}^{a}\;=\; \eta^{ac}\:J_{cb} =\:
\begin{bmatrix}
0 & +s \\
+s & 0 
\end{bmatrix}_{ab},
\end{equation*}

and consequently, we have

\begin{equation*}
(J^{2})_{\;\;b}^{a}\;=\;
\begin{bmatrix}
s^{2} & 0 \\
0 & s^{2}  
\end{bmatrix}_{ab} = -\delta_{\;\;b}^{a}.
\end{equation*}

But this implies that $s\: = \: \pm i$, so what \emph{should} have been the ``almost complex structure"
\begin{equation*}
J_{\;\;b}^{a}\;=\; \pm i
\begin{bmatrix}
0 & +1 \\
+1 & 0 
\end{bmatrix}
\end{equation*}

is actually pure imaginary and not real.  In contrast if we set 

\begin{equation*}
J_{\;\;b}^{a}\;=\; \pm i
\begin{bmatrix}
0 & +1 \\
-1 & 0 
\end{bmatrix},
\end{equation*}

then this would be an appropriate almost complex structure, but it does not extend to an almost Hermitian structure. 

\textbf{1+\:[2m-1] dimensions}

We can generalize this previous argument to the case of 1+\:[2m-1] dimensions.  Suppose we have an almost Hermitian structure, then

\begin{equation*}
J_{ab}\;=\;
\left[
\begin{array}{cc|cc}
0 & -s &  0 & 0 \\
+s & 0 & 0 & 0 \\
\hline
0 & 0 &  0 & -I_{m-1} \\
0 & 0 &  I_{m-1} & 0  
\end{array}
\right]_{ab}.
\end{equation*}

Just as in the previous case, we can choose coordinates to be set, so that 

\begin{equation*}
g_{ab}=\; \eta_{ab}\: =
\left[
\begin{array}{cc|cc}
-1 & 0 & 0 & 0\\
0 & +1 & 0 & 0 \\
\hline
0 & 0 & I_{m-1} & 0 \\
0 & 0 & 0 & I_{m-1} 
\end{array}
\right].
\end{equation*}

As a result of this, we find that

\begin{equation*}
J_{\;\;b}^{a}\;=\; \eta^{ac}\:J_{cb} =\:
\left[
\begin{array}{cc|cc}
0 & +s & 0 & 0 \\
+s & 0 & 0 & 0 \\
\hline
0 & 0 & 0 & -I_{m-1} \\
0 & 0 & I_{m-1} & 0   
\end{array}
\right]_{ab},
\end{equation*}

and thus

\begin{equation*}
(J^{2})_{\;\;b}^{a}\;=\;
\left[
\begin{array}{cc|cc}
s^{2} & 0 & 0 & 0 \\
0 & s^{2} & 0 & 0 \\
\hline
0 & 0 & -I_{m-1} & 0 \\
0 & 0 & 0 & -I_{m-1} 
\end{array}
\right]_{ab} = -\delta_{\;\;b}^{a}.
\end{equation*}

This implies that $s\: = \: \pm i$, so similarly what should have been the ``almost complex structure"

\begin{equation*}
J_{\;\;b}^{a}\;=\; 
\left[
\begin{array}{cc|cc}
0 & \pm i & 0 & 0 \\
\pm i & 0 & 0 & 0 \\
\hline
0 & 0 & 0 & -I_{m-1} \\
0 & 0 & I_{m-1} & 0 
\end{array}
\right]
\end{equation*}

is now partly imaginary but not pure real.

\textbf{(2 + [2m-2]) dimensions}

For our next case, consider again an almost Hermitian structure

\begin{equation*}
J_{ab}\;=\;
\left[
\begin{array}{cc|cc}
0 & -s & 0 & 0 \\
+s & 0 & 0 & 0 \\
\hline
0 & 0 & 0 & -I_{m-1} \\
0 & 0 & I_{m-1} & 0   
\end{array}
\right]_{ab},
\end{equation*}

and similarly we can choose coordinates such that 

\begin{equation*}
g_{ab}=\; \eta_{ab}\: =
\left[
\begin{array}{cc|cc}
-1 & 0 & 0 & 0\\
0 & -1 & 0 & 0 \\
\hline
0 & 0 & I_{m-1} & 0 \\
0 & 0 & 0 & I_{m-1}   
\end{array}
\right].
\end{equation*}

We find that 

\begin{equation*}
J_{\;\;b}^{a}\;=\; \eta^{ac}\:J_{cb} =\:
\left[
\begin{array}{cc|cc}
0 & +s & 0 & 0 \\
-s & 0 & 0 & 0 \\
\hline
0 & 0 & 0 & -I_{m-1} \\
0 & 0 & I_{m-1} & 0   
\end{array}
\right]_{ab}.
\end{equation*}

and consequently,

\begin{equation*}
(J^{2})_{\;\;b}^{a}\;=\;
\left[
\begin{array}{cc|cc}
-s^{2} & 0 & 0 & 0 \\
0 & -s^{2} & 0 & 0 \\
\hline
0 & 0 & -I_{m-1} & 0 \\
0 & 0 & 0 & -I_{m-1}   
\end{array}
\right]_{ab} = -\delta_{\;\;b}^{a}.
\end{equation*}

This implies $s=\pm1$ and we find that the ``almost complex structure" 

\begin{equation*}
J_{\;\;b}^{a}\;=\; 
\left[
\begin{array}{cc|cc}
0 & \pm 1 & 0 & 0 \\
\mp 1 & 0 & 0 & 0 \\
\hline
0 & 0 & 0 & -I_{m-1} \\
0 & 0 & I_{m-1} & 0    
\end{array}
\right],
\end{equation*}

fits the requirement.  Therefore, almost Hermitian structures are compatible with $(2 + [2m-2])$ dimensions.

\textbf{$(2m_{1} +2m_{2})$ dimensions}

More generally, it is now clear that things will work in ($2m_{1} +2m_{2}$) dimensions but fail whenever either the number of timelike dimensions is odd or the number of spacelike dimensions is odd.

To progress our discussion, it will be necessary to talk about a globally defined tetrad.  In any given coordinate patch, let $e_{A}^{\;\;a}$ be a tetrad with det$(e_{A}^{\;\;a}) \neq 0$ and let $e_{\;\;a}^{A}$ represent the inverse tetrad.  We let 

\begin{equation*}
J_{\;\;B}^{A} =
	\begin{bmatrix}
	0 & I_{m} \\
	-I_{m} & 0 
	\end{bmatrix}; \qquad \text{so that} \qquad J_{\;\;B}^{A}\:J_{\;\;C}^{B} = -\delta_{\;\;C}^{A}.
\end{equation*}  

We define an almost complex structure locally as 

\begin{equation*}
J_{\;\;b}^{a} \equiv J_{\;\;B}^{A}\:e_{A}^{\;\;a}\:e_{B}^{\;\;b} \qquad \text{so that} \qquad J_{\;\;b}^{a}\:J_{\;\;c}^{b} = -\delta_{\;\;c}^{a}. 
\end{equation*}

One can only extend this to a globally defined tensor field only when the tetrad is globally defined.  

We can summarize that, whenever a globally defined tetrad exists and the metric signature is ($2m_{1} +2m_{2}$), then globally defined almost Hermitian structures exist.

On the other hand, if the metric signature is not ($2m_{1} +2m_{2}$) but a globally defined tetrad exists, then globally defined almost Hermitian structures do not exist.  The best one can do is produce a structure that is part real and part imaginary.  

\textbf{(3+1) dimensions}

For this particular signature, we assume the existence of a global orthonormal tetrad.  Then

\begin{equation*}
g_{ab}\;=\; e_{\;\;a}^{A}\:e_{\;\;b}^{B}
\left[
\begin{array}{cc|cc}
-1 & 0 & 0 & 0 \\
0 & +1 & 0 & 0 \\
\hline
0 & 0 & 1 & 0 \\
0 & 0 & 0 & 1   
\end{array}
\right]_{AB},
\end{equation*}

and one can construct 

\begin{equation*}
J_{ab}\;=\; e_{\;\;a}^{A}\:e_{\;\;b}^{B}
\left[
\begin{array}{cc|cc}
0 & +i & 0 & 0 \\
-i & 0 & 0 & 0 \\
\hline
0 & 0 & 0 & 1 \\
0 & 0 & -1 & 0   
\end{array}
\right]_{AB},
\end{equation*}

so that

\begin{equation*}
J_{\;\;b}^{a}\;=\; e_{A}^{\;\;a}\:e_{\;\;b}^{B}
\left[
\begin{array}{cc|cc}
0 & -i & 0 & 0 \\
-i & 0 & 0 & 0 \\
\hline
0 & 0 & 0 & 1 \\
0 & 0 & -1 & 0 
\end{array}
\right]_{\;B}^{A}.
\end{equation*}

Accordingly, we find that $(J^{2})_{\;\;b}^{a} = -\delta_{\;\;b}^{a}$ and that this is globally defined.  Flaherty \cite{flaherty1976hermitian} names this as the modified almost complex structure.  

\textbf{Remarks}

Through our analysis, we see that a possibly cleaner way to implement the modification would be to view $(3+1)$ spacetime as a codimension 2 sub-manifold of a $(2+4)$ dimensional embedding space; then the $(2+4)$ embedding space can be given the standard almost Hermitian structure.

An alternative approach from Flaherty altogether is to use the standard almost complex structure and retain the consequences of such a construction.  An example of a standard almost complex structure would be 

\begin{equation*}
J_{\;\;b}^{a}\;=\; e_{A}^{\;\;a}\:e_{\;\;b}^{B}
\left[
\begin{array}{cc|cc}
0 & 1 & 0 & 0 \\
-1 & 0 & 0 & 0 \\
\hline
0 & 0 & 0 & 1 \\
0 & 0 & -1 & 0  
\end{array}
\right]_{\;B}^{A}
\end{equation*}

and the consequences are that it does not extend to a Hermitian structure.  A specific example of $J_{ab}$ is 

\begin{equation*}
J_{ab}\;=\; e_{\;\;a}^{A}\:e_{\;\;b}^{B}
\left[
\begin{array}{cc|cc}
0 & -1 & 0 & 0 \\
-1 & 0 & 0 & 0 \\
\hline
0 & 0 & 0 & 1 \\
0 & 0 & -1 & 0  
\end{array}
\right]_{AB}
\end{equation*}

and you can fix the choice of tetrad by demanding $g_{ab} = e_{\;\;a}^{A}\:e_{\;\;b}^{B} \:\eta_{AB}$.  This results in the most general almost complex structure being of form:

\begin{equation*}
J_{\;\;b}^{a}\;=\; e_{A}^{\;\;a}\:e_{\;\;b}^{B}\:J_{\;\;B}^{A}; \qquad J_{\;\;B}^{A} = X_{\;\;C}^{A} 
\left[
\begin{array}{cc|cc}
0 & 1 & 0 & 0 \\
-1 & 0 & 0 & 0 \\
\hline
0 & 0 & 0 & 1 \\
0 & 0 & -1 & 0 
\end{array}
\right]_{C}^{D} [X^{-1}]_{\;\;B}^{D}
\end{equation*}

with $X$ defined up to a local Lorentz transformation.  It follows that

\begin{equation*}
J_{ab}\;=\; e_{\;\;a}^{A}\:e_{\;\;b}^{B}\:J_{AB}; \quad J_{AB}\; = \; (\eta_{AE}\: X_{\;\;F}^{E} \; \eta^{FC}) 
\left[
\begin{array}{cc|cc}
0 & -1 & 0 & 0 \\
-1 & 0 & 0 & 0 \\
\hline
0 & 0 & 0 & 1 \\
0 & 0 & -1 & 0 
\end{array}
\right]_{CD}  [X^{-1}]_{\;\;B}^{D}.
\end{equation*}

We find that this is neither symmetric nor anti-symmetric.  The metric is then simply not Hermitian, in the standard sense.

\section{Discussion}

In this chapter, we have seen two approaches in the merger of Lorentzian metric manifolds with the theory of complex manifolds.  The first involved modifying spacetime metrics into Euclidean metrics with the additional calculation of turning them into complex Hermitian form.  

After that, we encountered an important theorem that a Lorentzian signature metric cannot admit an almost Hermitian structure.  This paved the way for Flaherty's modified almost Hermitian structure which is a complex-valued structure.

Finally some calculations that I performed (in collaboration with my supervisor), highlighted some simple and general cases and also showed alternative ways of looking at the situation. 

In the next chapter, the focus will be on a specific procedure that involves Lorentzian metric manifolds and complex variables.  It marks the highlight of the thesis by stating the intended research problem of this thesis.


\chapter{The Newman-Janis trick}

\begin{chapquote}{Philip Pullman, \textit{The Golden Compass}}
	``... an imaginary number, like the square root of minus one: you can never see any concrete proof that it exists, but if you include it in your equations, you can calculate all manner of things that couldn't be imagined without it.''
\end{chapquote}

This chapter presents an exposition of a procedure which highlights a powerful and mysterious application of using complex variables in the theory of General Relativity.  

This procedure, or the ``Newman-Janis trick" as it's commonly termed, was constructed in 1965 \cite{newman1965note}, shortly after the discovery of the Kerr metric \cite{kerr1963gravitational}. But since then, no one has fully understood why this method works \cite{adamo2009null} and most physicists consider it as an accidental trick.  Though there are a few that believe it to be a clue to a deeper structure.  

The Newman-Janis trick is an algorithm, or more correctly an ansatz, that ``derives" the Kerr metric from the Schwarzschild metric.  The main aspect of this ansatz is that it involves complex variables, in particular a complex coordinate transformation.  

The astonishing property of this procedure is that it involves steps that are in some respects few as well as elementary.  This is remarkable given that derivation of the Kerr metric took an astounding 50 years to derive from the field equations and involved incredible algebraic complexity \cite{wiltshire2009kerr}.  

In contrast, the Schwarzschild metric was derived from the field equations in less than a year after Einstein delivered his lecture on the final form of his theory of gravitation. Most standard textbooks in General Relativity provide a description of this Schwarzschild derivation (see e.g. Carroll \cite{carroll2004spacetime}).  

Relevant review articles, and ``explanations" of the Newman-Janis trick include \cite{adamo2014kerr, erbin2015demianski, whisker2008braneworld}.

\section{Newman-Janis trick}

In this section, we present the original version of the Newman-Janis trick, as described in the original paper in 1965 \cite{newman1965note}.  

The first step is to start with the familiar Schwarzschild metric in coordinates $(t, r, \theta, \phi)$,

\begin{equation}
ds^{2}= \Bigl(1-\frac{2\:m}{r}\Bigl)\:{dt}^{2}-\frac{1}{1-\frac{2\:m}{r}}dr^{2}-{r}^2({d\theta}^{2}+{{\sin}^{2}\theta}\:{d\phi}^2).
\end{equation}

We then perform a coordinate transformation, 

\begin{align}
\begin{split}
u &= t - r - 2m\:\text{ln}\:\Bigl(\frac{r}{2m}-1\Bigl), \\
r' &= r, \\
\theta' &= \theta, \\
\phi' &= \phi,
\end{split}
\end{align}

which results in the Schwarzschild metric being rewritten in advanced Eddington-Finkelstein coordinates (dropping the primes) as 

\begin{equation}\label{SchinEF}
ds^{2}= \Bigl(1-\frac{2\:m}{r}\Bigl)\:{du}^{2}+2\:du\:dr-{r}^2({d\theta}^{2}+{{\sin}^{2}\theta}\:{d\phi}^2).
\end{equation}

Recall, from the chapter on General Relativity, that a metric can be expressed in terms of a null tetrad.  Therefore, we can express the Eddington-Finkelstein form of the Schwarzschild metric (\ref{SchinEF}) in the following null tetrad

\begin{align}\label{Tetrad 1}
\begin{split}
l^{a}&={\partial}_{r}\\
n^{a}&={\partial}_{u}-{\frac{1}{2}}\Bigl(1-{\frac{2m}{r}\Bigl)} {\partial}_{r}\\
m^{a}&={\frac{1}{\sqrt{2}{r}}}\Bigl({\partial}_{\theta}+\frac{i}{\sin \theta}{\partial}_{\phi}\Bigl)\\
\bar{m}^{a}&={\frac{1}{\sqrt{2}{r}}}\Bigl({\partial}_{\theta}-\frac{i}{\sin \theta}{\partial}_{\phi}\Bigl).\\
\end{split}
\end{align}

One can easily check that these contravariant vectors satisfy the conditions of a null tetrad with respect to the Schwarzschild metric (\ref{SchinEF}). 

The trick starts by extending the coordinate $r$ take on complex values i.e. $r \in \mathbb{C}$.  In addition to this, certain terms involving $r$ are complex conjugated, while others are left alone.  This ambiguous step results in the following tetrad

\begin{align}\label{Tetrad 2}
\begin{split}
l^{a}&={\partial}_{r}\\
n^{a}&={\partial}_{u}-{\frac{1}{2}}(1-{\frac{m}{r}-\frac{m}{\bar{r}})} {\partial}_{r}\\
m^{a}&={\frac{1}{\sqrt{2}{\bar{r}}}}({\partial}_{\theta}+\frac{i}{\sin \theta}{\partial}_{\phi})\\
\bar{m}^{a}&={\frac{1}{\sqrt{2}{r}}}({\partial}_{\theta}-\frac{i}{\sin \theta}{\partial}_{\phi}).\\
\end{split}
\end{align}

The ambiguity in the previous step is reflected in that if the complex conjugation on $r$ was done in a different way, the desired result at the end of the procedure will not be derived.

After this,  we let the coordinate $u$ take on complex values, and perform the complex coordinate transformation

\begin{align} \label{NJcoordinate}
\begin{split}
u &\mapsto u'=u-{i}\:{a}\:{\cos\theta}\\
r &\mapsto r'=r+{i}\:{a}\:{\cos\theta}\\
\theta &\mapsto {\theta}'=\theta\\
\phi &\mapsto {\phi}'=\phi,
\end{split}
\end{align}

where $a$ is a constant.  This implies that the basis vectors transform as

\begin{align}
\begin{split}
{\partial}_{u}&= {\partial}_{u'}\\
{\partial}_{r}&= {\partial}_{r'}\\
{\partial}_{\theta}&= {\partial}_{{\theta}'}+{i}\:{a}\:{\sin\theta}({\partial}_{u'}-{\partial}_{r'})\\
{\partial}_{\phi}&={\partial}_{\phi '}
\end{split}
\end{align}

The original paper mentions a crucial point that part of the algorithm is to keep $l^{a}$ and $n^{a}$ real and $m^{a}$ and $\overline{m}^{a}$ the complex conjugates of each other.  This statement is extremely important for the discussion in our next chapter.

Applying this complex coordinate transformation to our null tetrad, we obtain 

	\begin{align}
	\begin{split}
	l^{a}&={\partial}_{r^{'}},\\
	n^{a}&={\partial}_{u^{'}}-{\frac{1}{2}}\Bigl(1-{\frac{2\:m\:r^{'}}{r^{'2}+a^{2}{\cos}^{2}\theta^{'}}}\Bigl) {\partial}_{r^{'}},\\
	m^{a}&={\frac{1}{\sqrt{2}({r^{'}+i\:a\:\cos\theta^{'}})}}\Bigl({\partial}_{\theta^{'}}+i\:a\:\sin\theta^{'}({\partial}_{u^{'}}-{\partial}_{r^{'}})+\frac{i}{\sin \theta^{'}}{\partial}_{\phi^{'}}\Bigl),\\
	\bar{m}^{a}&= {\frac{1}{\sqrt{2}({r^{'}-i\:a\:\cos\theta^{'}})}}\Bigl({\partial}_{\theta^{'}}-i\:a\:\sin\theta^{'}({\partial}_{u^{'}}-{\partial}_{r^{'}})-\frac{i}{\sin \theta^{'}}{\partial}_{\phi^{'}}\Bigl).\\
	\end{split}
	\end{align}
   
The final step of this procedure is to restrict the coordinates $u^{'}, r^{'}, \theta^{'}, \phi^{'}$ to be real and remove the primes to construct the final tetrad which is 

	\begin{align}\label{Tetrad 4}
	\begin{split}
	l^{a}&={\partial}_{r},\\
	n^{a}&={\partial}_{u}-{\frac{1}{2}}(1-{\frac{2\:m\:r}{r^{2}+a^{2}{\cos}^{2}\theta)}}) {\partial}_{r},\\
	m^{a}&={\frac{1}{\sqrt{2}({r+i\:a\:\cos\theta})}}({\partial}_{\theta}+i\:a\:\sin\theta({\partial}_{u}-{\partial}_{r})+\frac{i}{\sin \theta}{\partial}_{\phi}),\\
	\bar{m}^{a}&= {\frac{1}{\sqrt{2}({r-i\:a\:\cos\theta})}}({\partial}_{\theta}-i\:a\:\sin\theta({\partial}_{u}-{\partial}_{r})-\frac{i}{\sin \theta}{\partial}_{\phi}).\\
	\end{split}
	\end{align} 

Constructing the inverse metric corresponding to this this null tetrad, using $g^{ab} = l^{a}\:n^{b} + n^{a}\:l^{b} - m^{a}\:\overline{m}^{b} - \overline{m}^{a}\:m^{b}$ and then inverting this matrix yields 

		\begin{multline}
		ds^{2}= \Bigl(1-\frac{2\:m\:r}{r^{2}+a^{2}{\cos}^{2}\theta}\Bigl)\:{du}^{2}+2\:du\:dr+\frac{4\:m\:r\:a\sin^{2}\theta}{r^{2}+a^{2}{\cos}^{2}\theta}du\:d\phi-2\:a\:\sin^{2}\theta\:d\phi\:dr \\
		-(({r^{2}+a^{2}{\cos}^{2}\theta})\:a^{2}\sin^{2}\theta + 2\:m\:r\:a^{2}\sin^{2}\theta\\
		+({r^{2}+a^{2}{\cos}^{2}\theta})^{2})\frac{\sin^{2}\theta}{({r^{2}+a^{2}{\cos}^{2}\theta})}d\phi^{2} 
		-({r^{2}+a^{2}{\cos}^{2}\theta})\:{d\theta}^{2},
		\end{multline}
		
a real spacetime which is precisely the Kerr metric!

If I now set the parameter $a$ to zero, I recover the Schwarzschild metric.  In that sense, the Newman-Janis trick obtains a rotating solution from a non-rotating metric.

An interesting point to note is that if I apply the Newman-Janis trick on Kerr, I will obtain once again a Kerr metric.    

There are a number of unclear steps in this procedure, but a conventional interpretation of this ansatz is the remark that this special complex coordinate transformation (\ref{NJcoordinate}) (as opposed to a real coordinate transformation) allows one to move to a different geometry (i.e. Schwarzschild to Kerr).  We will see in the next chapters, there is more to this trick than just a complex coordinate transformation.

Interesting points to note are that: When one carries out this particular calculation using the Schwarzschild coordinates, as opposed to the advanced Eddington-Finkelstein coordinates, the trick does not work.  Another trivial calculation one can do is apply this particular procedure to flat space in advanced Eddington-Finkelstein cordinates.  This result is flat space again but in a different coordinate system, resulting in the metric:

	\begin{multline}
	ds^{2}= {du}^{2}+2\:du\:dr - 2\:a\sin^{2}\theta \: d\phi \: dr-(r^{2}+a^{2}{\cos}^{2}\theta) d{\theta}^2 \\
	-(a^{2} + r^{2}){\sin}^{2}{\theta}\:{d\phi}^2.
	\end{multline}

Finally, if one applies the complex coordinate transformation (\ref{NJcoordinate}) directly to the Schwarzschild metric (\ref{SchinEF}), we do not get the desired result.

\section{Extensions and Applications}

The success of the trick can be highlighted in that it has been used to find new metrics (e.g. Kerr-Newman \cite{newman1965metric}) as well as show that existing metrics can be obtained through this method.  In particular, it generates a rotating solution from corresponding static, spherically symmetric solutions.

In \cite{demianski1972new, demianski1966combined}, Demianski was able to obtain the most general vacuum solution from a complexified null tetrad of Reissner-Nordstr\"{o}m.  This general complex coordinate transformation obtains solutions with additional parameters such as the NUT charge.

Is there a general condition to specify when the Newman-Janis trick can be successfully applied?  The answer is ``not that we know of."  It turns out that so far, the trick has worked successfully only for a certain class of metrics known as Kerr-Schild metrics \cite{newman1965note, kerr2007discovering} which are of the form 

\begin{equation} \label{KerrSchild}
g_{ab} = \eta_{ab} + H\:l_{a}\:l_{b},
\end{equation}

where $H$ is a scalar function, $\eta$ is the Minkowski metric and $l_{a}$ is null with respect to $g$ and $\eta$. There has not been a proof though, that a Kerr-Schild metric is a necessary condition for the Newman-Janis trick to work.  

More precisely, the most general statement one can make is that it has been shown by Talbot \cite{talbot1969newman} that one can construct a generalized complex coordinate transformation for a certain sub-class of Kerr-Schild metrics (all known Kerr-Schild metrics fall under this sub-class) \cite{adamo2014kerr} and obtain a desired solution (such as Kerr and Demianski metrics from the Schwarzschild metric) \cite{adamo2014kerr}.  In particular, if we are given a solution in coordinates $(u, r, \zeta, \overline{\zeta}$), then this general complex coordinate transformation takes the form

\begin{align}
u^{'} = u + i\:S(\zeta, \overline{\zeta}), \\
r^{'} = r + i\:T(\zeta, \overline{\zeta}),
\end{align}

where $S$ and $T$ are real valued.

Variations of the Newman-Janis trick can be generalized in a way to give an algorithmic procedure to apply it directly to Weyl or Maxwell tensor components as shown in \cite{keane2014extension}.

It has also been extended for the 5-dimensional case \cite{lessner2008complex, erbin2014five}.  

Other notable applications include applying the trick to interior solutions  \cite{herrera1982complexification, drake1997application, ibohal2005rotating, viaggiu2006interior, azreg2014static, azreg2014generating, drake2000uniqueness, glass2004kottler}, dilaton-axion black holes \cite{yazadjiev2000letter},the BTZ black hole \cite{kim1999spinning}, D-dimensional Kerr Black holes \cite{dianyan1988exact}, Born-Infeld monopole \cite{lombardo2004newman}, gauge supergravity \cite{erbin2015demianski}, modified gravity theories such as the Brans-Dicke theory \cite{pirogov2013towards}, Lovelock gravity \cite{dadhich2013rotating}, $f(R)$ gravity \cite{ghosh2013radiating} and quadratic modified gravity \cite{hansen2013applicability}.  In addition to this, other work includes applying the trick to proposed theories of quantum gravity, such as in loop quantum gravity \cite{caravelli2010spinning} as well as in braneworld black holes \cite{whisker2008braneworld}.  A computer program has also been built for the method \cite{gutierrez2014computer}.

\section{Giampieri's method}

In an essay submitted to Gravity Research Foundation in 1990 \cite{giampieri}, the author, Giacomo Giampieri, was able to exhibit an ansatz to derive the Kerr metric directly from the Schwarzschild metric, similar to the Newman-Janis trick, (but avoiding the use of null tetrads.)  

The crucial difference is that this procedure involved working directly with the metric (as opposed to null tetrads) and embedding the spacetime into a 5-dimensional complex spacetime. 

Despite the similarities with the Newman-Janis trick, Giampieri's procedure has received almost no attention and there are only a few papers that mention his contribution \cite{erbin2015demianski}.   

Just like the Newman-Janis trick, Giampieri's method involves starting with the Schwarzschild metric in advanced Eddington-Finkelstein coordinates

\begin{equation}
ds^{2}= \Bigl(1-\frac{2\:m}{r}\Bigl)\:{du}^{2}+2\:du\:dr-{r}^2({d\theta}^{2}+{{\sin}^{2}\theta}\:{d\phi}^2).
\end{equation}

From this construction, we let $u$ and $r$ take on complex values and let some of the $r$ terms become complex conjugates of $r$ in a specific way to give us

\begin{equation}
ds^{2}= \Bigl(1-m \Bigl(\frac{1}{r}+\frac{1}{\bar{r}}\Bigl)\Bigl)\:{du}^{2}+2\:du\:dr-{r}\bar{r}({d\theta}^{2}+{{\sin}^{2}\theta}\:{d\phi}^2).
\end{equation}

We now perform a complex coordinate transformation similar to the one in the Newman-Janis trick (\ref{NJcoordinate}), except that we introduce a new real coordinate $\theta^{*}$, 

\begin{align*}
u &\mapsto u'=u-{i}\:{a}\:{\cos\theta^{*}},\\
r &\mapsto r'=r+{i}\:{a}\:{\cos\theta^{*}}.\\
\end{align*}

This step correlates to embedding our spacetime into a five dimensional one.  Substituting these coordinates into the previous metric and dropping the primes one obtains the metric 

	\begin{multline}
	ds^{2}= (1-\frac{2\:m\:r}{r^{2}+a^{2}{\cos}^{2}\theta^{*}})\:{du}^{2}+2\:du\:dr+\frac{4\:m\:r\:i\:a\sin\theta^{*}}{r^{2}+a^{2}{\cos}^{2}\theta^{*}}du\:d\theta^{*}\\
	-2\:i\:a\:\sin\theta^{*}\:d\theta^{*}\:dr+ 
	(1+\frac{2\:m\:r}{r^{2}+a^{2}{\cos}^{2}\theta^{*}})a^{2}\:{\sin}^{2}\theta^{*}d\theta^{*2} \\
	-({r^{2}+a^{2}{\cos}^{2}\theta^{*}})({d\theta}^{2}+{{\sin}^{2}\theta}\:{d\phi}^2).
	\end{multline}

The puzzling aspect of the next step is described by the fact that we have to perform the ansatz 

	\begin{equation} \label{giampieriansatz}
	i\:\frac{d\theta^{*}}{\sin\theta^{*}}=d\phi 
	\end{equation}
	
	followed by the substitution 
	
	\begin{equation}
	\theta=\theta^{*}.
	\end{equation}

	From this, Giampieri obtains the Kerr metric 
	
	\begin{multline}
	ds^{2}= (1-\frac{2\:m\:r}{r^{2}+a^{2}{\cos}^{2}\theta})\:{du}^{2}+2\:du\:dr+\frac{4\:m\:r\:a\sin^{2}\theta}{r^{2}+a^{2}{\cos}^{2}\theta}du\:d\phi \\
	-2\:a\:\sin^{2}\theta\:d\phi\:dr 
	-(({r^{2}+a^{2}{\cos}^{2}\theta})\:a^{2}\sin^{2}\theta + 2\:m\:r\:a^{2}\sin^{2}\theta\\
	+({r^{2}+a^{2}{\cos}^{2}\theta})^{2})\frac{\sin^{2}\theta}{({r^{2}+a^{2}{\cos}^{2}\theta})}d\phi^{2} 
	-({r^{2}+a^{2}{\cos}^{2}\theta})\:{d\theta}^{2}.
	\end{multline}

Giampieri argues that since his method does not involve introducing null tetrads, it is much simpler.  The reason for this is that, introducing null tetrads assumes certain extra structures on a spacetime such as a spinor structure \cite{flaherty1976hermitian, geroch1968spinor}.  

For our purposes, it is interesting to note that this method involves embedding the spacetime into a five-dimensional one and performing a mysterious ansatz (\ref{giampieriansatz}).  

In \cite{erbin2015demianski}, a generalization of this method is given for certain spacetimes other than Schwarzschild.

\section{An alternative version}

In this section, we present a new variation of Giampieri's method that I obtained (in collaboration with my supervisor) and which is not found in the literature.  Just like the Newman-Janis trick, this method obtains the Kerr metric from the Schwarzschild metric without the use of a null tetrad, but with the advantage of being computationally fast.  For this section, we shall work in the Lorentzian signature $(-+++)$.

One starts off with Schwarzschild solution in the Kerr-Schild form 

\begin{equation}
ds^{2} = -dt^{2} + dr^{2} + \frac{2m}{r}\:(dr + dt)^{2} + r^{2}\:(d\theta^{2} + \text{sin}^{2}\:\theta \:d\phi^{2}),
\end{equation}

and performs a coordinate transformation to coordinates ($t, r, s, \phi$) by  

\begin{align}
\begin{split}
s &= \text{sin}\:(\theta) \\
 ds &= \text{cos}\:(\theta)\: d\theta, \\
 d\theta^{2} &= \frac{ds^{2}}{1-s^{2}}.
\end{split}
\end{align}

Here Schwarzschild coordinates $t, r$ also undergo transformations so we can write the Schwarzschild metric as

\begin{equation} \label{MattSch}
ds^{2} = \Bigl(-1 + \frac{m}{r}\Bigl)\:dt^{2} + \frac{2m}{r}\:dt \: dr + (1 + \frac{m}{r})\:dr^{2} + \Bigl(\frac{r^{2}}{-s^{2}+1}\Bigl)\:ds^{2} + r^{2}\:s^{2}\:d\phi^{2}. 
\end{equation}

It is interesting to note that all the metric components are rational polynomial functions and also, one can verify that this is indeed the Schwarzschild metric by calculating the relevant geometric quantities.

Now we perform an ansatz 

\begin{align}  \label{MattAnsatz}
\frac{1}{r} \; \longrightarrow \; \text{Re}\Biggl(\frac{1}{(r+i\:a\:\sqrt{1-s^{2}})}\Biggl) \; = \; \frac{r}{r^{2} + a^{2}\:(1-s^{2})}, \\
r^{2} \; \longrightarrow \;  |r+i\:a\:\sqrt{1-s^{2}}|^{2} \; = \; r^{2} + a^{2}\:(1-s^{2}), \\
dr \; \longrightarrow \; dr + a\:s^{2}\:d\phi, 
\end{align}

and substitute the above transformation at each relevant term in (\ref{MattSch}), we obtain the Kerr metric!

The non-zero metric components of the Kerr metric in this coordinate system are:

\begin{equation}
g_{tt} = -1 + \frac{mr}{r^{2}+a^{2}\:(-s^{2} + 1)}, 
\end{equation} 

\begin{equation}
g_{rr} = 1 + \frac{mr}{r^{2}+a^{2}\:(-s^{2} + 1)}, 
\end{equation}

\begin{equation}
g_{ss} = \frac{r^{2}+a^{2}\:(-s^{2} + 1)}{-s^{2}+1}, 
\end{equation}

\begin{equation}
g_{\phi \phi} = a^{2}\:s^{4} + \frac{m\:r\:a^{2}\:s^{4}}{r^{2}+a^{2}\:(-s^{2} + 1)} + (r^{2}+a^{2}\:(-s^{2} + 1))\:s^{2},
\end{equation}

\begin{equation}
g_{tr} = \frac{m\:r}{r^{2}+a^{2}\:(-s^{2} + 1)},
\end{equation}

\begin{equation}
g_{t \phi} = \frac{m\:r\:a\:s^{2}}{r^{2}+a^{2}\:(-s^{2} + 1)},
\end{equation}

\begin{equation}
g_{r \phi} = a\:s^{2} + \frac{m\:r\:a\:s^{2}}{r^{2}+a^{2}\:(-s^{2} + 1)}.
\end{equation}

An interesting thing to note, is that all the components are rational polynomial functions.  In addition to that, we can see that the ansatz (\ref{MattAnsatz}) involves non-holomorphic transformations.  We will see in a later chapter this is not only a problem with this current method but in the Newman-Janis trick, as well as in Giampieri's trick.

An alternative version where one does not introduce tetrads nor complex components can be seen by starting with the Schwarzschild metric in Kerr-Schild coordinates

\begin{equation}
ds^{2} = -dt^{2} + dr^{2} + \frac{2m}{r}\:(dr + dt)^{2} + r^{2}\:(d\theta^{2} + \text{sin}^{2}\:\theta \:d\phi^{2}),
\end{equation}

and then performing the following transformations 

\begin{align}  
\frac{1}{r} \; \longrightarrow \; 
\frac{r}{r^{2}+a^{2}\:\text{cos}^{2}\theta} \\
r^{2} \; \longrightarrow \; r^{2}+a^{2}\:\text{cos}^{2}\theta  \\
dr \; \longrightarrow \; dr + a\:\text{sin}^{2}\:d\phi. 
\end{align}

Suprisingly, this results in the Kerr metric.

\section{Discussion}

In this chapter, we have seen a remarkable situation where the Schwarzschild metric and the Kerr metric are in some sense related by a complex coordinate transformation.  This is in contrast to a real coordinate transformation, where one would end up with the same spacetime.  

We also saw variations of the Newman-Janis trick such as Giampieri's trick which involved working directly with the metric.  This had the disadvantage of introducing an extra dimension but the advantage of not using a null tetrad structure.   

In addition to this, we introduced original variations of Giampieri's trick that were computationally fast.


\chapter{Various explanations}

\begin{chapquote}{Fox Mulder, \textit{The X-Files}}
	``Dreams are the answers to questions that we haven't figured out what to ask.''
\end{chapquote}

In some sense, the Newman-Janis trick is an answer to a question we don't yet know. It provides us with a direct path to obtain the ``right" answers, yet for 50 years since the inception of the trick, nobody knows why it works.

In this chapter, we will review the important partial explanations for the trick that have accumulated in the literature over the past 50 years.  We will also provide a contribution to this, with original work that will be highlighted in the next chapter.

\section{Kerr-Talbot explanation}

It was mentioned in the original paper on the Newman-Janis trick \cite{newman1965note}, that Roy Kerr had shown in private communication that the trick works on metrics in Kerr-Schild form (\ref{KerrSchild})

\begin{equation} \label{KerrSchild2}
g_{ab} = \eta_{ab} + H\:l_{a}\:l_{b}, \qquad g^{ab} = \eta^{ab} - H\:l^{a}\:l^{b},  
\end{equation}

where $\eta_{ab}$ is the Minkowski metric. 

Afterwards, Talbot's paper of 1969 \cite{talbot1969newman} showed explicit calculations which resulted in a general complex coordinate transformation for a subset of Kerr-Schild metrics. (Surprisingly, all known Kerr-Schild metrics lie in this subset.) 

The original Newman-Janis trick was a particular instance of this general transformation.  Hence the trick could be explained as a part of a general transformation that is sanctioned by the field equations. 

We aim to show the general argument for this explanation, in terms of the original Newman-Janis trick.  One first needs to put the relevant metrics in Kerr-Schild form and examine their relationship. 

\subsubsection*{Schwarzschild and Kerr in Kerr-Schild form}

The Kerr-Schild metrics have a property that they admit at least one Killing vector and if that vector is timelike, then the metric (\ref{KerrSchild2}), can be expressed as

\begin{equation} \label{KerrSchildtime}
ds^{2} = 2 \: du \: dv + 2\: d\zeta \:d\overline{\zeta} - 4 \sqrt{2} \; m \; \text{Re} \Biggl(\frac{1}{F,_{Y}}\Biggl)\Biggl[\frac{du + Y\:d\overline{\zeta} + \overline{Y}\:d\zeta - Y\:\overline{Y}\:dv}{1 + Y\:\overline{Y}} \Biggl]^{2},
\end{equation}

where $\eta_{ab} =2 \: du \: dv + 2\: d\zeta \:d\overline{\zeta}$.  In addition to this, the parameter $m$ is a real constant, $Y$ is a complex variable and $F$ is given by 

\begin{equation}
F(Y, \zeta, \overline{\zeta}, u+v) = \phi (Y) + [Y^{2}\:\overline{\zeta} - \zeta + (u+v)\:Y],
\end{equation}

where $\phi (Y)$ is an arbitrary holomorphic function of $Y$.  Furthermore, $Y$ is implicitly described by the function 

\begin{equation} \label{Ydef}
\phi (Y) = -Y^{2} \:\overline{\zeta} + \zeta - (u+v)\: Y.
\end{equation}

Therefore $F=0$, and one can solve for $Y$.

The Schwarzschild spacetime can be elegantly expressed in terms of (\ref{KerrSchildtime}), by setting $\phi (Y) = 0$.  Similarly, the Kerr spacetime is can be obtained by setting $\phi (Y) = -\sqrt{2}\:i\:a\:Y$.

\subsubsection*{Explanation \& Problem}

Given that the Schwarzschild metric corresponds to $\phi (Y) = 0$, one can perform the following operation to obtain a new function (corresponding to the Kerr metric)

\begin{equation}
\hat{\phi} (Y) = \phi(Y) - \sqrt{2}\:i\:a\:Y.
\end{equation}

Expressing this new function in terms of (\ref{Ydef}), we find the expression

\begin{equation}
\phi(Y) - \sqrt{2}\:i\:a\:Y = -Y^{2} \:\overline{\zeta} + \zeta - (u+v)\: Y.
\end{equation} 

Rearranging one finds that 

\begin{equation}
\phi(Y) = -Y^{2} \:\overline{\zeta} + \zeta - (u+v)\: Y + \sqrt{2}\:i\:a\:Y,
\end{equation} 

and therefore

\begin{equation}
\phi(Y) = -Y^{2} \:\overline{\zeta} + \zeta - [(u+ \sqrt{2}\:i\:a )+ v]\: Y.
\end{equation}

Hence, the Kerr metric can be derived from the Schwarzschild metric by performing the complex coordinate transformation 

\begin{equation} \label{KerrSchildcoordinate}
u \; \rightarrow \; u + \sqrt{2}\:i\:a.
\end{equation}
  
Though this explanation seems satisfactory on first sight, there are some underlying problems.  

Suppose that the Schwarzschild metric is expressed in coordinates  $(u, v, \zeta, \overline{\zeta})$.  From there one complexifies the coordinates, and then performs the transformation (\ref{KerrSchildcoordinate}).  Surprisingly, the result would not be the Kerr solution if one took a real slice.  

The source of this problem lies in the fact that while $Y(x^{a})$ from Schwarzschild transforms to $Y(x^{a^{'}} )$ for Kerr, the transformation produces ``mistakes" for $\overline{Y}(x^{a})$.  What one needs to do, to produce the Kerr metric, is rewrite $\overline{Y}(u)$ as $\overline{Y}(\overline{u})$, which is a non-holomorphic transformation, i.e. $u \rightarrow \overline{u}$. 

We shall find that non-holomorphic transformations play the central role of the problem regarding the Newman-Janis trick.

\section{Newman's explanation}

Following the construction of the Newman-Janis trick, in \cite{newman1973complex}, Newman provided a partial explanation for the trick from the perspective of complex Minkowski space.

\subsubsection*{Explanation}

One starts by writing the Schwarzschild metric in the Kerr-Schild form (\ref{KerrSchild2}) 

\begin{equation} \label{KerrSchild3}
g^{ab} = \eta^{ab} - H\:l^{a}\:l^{b},
\end{equation}

with coordinates such that $l^{a} = (0,1,0,0)$.  In addition to this, the Minkowski metric is given by 

\begin{equation} \label{Minkowski}
\eta_{ab}\:dx^{a}\:dx^{b} = du^{2} + 2\:du\:dr - r^{2}(d\theta^{2}+sin^{2}\theta\:d\phi^{2}).
\end{equation}

The tetrad for (\ref{Minkowski}) is expressed as 

\begin{align}\label{MinkowskiTetrad}
\begin{split}
l^{a}&={\partial}_{r},\\
n^{a}&={\partial}_{u}-{\frac{1}{2}} {\partial}_{r},\\
m^{a}&={\frac{1}{\sqrt{2}{r}}}({\partial}_{\theta}+\frac{i}{\sin \theta}{\partial}_{\phi}),\\
\bar{m}^{a}&={\frac{1}{\sqrt{2}{r}}}({\partial}_{\theta}-\frac{i}{\sin \theta}{\partial}_{\phi}).\\
\end{split}
\end{align}

From this, one can construct a complex Minkowski space by letting \emph{all} the coordinates become complex-valued.  We then perform a complex coordinate transformation (which is similar to (\ref{NJcoordinate}), but not equivalent) 

\begin{align} \label{Newmanstransformation}
\begin{split}
u&=u'+{i}\:{a}\:{\cos\theta},\\
r&=r'-{i}\:{a}\:{\cos\theta},\\
\text{cos} \: \theta &=\frac{r' \: \text{cos} \: \theta^{'} -i\:a}{r' - i\:a\:\text{cos} \:\theta'},\\
\text{cos}\:2(\phi - \phi^{'}) &= \frac{r'^{2}-a^{2}}{r'^{2}+a^{2}}.
\end{split}
\end{align}

The resulting metric takes the form

\begin{align} \label{MinkowksiTransformed}
\begin{split}
ds^{2} = du'^{2} + 2\:du'\:dr' - 2\:a\:\text{sin}^{2} \:\theta^{'}\:dr'\:d\phi^{'}\\
 - (r'^{2} +a^{2}  \:\text{cos}^{2}\theta^{'})(d\theta'^{2} + \text{sin}^{2}\theta' d\phi'^{2}) \\
  - a^{2}\:\text{sin}^{4}\theta' \: d\phi'^{2}.
\end{split}
\end{align}
  
The null tetrad for (\ref{MinkowksiTransformed}) is provided by

	\begin{align}\label{MinkowskiTransformedTetrad}
	\begin{split}
	l^{a}&={\partial}_{r},\\
	n^{a}&={\partial}_{u}-{\frac{1}{2}} {\partial}_{r},\\
	m^{a}&={\frac{1}{\sqrt{2}({r'+i\:a\:\cos\theta'})}}({\partial}_{\theta}+i\:a\:\sin\theta'({\partial}_{u}-{\partial}_{r})+\frac{i}{\sin \theta'}{\partial}_{\phi}),\\
	\bar{m}^{a}&= {\frac{1}{\sqrt{2}({r'-i\:a\:\cos\theta'})}}({\partial}_{\theta}-i\:a\:\sin\theta'({\partial}_{u}-{\partial}_{r})-\frac{i}{\sin \theta'}{\partial}_{\phi}).\\
	\end{split}
	\end{align}

From this step, the coordinates $u', r', \theta', \phi'$ are restricted to real values and the metric (\ref{MinkowksiTransformed}) turns out to be a real Minkowski metric.  This is expressed in the Kerr-type coordinates found at the end of the Newman-Janis trick.

One can verify that the transformed null tetrads (\ref{MinkowskiTransformedTetrad}) can be obtained by a complex Lorentz transformation from the null tetrad (\ref{MinkowskiTetrad}).

For Schwarzschild, the function $H$ in (\ref{KerrSchild3}) is given by 

\begin{equation} \label{SchH}
H = \frac{2m}{r}
\end{equation} 

in coordinates $u, r, \theta, \phi$.  To obtain the $H$ value for the Kerr metric, one needs to rewrite (\ref{SchH}) into the form

\begin{equation} \label{rewrite}
H = m\:\Bigl(\frac{1}{r} + \frac{1}{\overline{r}} \Bigl),
\end{equation}

and then perform the complex coordinate transformation (\ref{Newmanstransformation}).  This results in the $H$ value of the Kerr metric given by

\begin{equation}
H = \frac{2mr'}{r'^{2}+a^{2}\:\text{cos}^{2}\:\theta'}.
\end{equation}

In addition to this, one can verify that the Weyl tensor can be interpreted as a field on the complex Minkowski space and chosen in a way that its components with respect to tetrad (\ref{MinkowskiTetrad}) is $\Phi_{2} = -m/r^{3}$.  This is the Schwarzschild value.  The Kerr value can be obtained applying the same procedure to tetrad (\ref{MinkowskiTransformedTetrad}).

\subsubsection*{Problem}

One can argue that the above transformation has ability to relate the different Kerr-Schild metrics and provides an explanation to why the Newman-Janis trick is successful.  

In addition to this, this explanation has a geometric interpretation (i.e. complex Minkowski space) which the Kerr-Talbot explanation lacks.  

Nonetheless, it seems that the problem with Newman's interpretation is similar to the problems experienced by the Kerr-Talbot interpretation, which is the conjugation process of certain coordinates.  

In particular, rewriting (\ref{SchH}) to (\ref{rewrite}) presents an arbitrariness of writing functions of $r$ which cannot be explained by Newman's interpretation (one could as well have replaced (\ref{SchH}) in another way). If one did not do this step, then the desired result for $H$ cannot be obtained.  

The non-holomorphic transformation $r \rightarrow \overline{r}$ and the arbitrariness involved, plays the central role in misaligning this interpretation with why the Newman-Janis trick works.

\subsubsection*{Further remarks}

The question one would like to ask is whether there is a way of bypassing the process of arbitrarily conjugating certain coordinates.  More precisely can we relate the complexified Schwarzschild metric and the complexified Kerr metric by means of a holomorphic transformation. 

An argument for why such a transformation cannot exist is provided by Newman and Winicour through private correspondance as mentioned in \cite{flaherty1976hermitian}.  

Suppose a holomorphic transformation existed between the complexified Schwarz\-schild metric and the complexified Kerr metric.  We know that the Schwarzschild metric has three $\mathbb{R}$-linearly independent Killing vectors, $U, V$ and $W$.  Under a holomorphic transformation, these would be mapped to three Killing complex vectors on Kerr, $U_{1} + i\:U_{2}$,  $V_{1} + i\:V_{2}$ and $W_{1} + i\:W_{2}$.  

Since $U, V$ and $W$  are $\mathbb{R}$-linearly independent, this implies that $U_{1}, V_{1}$ and $W_{1}$ are $\mathbb{R}$-linearly independent Killing vectors in Kerr.  Taking the real part of the complexified Kerr metric would result in the Kerr metric.  This presents a contradiction, since we know that Kerr has two $\mathbb{R}$-linearly independent Killing vectors. Therefore such a holomorphic transformation could not exist.

Further work was developed by Newman and collaborators on complex Minkowski spaces and its application to the Newman-Janis trick.  This resulted in the theory of ``Heavenly" spacetimes. \cite{newman1976heaven, adamo2009null}.  

In this construction, the Schwarzschild and Kerr spacetime have an associated complex Minkowski space (called H-space, in this context).  A complex center of mass line can be constructed on these associated spaces.  Furthermore, it can be shown that the complex center of mass lines corresponding to Schwarzschild differs from Kerr by an imaginary translation.

The physical interpretation of such a translation is that it corresponds to an intrinsic spin of a system.  Therefore, one can interpret the angular momentum of the Kerr spacetime as being in a direct relationship with the intrinsic spin of its associated complex Minkowski space.
  
Though these results hint at a deeper structure, the direct relationship to steps of the Newman-Janis trick is still lacking.

\section{Flaherty's explanation}

A major component of Flaherty's \cite{flaherty1976hermitian} work involved finding a mathematically elegant way to express the Newman-Janis trick.  

He suggests this can be done by considering a four-complex-dimensional complex manifold $\mathbb{C}M$ with coordinates $z^{a} = (u, r, \theta, \phi)$.  From there, a Hermitian metric $g_{a\overline{b}}$ is put on the manifold and expressed via contravariant components

\begin{align} \label{FlahertyHermitianmetric}
\begin{split}
\Bigl(\frac{\partial}{\partial s}\Bigl)^{2} = g^{a\overline{b}}\: \frac{\partial}{\partial z^{a}} \frac{\partial}{\partial\overline{ z^{b}}} + g^{\overline{a}b} \frac{\partial}{\partial \overline{z^{a}}} \frac{\partial}{\partial z^{b}} \\ 
= \partial_{r}\:\partial_{\overline{u}} +  \partial_{u}\:\partial_{\overline{r}} \: - \:\Bigl(1 - \frac{m}{r} - \frac{m}{\overline{r}} \Bigl) \partial_{r}\:\partial_{\overline{r}}  -\frac{1}{r\:\overline{r}}\:\partial_{\theta}\:\partial_{\overline{\theta}} 
- \frac{1}{r\: \text{sin} \: \theta \; \overline{r}\:\text{sin}\overline{\theta}}\:\partial_{\phi}\:\partial_{\overline{\phi}}\\
 + \frac{2\:i}{r\:\overline{r}\:(\text{sin}\:\theta + \text{sin}\:\overline{\theta})}\:\partial_{\theta}\:\partial_{\overline{\phi}} 
- \frac{2\:i}{r\:\overline{r}\:(\text{sin}\:\theta + \text{sin}\:\overline{\theta})}\:\partial_{\phi}\:\partial_{\overline{\theta}},
\end{split}
\end{align}

where $m$ is a real constant.  This is a Hermitian metric since $\overline{g^{a\overline{b}}} = g^{\overline{a}b}$.  From such a construction, if one sets the reality conditions $u=\overline{u}$, $r=\overline{r}$, $\theta = \overline{\theta}$, $\phi = \overline{\phi}$, then one is able to obtain a real metric on $M$ where $\mathbb{C}M$ was the complex extension of $M$.  This metric is given by

\begin{equation}
\Bigl(\frac{\partial}{\partial s} \Bigl)_{M}^{2} = 2\:  \partial_{u} \partial_{r} - \Bigl(1 - \frac{2m}{r}\Bigl) \:\partial_{r}\: \partial_{r} - \frac{1}{r^{2}}\:\partial_{\theta}\:\partial_{\theta} - \frac{1}{r^{2}\:\text{sin}^{2}\:\theta} \partial_{\phi}\:\partial_{\phi}, 
\end{equation}

which turns out to be the Schwarzschild metric.  Turning our attention back to the complex manifold $\mathbb{C}M$, and performing the complex coordinate transformation (which is the same as the original Newman-Janis trick with $a$ being a real constant), 

\begin{align}
\begin{split}
u' &= u - i\:a\:\text{cos}\theta, \\
r' &= r + i\:a\:\text{cos}\theta, \\
\theta' &= \theta, \\
\phi' &= \phi,
\end{split}
\end{align}  

\begin{align}
\begin{split}
\overline{u}' &= \overline{u} + i\:a\:\text{cos}\overline{\theta}, \\
\overline{r}' &= \overline{r} - i\:a\:\text{cos}\overline{\theta}, \\
\overline{\theta}' &= \overline{\theta}, \\
\overline{\phi}' &= \overline{\phi},
\end{split}
\end{align} 

the Hermitian metric (\ref{FlahertyHermitianmetric}) becomes

\begin{align}
\begin{split}
\Bigl(\frac{\partial}{\partial s}\Bigl)^{2} = \partial_{u'}\partial_{\overline{r}'} \: + \: \partial_{r'}\:\partial_{\overline{u}'} - \Bigl(1 - \frac{m}{r' - i\:a\:\text{cos}\:\theta '} - \frac{m}{\overline{r}' + i\:a\:\text{cos}\:\overline{\theta} '}  \Bigl)\:\partial_{r'}\:\partial_{\overline{r}'} \\
- \frac{1}{(r' - i\:a\:\text{cos}\:\theta ')\:(\overline{r}' + i\:a\:\text{cos}\:\overline{\theta} ')} \:\Bigl[(\partial_{\theta'}  + i\:a\:\text{sin}\:\theta'\:\partial_{u'} - i\:a\:\text{sin}\:\theta'\:\partial_{r'})  \\
(\partial_{\overline{\theta}'}  - i\:a\:\text{sin}\:\overline{\theta}'\:\partial_{\overline{u}'} + i\:a\:\text{sin}\:\overline{\theta}'\:\partial_{\overline{r}'})   \Bigl] \: - \: \frac{1}{(r' - i\:a\:\text{cos}\:\theta ')\:\text{sin}\:\theta^{'}(\overline{r}' + i\:a\:\text{cos}\:\overline{\theta} ')\:\text{sin}\:\overline{\theta^{'}}} \partial_{\phi^{'}} \: \partial_{\overline{\phi '}} \\
+ \frac{2\:i}{(r' - i\:a\:\text{cos}\:\theta ')\:(\overline{r}' + i\:a\:\text{cos}\:\overline{\theta} ')\:(\text{sin}\:\theta' + \text{sin}\:\overline{\theta'})}  \\
\Bigl[(\partial_{\theta'}  + i\:a\:\text{sin}\:\theta'\:\partial_{u'} - i\:a\:\text{sin}\:\theta'\:\partial_{r'}) \:\partial_{\overline{\phi'}} \: - \: \partial_{\phi '} \: 
(\partial_{\overline{\theta}'}  - i\:a\:\text{sin}\:\overline{\theta}'\:\partial_{\overline{u}'} + i\:a\:\text{sin}\:\overline{\theta}'\:\partial_{\overline{r}'})   \Bigl].
\end{split}
\end{align}

If we now impose the reality conditions $u' = \overline{u}'$, $r' = \overline{r}'$, $\theta' = \overline{\theta}'$, $\phi' = \overline{\phi}'$, then one can find this is reduced to the Kerr metric on a real slice $M'$ of $\mathbb{C}M$.

Flaherty was able to generalize this geometric formulation to special Kerr-Schild metrics with the Kerr-Talbot complex coordinate transformation.   

To summarize, we find that the Newman-Janis trick can be put on a more elegant mathematical footing by using a four-complex dimensional Hermitian manifold. 

Nonetheless, the weakness of this explanation lies in that it does not explain the ambiguity in complex conjugating the $r$ coordinate in certain terms, of the original Newman-Janis trick (\ref{Tetrad 2}).  In addition to this, very little can be said of what the physics of the Newman-Janis trick is, from Flaherty's geometric formulation.

\section{Schiffer et al. explanation}

In a paper by Schiffer et al \cite{schiffer1973kerr}, it was shown the Kerr metric can be considered as a complexification of the Schwarzschild metric in a completely different way.  The exact relationship between this approach and the Newman-Janis trick is not clear.

A particular subset of Kerr-Schild metrics can be calculated from a complex potential function $\gamma$ in flat 3-space.  This potential is harmonic and simultaneously satisfies the equations 

\begin{equation}
\nabla^{2} \gamma = 0, \qquad (\nabla \gamma)^{2} = \gamma^{4}.
\end{equation} 

The Schwarzschild metric can be generated by the potential 

\begin{equation}
\gamma = \frac{1}{\sqrt{x^{2} + y^{2} + z^{2}}}, 
\end{equation}

and performing the complex coordinate transformation $z \rightarrow z-i\:a$, constructs the Kerr solution!

Generalizations of this method for the Reissner-Nordstr\"{o}m and Kerr-Newman geometries can be seen in \cite{finkelstein1975general}.

\section{Drake-Szekeres' explanation}

In the paper by S.P. Drake and P. Szekeres \cite{szekeres1998explanation}, a number of important results were mathematically proven for a special case of metrics.  Their analysis involved starting with metrics of the form

\begin{equation} \label{seedmetrics}
ds^{2} = -e^{2\phi(r)}\: du^{2} - 2\:e^{\lambda(r) + \phi(r)}\:du \: dr + r^{2}\:(d\theta^{2} + \text{sin}^{2}\theta d\phi^{2}),
\end{equation}

which is in Eddington-Finkelstein coordinates.  They also worked with the specific complex coordinate transformation 

\begin{align}
\begin{split}
u' &= u - i\:a\:\text{cos}\theta, \\
r' &= r + i\:a\:\text{cos}\theta, \\
\theta' &= \theta, \\
\phi' &= \phi,
\end{split}
\end{align}

for a Newman-Janis trick on metrics of the form given by (\ref{seedmetrics}). The following results were proven.

\begin{theorem}
	The only perfect fluid generated by the Newman-Janis trick is the vacuum (i.e. the Kerr metric).  (See Drake and Szekeres \cite{szekeres1998explanation}.)  
\end{theorem}

For the next theorem, it is worth noting that the Schwarzschild and the Kerr metrics are spacetimes of type D.  For more information on this algebraic classification scheme, refer to \cite{stephani2003exact}.

\begin{theorem}
	The only algebraically special spacetimes generated by the Newman-Janis trick are Petrov type D. (See Drake and Szekeres \cite{szekeres1998explanation}.)  
\end{theorem}

\begin{theorem}
	The only Petrov type D spacetime generated by the Newman-Janis algorithm with a vanishing Ricci scalar is the Kerr-Newman spacetime.  (See Drake and Szekeres \cite{szekeres1998explanation}.)  
\end{theorem}

\section{Discussion}

In this chapter, we have been presented with a wide range of partial explanations and analysis.  The most important point to note is that the analysis so far has suffered from explaining the ambiguity involved in conjugating certain coordinate terms.  This has been evident in both Kerr-Talbot's explanation as well as in Newman's explanation.

\chapter{Original contribution}

In this chapter, we present some results that I obtained (in collaboration with my supervisor) regarding a wide range of different issues about the Newman-Janis trick.

\section[Newman-Janis versus Giampieri]{Equivalence between the Newman-Janis trick and Giampieri's method}

In this section, we show the equivalence between the Newman-Janis trick and Giampieri's method.  The former involves the complexification of null tetrads while the latter involves embedding the spacetime in a 5-dimensional manifold and performing an arbitrary ansatz.  On first sight, these two approaches do not seem to have a direct relationship.  

In \cite{erbin2015demianski, erbin2014janis}, it was mentioned that these two approaches were completely equivalent and a few details were given.  Here we provide a different way to look at the equivalence and present steps to see the direct correspondance between the Newman-Janis trick and Giampieri's method. 

The main ingredient for the following argument is that the Newman-Janis trick involves a hidden tetrad, that was not explicitly highlighted in the original presentation of the trick. 

To see this, let us review briefly the steps of the Newman-Janis trick with the hidden tetrad explicitly spelled out.  One first starts with the Schwarzschild tetrad, which we shall call Tetrad 1: 

\begin{align}\label{NJTetrad 1}
l^{a}&={\partial}_{r},\nonumber\\
n^{a}&={\partial}_{u}-{\frac{1}{2}}\Bigl(1-{\frac{2m}{r}\Bigl)} {\partial}_{r}\nonumber,\\
m^{a}&={\frac{1}{\sqrt{2}{r}}}\Bigl({\partial}_{\theta}+\frac{i}{\sin \theta}{\partial}_{\phi}\Bigl)\nonumber,\\
\bar{m}^{a}&={\frac{1}{\sqrt{2}{r}}}\Bigl({\partial}_{\theta}-\frac{i}{\sin \theta}{\partial}_{\phi}\Bigl)\nonumber.\\
\end{align}  

Notice the trivial statement that Tetrad One has vectors $m^{a}$ and  ${\bar{m}}^{a}$ which are complex conjugates of each other. It is these specific vectors of a null tetrad that one should keep the focus on while reading the following steps.

The Newman-Janis trick proceeds to let $r$ take complex values and then introduce the complex conjugates of $r$.  We will introduce the following such tetrad as Tetrad Two

	\begin{align}\label{NJTetrad 2}
	l^{a}&={\partial}_{r},\nonumber\\
	n^{a}&={\partial}_{u}-{\frac{1}{2}}\Bigl(1-{\frac{m}{r}-\frac{m}{\bar{r}}\Bigl)} {\partial}_{r}\nonumber,\\
	m^{a}&={\frac{1}{\sqrt{2}{\bar{r}}}}\Bigl({\partial}_{\theta}+\frac{i}{\sin \theta}{\partial}_{\phi}\Bigl)\nonumber,\\
	\bar{m}^{a}&={\frac{1}{\sqrt{2}{r}}}\Bigl({\partial}_{\theta}-\frac{i}{\sin \theta}{\partial}_{\phi}\Bigl)\nonumber.\\
	\end{align}

As mentioned before, this step is ambiguous as to which terms involving $r$ coordinates get complex conjugated, but notice again that the vectors $m^{a}$ and  ${\bar{m}}^{a}$ are complex conjugates of each other. Hence they satisfy the requirements of being part of a null tetrad.

The next step involves performing the complex coordinate transformation

\begin{align} \label{Complexcoordinatetransformation}
u &\mapsto u'=u-{i}\:{a}\:{\cos\theta},\\
r &\mapsto r'=r+{i}\:{a}\:{\cos\theta},\\
\theta &\mapsto {\theta}'=\theta,\\
\phi &\mapsto {\phi}'=\phi.
\end{align}

This implies that the basis vectors transform as

\begin{align}
{\partial}_{u}&= {\partial}_{u'},\\
{\partial}_{r}&= {\partial}_{r'},\\
{\partial}_{\theta}&= {\partial}_{{\theta}'}+{i}\:{a}\:{\sin\theta}({\partial}_{u'}-{\partial}_{r'}),\\
{\partial}_{\phi}&={\partial}_{\phi '}.
\end{align}

If we now perform this complex coordinate transformation on Tetrad Two, and replace the primed letters with unprimed letters (this is usually done later in the NJ trick but we choose to do it now for a cleaner notation for the purposes of this subsection), we obtain what can be called Tetrad Three

\begin{align}\label{NJTetrad 3}
l^{a}&={\partial}_{r},\nonumber\\
n^{a}&={\partial}_{u}-{\frac{1}{2}}\Bigl(1-{\frac{2\:m\:r}{r^{2}+a^{2}{\cos}^{2}\theta}}\Bigl) {\partial}_{r}\nonumber,\\
m^{a}&={\frac{1}{\sqrt{2}({r+i\:a\:\cos\theta})}}({\partial}_{\theta}+i\:a\:\sin\theta({\partial}_{u}-{\partial}_{r})+\frac{i}{\sin \theta}{\partial}_{\phi})\nonumber,\\
\bar{m}^{a}&= {\frac{1}{\sqrt{2}({r-i\:a\:\cos\theta})}}({\partial}_{\theta}+i\:a\:\sin\theta({\partial}_{u}-{\partial}_{r})-\frac{i}{\sin \theta}{\partial}_{\phi})\nonumber.\\
\end{align}

The most crucial piece of this subsection is Tetrad Three, which is the hidden tetrad.  This tetrad is not mentioned at all in the literature.  What usually happens is that the literature on the Newman-Janis trick moves straight to Tetrad Four (which is what we shall come to in the next step).  It somehow doesn't see there is an extra operation involved, thereby missing out on explicitly mentioning Tetrad Three.  

To be more precise, the part of the way the original paper presents the Newman-Janis trick, is to say that the trick involves keeping $m^{a}$ and $\overline{m}^{a}$ complex conjugates of each other, throughout the procedure.  But from a strict mathematical perspective, we see that this should constitute an extra step.  

Therefore, applying the Newman-Janis complex coordinate transformation on Tetrad Two does not produce Kerr (more accurately, a complexified Kerr)!  It produces something which is not a null tetrad.  There needs to be an extra step outlined which is that one needs to change the vector $\bar{m}^{a}$ to be the actual complex conjugate to the vector $m^{a}$.  This is done by changing the plus sign to a minus sign in front of the term $i\:a\:\sin\theta({\partial}_{u}-{\partial}_{r})$ in the $\bar{m}^{a}$ vector.

Once we do that, we get Tetrad Four

\begin{align}\label{NJTetrad 4}
l^{a}&={\partial}_{r},\nonumber\\
n^{a}&={\partial}_{u}-{\frac{1}{2}}(1-{\frac{2\:m\:r}{r^{2}+a^{2}{\cos}^{2}\theta)}}) {\partial}_{r},\nonumber\\
m^{a}&={\frac{1}{\sqrt{2}({r+i\:a\:\cos\theta})}}({\partial}_{\theta}+i\:a\:\sin\theta({\partial}_{u}-{\partial}_{r})+\frac{i}{\sin \theta}{\partial}_{\phi}),\nonumber\\
\bar{m}^{a}&= {\frac{1}{\sqrt{2}({r-i\:a\:\cos\theta})}}({\partial}_{\theta}-i\:a\:\sin\theta({\partial}_{u}-{\partial}_{r})-\frac{i}{\sin \theta}{\partial}_{\phi}).\nonumber\\
\end{align}

A more mathematically precise way to express this sign change is to perform the following tetrad leg transformation on Tetrad 3, 

\begin{align} 
\begin{split}
l^{a} \; &\longrightarrow \; \hat{l}^{a} = l^{a}, \\
n^{a} \; &\longrightarrow \; \hat{n}^{a} = n^{a},\\
m^{a} \; &\longrightarrow \; \hat{m}^{a} = m^{a}, \\
\overline{m}^{a} \; &\longrightarrow \; \hat{\overline{m}}^{a} = \frac{-2\:i\:a\:\text{sin}\:\theta}{\sqrt{2}\:(r-i\:a\:\text{cos}\:\theta)} \Bigl(n^{a} - \frac{1}{2}\Bigl(1 + \frac{2mr}{r^{2}+a^{2}\:\text{cos}^{2}\:\theta}\Bigl)\:l^{a}\Bigl)
 + \;\overline{m}^{a}.
\end{split}
\end{align}

Typically in the literature, Tetrad Four is the tetrad that is shown to be the tetrad after the complex coordinate transformation is performed on Tetrad Two.  

From a strict mathematical perspective, we see that this cannot be true.  In fact, one needs  to somehow ``fix up" the vector $\bar{m}^{a}$.  This is needed, so to be able to obtain Tetrad Four which is the Kerr metric (once the coordinates are made real).

We shall now show that Tetrad Three is the key to explaining Giampieri's method, from the perspective of the Newman-Janis trick. 

One can think of Giampieri's method in a way where one does not need to introduce a fifth term (thereby avoid embedding into a five dimensional spacetime) and perform the ansatz on specific $d\theta$ terms of the metric.  This is completely equivalent to the original procedure.

We start off with the usual Schwarzschild metric which we shall call Metric One  

\begin{equation} \label{Metric1}
ds^{2}= \Bigl(1-\frac{2\:m}{r}\Bigl)\:{du}^{2}+2\:du\:dr-{r}^2({d\theta}^{2}+{{\sin}^{2}\theta}\:{d\phi}^2).
\end{equation}

We can easily verify that Tetrad One from the previous section is the corresponding null tetrad for this metric.

We then proceed to the next metric in Giampieri's method, which we name as Metric Two (so far everything has been the same as the original Giampieri's method)

\begin{equation}
ds^{2}= \Bigl(1-\frac{2\:m}{2}\Bigl(\frac{1}{r}+\frac{1}{\bar{r}}\Bigl)\Bigl)\:{du}^{2}+2\:du\:dr-{r}\bar{r}({d\theta}^{2}+{{\sin}^{2}\theta}\:{d\phi}^2).
\end{equation}

One can easily verify that the corresponding null tetrad for Metric Two is Tetrad Two.

Giampieri introduces a new coordinate $\theta^{*}$ in the complex coordinate transformation, but we will not do this and rather, put $\theta$ in the complex coordinate transformation.  Giampieri's fifth coordinate is unnecessary and we'll find we get the same result by rather performing

\begin{align}
u &\mapsto u'=u-{i}\:{a}\:{\cos\theta},\\
r &\mapsto r'=r+{i}\:{a}\:{\cos\theta}.\\
\end{align}

Notice that this is now exactly the complex coordinate transformation, in the Newman-Janis trick.

From there, we derive Metric Three given by

\begin{multline}\label{Metric Three}
ds^{2}= (1-\frac{2\:m\:r}{r^{2}+a^{2}{\cos}^{2}\theta})\:{du}^{2}+2\:du\:dr+\frac{4\:m\:r\:i\:a\sin\theta}{r^{2}+a^{2}{\cos}^{2}\theta}du\:d\theta\\
-2\:i\:a\:\sin\theta\:d\theta\:dr+ 
(1+\frac{2\:m\:r}{r^{2}+a^{2}{\cos}^{2}\theta})a^{2}\:{\sin}^{2}\theta d\theta^{2} \\
-({r^{2}+a^{2}{\cos}^{2}\theta})({d\theta}^{2}+{{\sin}^{2}\theta}\:{d\phi}^2)
\end{multline}

This metric is exactly like the Giampieri's metric after the complex coordinate transformation, but with terms involving $\theta^{*}$ replaced by $\theta$.  One way to think about this compared to Giampieri's corresponding metric is that it's simply a notational difference, not a mathematical difference.

What is striking is that this metric corresponds to the hidden tetrad we found in the previous paragraphs, i.e. Tetrad Three!   

When one simplifies Metric Three, we see that it can be written as

\begin{multline}
ds^{2}= (1-\frac{2\:m\:r}{r^{2}+a^{2}{\cos}^{2}\theta})\:{du}^{2}+2\:du\:dr+\frac{4\:m\:r\:i\:a\sin\theta}{r^{2}+a^{2}{\cos}^{2}\theta}du\:d\theta\\
-2\:i\:a\:\sin\theta\:d\theta\:dr+ 
((1+\frac{2\:m\:r}{r^{2}+a^{2}{\cos}^{2}\theta})a^{2}\:{\sin}^{2}-({r^{2}+a^{2}{\cos}^{2}\theta}))d\theta^{2} \\
-({r^{2}+a^{2}{\cos}^{2}\theta})({{\sin}^{2}\theta}){d\phi}^2.
\end{multline}

From this, it easy to see that this metric can be built up from Tetrad Three following the standard formula

\begin{equation}
g_{ab}=l_{a}\:n_{b}+n_{a}\:l_{b}-m_{a}\:\bar{m}_{b}-\bar{m}_{a}\:m_{b}.
\end{equation}

In Giampieri's method, we didn't really understand the significance of this metric and now we have the understanding that this is directly related to the Newman-Janis trick, with the explicit connection being Tetrad Three.

Going back to our modified but equivalent version of Giampieri's method we perform the ansatz on Metric Three with a slight difference.  We're going to perform the ansatz on every applicable term except the last term in (\ref{Metric Three}).  This would be exactly the same as the original Giampieri's trick where we perform the ansatz on only the $d\theta^{*}$ terms (and not on the $d\theta$ terms).  The ansatz is given by

\begin{equation}
i\:\frac{d\theta}{\sin\theta}=d\phi. 
\end{equation}

Substituting this all in, we obtain the Kerr metric which we call Metric Four

\begin{multline}
ds^{2}= (1-\frac{2\:m\:r}{r^{2}+a^{2}{\cos}^{2}\theta})\:{du}^{2}+2\:du\:dr+\frac{4\:m\:r\:a\sin^{2}\theta}{r^{2}+a^{2}{\cos}^{2}\theta}du\:d\phi-2\:a\:\sin^{2}\theta\:d\phi\:dr \\
-(({r^{2}+a^{2}{\cos}^{2}\theta})\:a^{2}\sin^{2}\theta + 2\:m\:r\:a^{2}\sin^{2}\theta\\
+({r^{2}+a^{2}{\cos}^{2}\theta})^{2})\frac{\sin^{2}\theta}{({r^{2}+a^{2}{\cos}^{2}\theta})}d\phi^{2} 
-({r^{2}+a^{2}{\cos}^{2}\theta})\:{d\theta}^{2}
\end{multline}

Unsurprisingly, this metric can be built from Tetrad Four as the corresponding null tetrad.

To summarize, for our version of Giampieri's method (which is equivalent to the original Giampieri's method), the only ambiguity (we don't have a fifth dimension) is in the ansatz given by 

\begin{equation*}
i\:\frac{d\theta}{\sin\theta}=d\phi. 
\end{equation*} 

The effect of this ansatz, is that it takes us from Metric Three to Metric Four.  In other words from Tetrad Three to Tetrad Four. 

To go from Tetrad Three to Tetrad Four we have to ``fix up" the $\bar{m}^{a}$ term to make it into the actual complex conjugate of $m^{a}$.  This was highlighted in the above paragraphs.

Therefore, Giampieri's ansatz on Metric Three corresponds to the operation of ``fixing up" the $\bar{m}^{a}$ term in Tetrad Three.  

This concludes the subsection on explaining Giampieri's method with its arbitrary ansatz, from the viewpoint of the Newman-Janis trick.

For the purposes of the next section, it is very important to note that in the original Newman-Janis trick it was mentioned to keep $m^{a}$ and $\overline{m}^{a}$ as complex conjugates of each other during the whole trick.  \emph{This assumption is very restrictive, and strictly speaking should constitute an extra step.}  As we have seen, this was useful in terms of showing the equivalence between the Newman-Janis trick and Giampieri's method.

When one deconstructs the Newman-Janis trick, it can seen that it is not a very clean mathematical procedure.  One has to to an arbitrary conjugation process, then do a complex coordinate transformation, and then finally perform a tetrad leg transformation on $\overline{m}^{a}$ to make the necessary sign change.  

Contrary to the popular account of the trick, it is not as clean as performing a complex coordinate transformation.

One can of course make the Newman-Janis trick more elegant by introducing a  $5$th coordinate term which would be a complex conjugate to $\theta$.  Performing the complex coordinate transformation and then plugging the relevant basis vectors in, would produce the desired output without any procedures of manually changing signs.  The disadvantage of this approach would be that one would have to introduce extra dimensionality into the problem.  We will not pursue such an direction.

\section{Non-holomorphic problems}

The conventional focus regarding the Newman-Janis trick is the complex coordinate transformation (holomorphic transformation) (\ref{Complexcoordinatetransformation}), but we will argue in this section that the main focus should be the non-holomorphic facets of the trick.  In particular, these non-holomorphic aspects are what makes the Newman-Janis trick unique. 

\subsubsection*{Non-holomorphic transformation}

The first problem can be seen in the transformation where Tetrad 2 (\ref{NJTetrad 2}) is obtained from Tetrad 1 (\ref{NJTetrad 1}).  Not only is there the ambiguity in which of the $r$ coordinates are to be complex conjugated, but the more serious problem is that this process involves a non-holomorphic transformation via $r \rightarrow \overline{r}$.  

Therefore to go from Tetrad 1 to Tetrad 2 is a transformation which cannot be explained using any of the tools of standard complex manifold theory.  This is due to the fact that complex manifolds by their definition require their transition functions to be holomorphic.

\subsubsection*{Non-holomorphic geometric quantities}

Tetrad 2 (\ref{NJTetrad 2}) is the crucial piece of the Newman-Janis trick and the ambiguity in what $r$ terms are to be complex conjugated is what has resulted in the inability to explain the trick, as mentioned in the previous chapter.  

But what is of more direct concern is that Tetrad 2 involves geometric quantities which are not mathematically well-defined within the context of standard complex analysis.  Specifically, if one were to calculate out the Riemann tensor, the Ricci tensor and the Ricci scalar of this tetrad using 

\begin{equation} \label{nulltetradmetricinverse}
g^{ab} = l^{a}\:n^{b} + n^{a}\:l^{b} - m^{a}\:\overline{m}^{b} - \overline{m}^{a}\:m^{b},
\end{equation} 

and the inverse metric, then one would find quantities which are \emph{derivatives of the modulus of the complex coordinate $r$.}  This means that a lot of mathematics that was developed by complex analysis and complex geometry would not seem to be very useful for analyzing the Newman-Janis trick.  

\textbf{This tetrad is by far the central problem of the Newman-Janis trick.}

Any hope of deriving the physics from the Newman-Janis trick, must first solve the issue:  What is the physical significance of Tetrad 2 , equation (\ref{NJTetrad 2})?

\section{A new approach}

In this section, we present a novel (modified) alternative approach to the Newman-Janis trick.  There are three principles in developing this modified trick:

(1) We do not introduce complex conjugates for the coordinates and hence, any of the ambiguities related to this.   

(2) All our metrics and associated tetrads, throughout the process satisfy the Einstein equations.  This is in contrast to the original Newman-Janis trick where there is significant problems with Tetrad 2.

(3) The original Newman-Janis trick mentions to keep the vectors $m^{a}$ and $\overline{m}^{a}$ as complex conjugates during the whole trick.  But we have seen, strictly speaking, this would involve an extra step -- which is that one had to perform a sign change in one of the terms of Tetrad 3 (\ref{NJTetrad 3}) to obtain Tetrad 4 (\ref{NJTetrad 4}).  The Newman-Janis trick has this arbitrariness in it, and we exposed it in the last section.  We aim to make full use of this ``sign change" transformation and exploit it.

The result will be a mathematically well defined version of a Newman-Janis trick which provides us with a framework to consider the physics of the situation.

\subsubsection*{Standard Newman-Janis trick on flat space}

We start with the simple example of Minkowski space and describe the standard Newman-Janis trick on it, before moving to our modified version.  

The only difference between this version and the original version is that we explicitly spell out the hidden tetrad (i.e. ``Tetrad 3" of the flat space version).  The Minkowski metric in advanced Eddington-Finkelstein coordinates is given by 

\begin{equation} \label{MinkMetric1}
ds^{2}= {du}^{2}+2\:du\:dr-{r}^2({d\theta}^{2}+{{\sin}^{2}\theta}\:{d\phi}^2).
\end{equation}

A null tetrad for this metric is given by 

\begin{eqnarray}\label{MinkTetrad1}
\begin{split}
l^{a}&={\partial}_{r},\nonumber\\
n^{a}&={\partial}_{u}-{\frac{1}{2}} {\partial}_{r}\nonumber,\\
m^{a}&={\frac{1}{\sqrt{2}{r}}}\Bigl({\partial}_{\theta}+\frac{i}{\sin \theta}{\partial}_{\phi}\Bigl)\nonumber,\\
\bar{m}^{a}&={\frac{1}{\sqrt{2}{r}}}\Bigl({\partial}_{\theta}-\frac{i}{\sin \theta}{\partial}_{\phi}\Bigl)\nonumber.\\
\end{split}
\end{eqnarray}

Notice that we would be able to obtain this tetrad if we set $m=0$ for the Schwarzschild tetrad, Tetrad 1 (\ref{NJTetrad 1}).  From there we let $u$ and $r$ become complex-valued coordinates and perform a non-holomorphic transformation $r \rightarrow \overline{r}$, to obtain

\begin{align}\label{MinkTetrad2}
l^{a}&={\partial}_{r},\nonumber\\
n^{a}&={\partial}_{u}-{\frac{1}{2}} {\partial}_{r}\nonumber,\\
m^{a}&={\frac{1}{\sqrt{2}{\overline{r}}}}\Bigl({\partial}_{\theta}+\frac{i}{\sin \theta}{\partial}_{\phi}\Bigl)\nonumber,\\
\bar{m}^{a}&={\frac{1}{\sqrt{2}{r}}}\Bigl({\partial}_{\theta}-\frac{i}{\sin \theta}{\partial}_{\phi}\Bigl)\nonumber.\\
\end{align}

Just like in the case of the standard Newman-Janis trick with Tetrad 2 (\ref{NJTetrad 2}), we find that this corresponding tetrad would be geometrically undefined.  To be more precise, if one were to calculate out the Riemann tensor, Ricci tensor and Ricci scalar, one would find they would depend on terms involving derivatives of the modulus of the complex coordinate $r$.

The next step would be to perform the complex coordinate transformation 

\begin{align} 
u \mapsto u'&=u-{i}\:{a}\:{\cos\theta},\\
r \mapsto r'&=r+{i}\:{a}\:{\cos\theta},\\
\theta \mapsto {\theta}'&=\theta,\\
\phi \mapsto {\phi}'&=\phi.
\end{align}

which results in the ``hidden" tetrad (analogous to (\ref{NJTetrad 3}) and dropping the primes),

\begin{align}\label{MinkTetrad 3}
l^{a}&={\partial}_{r},\nonumber\\
n^{a}&={\partial}_{u}-{\frac{1}{2}} {\partial}_{r}\nonumber,\\
m^{a}&={\frac{1}{\sqrt{2}({r+i\:a\:\cos\theta})}}({\partial}_{\theta}+i\:a\:\sin\theta({\partial}_{u}-{\partial}_{r})+\frac{i}{\sin \theta}{\partial}_{\phi})\nonumber,\\
\bar{m}^{a}&= {\frac{1}{\sqrt{2}({r-i\:a\:\cos\theta})}}({\partial}_{\theta}+i\:a\:\sin\theta({\partial}_{u}-{\partial}_{r})-\frac{i}{\sin \theta}{\partial}_{\phi})\nonumber.\\
\end{align}

Notice that in this tetrad, $m^{a}$ and $\overline{m}^{a}$ are \emph{not} complex conjugates of each other.  Therefore, one needs to change the plus sign into a minus sign in front of the term $i\:a\:\sin\theta(\partial_{u} - \partial_{r})$ in the vector $\overline{m}^{a}$.  The resulting tetrad will be 

\begin{align}\label{MinkTetrad4}
l^{a}&={\partial}_{r},\nonumber\\
n^{a}&={\partial}_{u}-{\frac{1}{2}} {\partial}_{r}\nonumber,\\
m^{a}&={\frac{1}{\sqrt{2}({r+i\:a\:\cos\theta})}}({\partial}_{\theta}+i\:a\:\sin\theta({\partial}_{u}-{\partial}_{r})+\frac{i}{\sin \theta}{\partial}_{\phi})\nonumber,\\
\bar{m}^{a}&= {\frac{1}{\sqrt{2}({r-i\:a\:\cos\theta})}}({\partial}_{\theta}-i\:a\:\sin\theta({\partial}_{u}-{\partial}_{r})-\frac{i}{\sin \theta}{\partial}_{\phi})\nonumber,\\
\end{align}

which can be seen to be Minkowski metric in oblate spheroidal coordinates, if all the coordinates are set to be real. 

The last step of a sign change is equivalent to performing the following tetrad leg transformation

\begin{align} 
\begin{split}
l^{a} \; &\longrightarrow \; \hat{l}^{a} = l^{a}, \\
n^{a} \; &\longrightarrow \; \hat{n}^{a} = n^{a}\\
m^{a} \; &\longrightarrow \; \hat{m}^{a} = m^{a} \\
\overline{m}^{a} \; &\longrightarrow \; \hat{\overline{m}}^{a} = \frac{-2\:i\:a\:\text{sin}\:\theta}{\sqrt{2}\:(r-i\:a\:\text{cos}\:\theta)} \Bigl(n^{a} - \frac{1}{2}\:l^{a}\Bigl)
+ \;\overline{m}^{a}.
\end{split}
\end{align}

To summarize, the standard Newman-Janis has many unwanted features such as a tetrad which is geometrically undefined, a complex coordinate transformation and furthermore, we expose a tetrad leg transformation which again adds more arbitrariness.  It seems very unclean as a method.

\subsubsection*{Modified Newman-Janis trick on flat space}

Our modified Newman-Janis for flat space starts with the tetrad (\ref{MinkTetrad1}) 

\begin{align} \label{ModifiedMinkTetrad1}
l^{a}&={\partial}_{r},\nonumber\\
n^{a}&={\partial}_{u}-{\frac{1}{2}} {\partial}_{r}\nonumber,\\
m^{a}&={\frac{1}{\sqrt{2}{r}}}({\partial}_{\theta}+\frac{i}{\sin \theta}{\partial}_{\phi})\nonumber,\\
\bar{m}^{a}&={\frac{1}{\sqrt{2}{r}}}({\partial}_{\theta}-\frac{i}{\sin \theta}{\partial}_{\phi})\nonumber,\\
\end{align} 

just like the standard version.  From there we make the statement: let coordinates $u$ and $r$ become complex.  This is also part of the standard version of the trick.  But here we stop, and observe something incredibly crucial.  

After letting $u$ and $r$ become complex, notice that tetrad above (\ref{ModifiedMinkTetrad1}) is not a null tetrad, since the vectors $m^{a}$ and $\overline{m^{a}}$ are not complex conjugates of each other any more.  To make this a null tetrad, we can make the $r$ variable in $m^{a}$ become $\overline{r}$ to give us     

\begin{align} \label{WrongMinkTetrad1}
l^{a}&={\partial}_{r},\nonumber\\
n^{a}&={\partial}_{u}-{\frac{1}{2}} {\partial}_{r}\nonumber,\\
m^{a}&={\frac{1}{\sqrt{2}{\overline{r}}}}({\partial}_{\theta}+\frac{i}{\sin \theta}{\partial}_{\phi})\nonumber,\\
\bar{m}^{a}&={\frac{1}{\sqrt{2}{r}}}({\partial}_{\theta}-\frac{i}{\sin \theta}{\partial}_{\phi})\nonumber.\\
\end{align} 

But we then face the serious problem of having a tetrad (\ref{WrongMinkTetrad1}) becoming geometrically very difficult to work.  If I calculate out the Riemann tensor, Ricci tensor or Ricci scalar of this tetrad, I will find these quantities which depend on derivatives of the modulus of the complex coordinate $r$.  

So in our modified Newman-Janis trick, we simply let $u$ and $r$ become complex and let our tetrad stay as (\ref{ModifiedMinkTetrad1}).  It may not be a null tetrad anymore, but it still produces a well-defined flat space metric.  \emph{ Even in the standard Newman-Janis trick, one can see that the hidden Tetrad 3 was not a null tetrad.  We are simply exploiting this type of property.} 

In fact, it may be more appropriate to complexify all the coordinates in our modified Newman-Janis trick.  The reason for this, is that it would provide a more effective approach due to the comparisons we can make when considering complex spacetimes.

We then perform the Newman-Janis complex coordinate transformation

\begin{align} 
u \mapsto u'&=u-{i}\:{a}\:{\cos\theta},\\
r \mapsto r'&=r+{i}\:{a}\:{\cos\theta},\\
\theta \mapsto {\theta}'&=\theta,\\
\phi \mapsto {\phi}'&=\phi.
\end{align}

on (\ref{ModifiedMinkTetrad1}), (rather than on (\ref{WrongMinkTetrad1})), to give us (dropping primes)

\begin{align}\label{ModifiedMinkTetrad2a}
l^{a}&={\partial}_{r},\nonumber\\
n^{a}&={\partial}_{u}-{\frac{1}{2}} {\partial}_{r}\nonumber,\\
m^{a}&={\frac{1}{\sqrt{2}({r-i\:a\:\cos\theta})}}({\partial}_{\theta}+i\:a\:\sin\theta({\partial}_{u}-{\partial}_{r})+\frac{i}{\sin \theta}{\partial}_{\phi})\nonumber,\\
\bar{m}^{a}&= {\frac{1}{\sqrt{2}({r-i\:a\:\cos\theta})}}({\partial}_{\theta}+i\:a\:\sin\theta({\partial}_{u}-{\partial}_{r})-\frac{i}{\sin \theta}{\partial}_{\phi})\nonumber.\\
\end{align}

Notice again that this tetrad is not a null tetrad since $m^{a}$ and $\overline{m}^{a}$ are not complex conjugates of each other.

But the most remarkable aspect regarding this tetrad, is that when one calculates out the corresponding metric using 

\begin{equation} \label{nulltetradmetric}
g^{ab} = l^{a}\:n^{b} + n^{a}\:l^{b} - m^{a}\:\overline{m}^{b} - \overline{m}^{a}\:m^{b}
\end{equation} 

and the inverse metric, one finds it has a vanishing Ricci scalar, Ricci tensor and Riemann tensor.  Essentially the resulting metric is not real, but it is Ricci flat! 

To obtain the desired tetrad, (the same tetrad as one would obtain at the end of the standard Newman-Janis trick), one would need to change certain minus signs into a plus signs and certain plus sign into a minus sign, on the tetrad (\ref{ModifiedMinkTetrad2a}).  In particular

\begin{equation}
m^{a}={\frac{1}{\sqrt{2}({r-i\:a\:\cos\theta})}}({\partial}_{\theta}+i\:a\:\sin\theta({\partial}_{u}-{\partial}_{r})+\frac{i}{\sin \theta}{\partial}_{\phi}), 
\end{equation}

would become 

\begin{equation}
m^{a}={\frac{1}{\sqrt{2}({r+i\:a\:\cos\theta})}}({\partial}_{\theta}+i\:a\:\sin\theta({\partial}_{u}-{\partial}_{r})+\frac{i}{\sin \theta}{\partial}_{\phi}), 
\end{equation}

and 

\begin{equation}
\bar{m}^{a}= {\frac{1}{\sqrt{2}({r-i\:a\:\cos\theta})}}({\partial}_{\theta}+i\:a\:\sin\theta({\partial}_{u}-{\partial}_{r})-\frac{i}{\sin \theta}{\partial}_{\phi}),
\end{equation}

would become

\begin{equation}
\bar{m}^{a}= {\frac{1}{\sqrt{2}({r-i\:a\:\cos\theta})}}({\partial}_{\theta}-i\:a\:\sin\theta({\partial}_{u}-{\partial}_{r})-\frac{i}{\sin \theta}{\partial}_{\phi}).
\end{equation}

This results in the final tetrad of this modified Newman-Janis trick to be the same tetrad as one would obtain at the end of the standard Newman-Janis trick, namely

\begin{align} \label{ModifiedMinkTetrad4}
l^{a}&={\partial}_{r},\nonumber\\
n^{a}&={\partial}_{u}-{\frac{1}{2}} {\partial}_{r}\nonumber,\\
m^{a}&={\frac{1}{\sqrt{2}({r+i\:a\:\cos\theta})}}({\partial}_{\theta}+i\:a\:\sin\theta({\partial}_{u}-{\partial}_{r})+\frac{i}{\sin \theta}{\partial}_{\phi})\nonumber,\\
\bar{m}^{a}&= {\frac{1}{\sqrt{2}({r-i\:a\:\cos\theta})}}({\partial}_{\theta}-i\:a\:\sin\theta({\partial}_{u}-{\partial}_{r})-\frac{i}{\sin \theta}{\partial}_{\phi})\nonumber,\\
\end{align}

which is flat space as desired.  

Notice that each metric in this modified version has well-defined geometric quantities.  In fact, they are all flat space metrics!

The last step of changing signs can be made mathematically precise, by saying after one obtains the tetrad (\ref{ModifiedMinkTetrad2a}), one can then perform a transformation

\begin{align} \label{LorentzlikeMink}
\begin{split}
l^{a} \; &\longrightarrow \; \hat{l}^{a} = l^{a}, \\
n^{a} \; &\longrightarrow \; \hat{n}^{a} = n^{a}, \\
m^{a} \; &\longrightarrow \; \hat{m}^{a} = \Bigl(\frac{r-i\:a\:\text{cos}\:\theta}{r+i\:a\:\text{cos}\:\theta} \Bigl)\; m^{a}, \\
\overline{m}^{a} \; &\longrightarrow \; \hat{\overline{m}}^{a} = \frac{-2\:i\:a\:\text{sin}\:\theta}{\sqrt{2}\:(r-i\:a\:\text{cos}\:\theta)} \Bigl(n^{a} - \frac{1}{2}\:l^{a}\Bigl) + \overline{m}^{a}.
\end{split}
\end{align}

This transformation has the exact same effect as changing the required signs on (\ref{ModifiedMinkTetrad2a}) to obtain (\ref{ModifiedMinkTetrad4}),  but now we have a mathematical precise way of expressing it.  Not only that, but there is a close resemblance between these transformations and a Lorentz transformation.  

In particular, a Lorentz transformation with a ``boost" in the $l^{a}-n^{a}$ plane and a rotation in the $m^{a}-\overline{m}^{a}$ plane, can be represented as

\begin{align} \label{Lorentz}
\begin{split}
l^{a} \; &\longrightarrow \; \hat{l}^{a} = A^{-1}\:l^{a}, \\
n^{a} \; &\longrightarrow \; \hat{n}^{a} = A\:n^{a}, \\
m^{a} \; &\longrightarrow \; \hat{m}^{a} =e^{i\:\phi}\:m^{a}, \\
\overline{m}^{a} \; &\longrightarrow \; \hat{\overline{m}}^{a} =e^{-i\:\phi}\: \overline{m}^{a},
\end{split}
\end{align}

where $A$ and $\phi$ are real arbitrary functions.  To make a comparison to our particular transformation, (\ref{LorentzlikeMink}), we can see that if we set

\begin{equation}
A=1, \qquad \phi = -i\; \text{ln} \; \Bigl(\frac{r-i\:a\:\text{cos}\:\theta}{r+i\:a\:\text{cos}\:\theta}  \Bigl),
\end{equation}

into {\ref{Lorentz}}, (notice that $\phi$ is a complex function) then this results in

\begin{align} \label{Lorentzlike}
\begin{split}
l^{a} \; &\longrightarrow \; \hat{l}^{a} = l^{a}, \\
n^{a} \; &\longrightarrow \; \hat{n}^{a} = n^{a}, \\
m^{a} \; &\longrightarrow \; \hat{m}^{a} = \Bigl(\frac{r-i\:a\:\text{cos}\:\theta}{r+i\:a\:\text{cos}\:\theta} \Bigl)\; m^{a}, \\
\overline{m}^{a} \; &\longrightarrow \; \hat{\overline{m}}^{a} =  \Bigl(\frac{r+i\:a\:\text{cos}\:\theta}{r-i\:a\:\text{cos}\:\theta}  \Bigl)\overline{m}^{a}.
\end{split}
\end{align}

This has a very close resemblance to our transformation (\ref{LorentzlikeMink}), and in particular the difference is due to the vector $\overline{m}^{a}$.  

To summarize, our modified procedure does not involve any non-holomorphic transformations or geometric quantities which are not well-defined.  We work with tetrads which are not null tetrads (just like in the standard Newman-Janis trick via Tetrad 3), but all along our method involves only working in flat space.  We perform holomorphic coordinate transformations and then we perform a Lorentz-like transformation at the end.  

This modified Newman-Janis trick can be extended for the case of Schwarzschild to Kerr.  

\subsubsection*{Modified Newman-Janis trick on Schwarzschild}

Our modified Newman-Janis trick begins with the Schwarzschild tetrad (\ref{NJTetrad 1})

\begin{align}\label{ModifiedTetrad 1}
l^{a}&={\partial}_{r},\nonumber\\
n^{a}&={\partial}_{u}-{\frac{1}{2}}\Bigl(1-{\frac{2m}{r}\Bigl)} {\partial}_{r}\nonumber,\\
m^{a}&={\frac{1}{\sqrt{2}{r}}}\Bigl({\partial}_{\theta}+\frac{i}{\sin \theta}{\partial}_{\phi}\Bigl)\nonumber,\\
\bar{m}^{a}&={\frac{1}{\sqrt{2}{r}}}\Bigl({\partial}_{\theta}-\frac{i}{\sin \theta}{\partial}_{\phi}\Bigl)\nonumber.\\
\end{align}  

but we would write it in the equivalent form 

\begin{align}\label{ModifiedTetrad 1 Part 2}
l^{a}&={\partial}_{r},\nonumber\\
n^{a}&={\partial}_{u}-{\frac{1}{2}}{\Bigl(1-\frac{m}{r} - \frac{m}{r}\Bigl)} {\partial}_{r}\nonumber,\\
m^{a}&={\frac{1}{\sqrt{2}{r}}}\Bigl({\partial}_{\theta}+\frac{i}{\sin \theta}{\partial}_{\phi}\Bigl)\nonumber,\\
\bar{m}^{a}&={\frac{1}{\sqrt{2}{r}}}\Bigl({\partial}_{\theta}-\frac{i}{\sin \theta}{\partial}_{\phi}\Bigl)\nonumber.\\
\end{align}  

Just like in the standard Newman-Janis trick, we let coordinates $u$ and $r$ become complex valued (in fact let all the coordinate become complex-valued just like in our modified flat space case).   This is then a complexified version of the Schwarzschild metric.   Notice that after that statement, this tetrad (\ref{ModifiedTetrad 1 Part 2}) is not a null tetrad, since $m^{a}$ and $\overline{m}^{a}$ are not complex conjugates of each other. 

To fix this, would require the introduction replacing certain $r$ terms with the variable $\overline{r}$.  But this procedure would result in metrics which are not well-defined geometrically from the case of standard complex analysis.  Hence we will stick with the tetrad (\ref{ModifiedTetrad 1 Part 2}) and perform the Newman-Janis complex coordinate transformation 

\begin{align} 
u &\mapsto u'=u-{i}\:{a}\:{\cos\theta},\\
r &\mapsto r'=r+{i}\:{a}\:{\cos\theta},\\
\theta &\mapsto {\theta}'=\theta,\\
\phi &\mapsto {\phi}'=\phi,
\end{align}

which results in the tetrad (dropping primes)

\begin{align}\label{ModifiedNJTetrad 3}
l^{a}&={\partial}_{r},\nonumber\\
n^{a}&={\partial}_{u}-{\frac{1}{2}}\:\Bigl(1-\frac{m}{r-i\:a\:\text{cos}\:\theta}-\frac{m}{r-i\:a\:\text{cos}\:\theta}\Bigl) {\partial}_{r}\nonumber,\\
m^{a}&={\frac{1}{\sqrt{2}({r-i\:a\:\cos\theta})}}({\partial}_{\theta}+i\:a\:\sin\theta({\partial}_{u}-{\partial}_{r})+\frac{i}{\sin \theta}{\partial}_{\phi})\nonumber,\\
\bar{m}^{a}&= {\frac{1}{\sqrt{2}({r-i\:a\:\cos\theta})}}({\partial}_{\theta}+i\:a\:\sin\theta({\partial}_{u}-{\partial}_{r})-\frac{i}{\sin \theta}{\partial}_{\phi})\nonumber.\\
\end{align}

As one can see, this is not a null tetrad since the vectors $m^{a}$ and $\overline{m}^{a}$ are not complex conjugates of each other.  But if one uses the formula

\begin{equation} 
g^{ab} = l^{a}\:n^{b} + n^{a}\:l^{b} - m^{a}\:\overline{m}^{b} - \overline{m}^{a}\:m^{b}
\end{equation} 

and inverse metric, one finds this complex metric has a vanishing Ricci tensor, hence it satisfies the vacuum equations!  

The Riemann tensor has complex valued components which are well-defined. If one sets $m=0$, then all the components of the Riemann tensor vanish.  Similarly, if one sets $a=0$, the Riemann tensor matches the Riemann tensor of the real Schwarzschild metric.  This is an indication of simply taking the real part since $a$ was introduced as part of the imaginary part.  

In some sense, we had a complex Schwarzschild metric, and then we performed a holomorphic transformation on it.  Hence we have some version of a complex Schwarzschild metric, that has a Kerr-like property that when $a=0$, we get a Riemann tensor that matches the real Schwarzschild metric. 

We need to now obtain the desired tetrad, which is the final tetrad of the standard Newman-Janis trick, i.e. Kerr.  This would involve changing relevant plus signs into minus signs and certain minus signs into plus signs to obtain the Kerr tetrad

\begin{align}\label{ModifiedNJTetrad 4}
l^{a}&={\partial}_{r},\nonumber\\
n^{a}&={\partial}_{u}-{\frac{1}{2}}\:\Bigl(1-\frac{m}{r-i\:a\:\text{cos}\:\theta}-\frac{m}{r+i\:a\:\text{cos}\:\theta}\Bigl) {\partial}_{r}\nonumber,\\
m^{a}&={\frac{1}{\sqrt{2}({r+i\:a\:\cos\theta})}}({\partial}_{\theta}+i\:a\:\sin\theta({\partial}_{u}-{\partial}_{r})+\frac{i}{\sin \theta}{\partial}_{\phi})\nonumber,\\
\bar{m}^{a}&= {\frac{1}{\sqrt{2}({r-i\:a\:\cos\theta})}}({\partial}_{\theta}-i\:a\:\sin\theta({\partial}_{u}-{\partial}_{r})-\frac{i}{\sin \theta}{\partial}_{\phi})\nonumber.\\
\end{align}

A more precise mathematical statement would be that to obtain the final tetrad (\ref{ModifiedNJTetrad 4}) from tetrad (\ref{ModifiedNJTetrad 3}) is to perform the following transformation on the tetrad legs

\begin{align} \label{LorentzlikeSch}
\begin{split}
l^{a} \; &\longrightarrow \; \hat{l}^{a} = l^{a}, \\
n^{a} \; &\longrightarrow \; \hat{n}^{a} = n^{a} + \Bigl(\frac{-\:m\:i\:a\:\text{cos}\:\theta}{r^{2}+a^{2}\:\text{cos}^{2}\:\theta} \Bigl)\:l^{a}, \\
m^{a} \; &\longrightarrow \; \hat{m}^{a} = \Bigl(\frac{r-i\:a\:\text{cos}\:\theta}{r+i\:a\:\text{cos}\:\theta} \Bigl)\; m^{a}, \\
\overline{m}^{a} \; &\longrightarrow \; \hat{\overline{m}}^{a} = \frac{-2\:i\:a\:\text{sin}\:\theta}{\sqrt{2}\:(r-i\:a\:\text{cos}\:\theta)} \Bigl(n^{a} - \Bigl(\frac{1}{2} + \frac{m}{r-i\:a\:\text{cos}\:\theta}\Bigl)\:l^{a}\Bigl) + \overline{m}^{a}.
\end{split}
\end{align}

The crucial point is that the last step of the standard Newman-Janis trick involves this type of tetrad leg transformation as well. We are simply exploiting such a property in this modified trick.   

Perhaps a more cleaner tetrad leg transformation is perform the following following transformation on (\ref{ModifiedNJTetrad 3}), 

\begin{align} \label{LorentzlikeSch}
\begin{split}
l^{a} \; &\longrightarrow \; \hat{l}^{a} = l^{a}, \\
n^{a} \; &\longrightarrow \; \hat{n}^{a} = n^{a} + \Bigl(\frac{-\:m\:i\:a\:\text{cos}\:\theta}{r^{2}+a^{2}\:\text{cos}^{2}\:\theta} \Bigl)\:l^{a}, \\
m^{a} \; &\longrightarrow \; \hat{m}^{a} = m^{a}, \\
\overline{m}^{a} \; &\longrightarrow \; \hat{\overline{m}}^{a} = \frac{-2\:i\:a\:\text{sin}\:\theta}{\sqrt{2}\:(r+i\:a\:\text{cos}\:\theta)} \Bigl(n^{a} - \Bigl(\frac{1}{2} + \frac{m}{r-i\:a\:\text{cos}\:\theta}\Bigl)\:l^{a}\Bigl) + \Bigl(\frac{r-i\:a\:\text{cos}\:\theta}{r+i\:a\:\text{cos}\:\theta} \Bigl)\; \overline{m}^{a}.
\end{split}
\end{align}

resulting in the tetrad 

\begin{align}
l^{a}&={\partial}_{r},\nonumber\\
n^{a}&={\partial}_{u}-{\frac{1}{2}}\:\Bigl(1-\frac{m}{r-i\:a\:\text{cos}\:\theta}-\frac{m}{r+i\:a\:\text{cos}\:\theta}\Bigl) {\partial}_{r}\nonumber,\\
m^{a}&={\frac{1}{\sqrt{2}({r-i\:a\:\cos\theta})}}({\partial}_{\theta}+i\:a\:\sin\theta({\partial}_{u}-{\partial}_{r})+\frac{i}{\sin \theta}{\partial}_{\phi})\nonumber,\\
\bar{m}^{a}&= {\frac{1}{\sqrt{2}({r+i\:a\:\cos\theta})}}({\partial}_{\theta}-i\:a\:\sin\theta({\partial}_{u}-{\partial}_{r})-\frac{i}{\sin \theta}{\partial}_{\phi})\nonumber,\\
\end{align}

which is also the Kerr metric, being the $\theta \rightarrow -\theta$ transformation of (\ref{ModifiedNJTetrad 3}), with $m^{a}$ and $\overline{m}^{a}$ interchanged.

\subsubsection*{Standard Newman-Janis vs. Modified Newman-Janis}

Let us review the standard Newman-Janis trick which includes the hidden tetrad, namely Tetrad 3.  From this one can conclude that the trick is not very clean as a method.  

To elaborate on this, we can see that Tetrad 2 has geometric quantities which are not well-defined in terms of standard complex analysis and hence very little physics can be associated to it. Furthermore, Tetrad 3 does not satisfy the requirements of being a null tetrad since $m^{a}$ and $\overline{m}^{a}$ are not complex conjugates of each other.  The step at the end of the procedure involves an arbitrary tetrad leg transformation which turns the vector $\overline{m}^{a}$ to be the actual complex conjugate of $m^{a}$. 

In fact, it may be correct to argue that in the standard Newman-Janis trick, the conjugate of $r$ is introduced in Tetrad 2 merely to provide the relevant term in $n^{a}$ vector of Tetrad 3. Specifically, this term would be the coefficient of the mass parameter, after the complex coordinate transformation.  This means that the term $2m/r$ needs to get turned into $2mr/(r^{2}+a^{2}\:\text{cos}^{2}\:\theta)$.  And conjugating $r$ in a certain way allows us to obtain the latter expression from the former after a complex coordinate transformation.

The vectors $m^{a}$ and $\overline{m}^{a}$ in Tetrad 2 do not really need the conjugate term since the last step of the arbitrary tetrad leg transformation can be extended, so to fix up the vector $\overline{m}^{a}$ to be the complex conjugate of $m^{a}$.  Following in this line of thought, \emph{the idea of introducing the conjugate of $r$ in the standard Newman-Janis trick, is really there for a notational purpose to get the right coefficient for the mass term as opposed to anything of direct physical relevance.}
  
We argue that our modified version of the trick has the hope of finding the physics but includes the disadvantage of the standard Newman-Janis trick of having an arbitrary tetrad leg transformation at the end.  Other than that, it a much cleaner method.

In the modified case, we have the luxury of having well-defined tetrads throughout the procedure, without any conjugation process or any of the arbitrariness associated with it.

The tetrad leg transformation at the end also shows a hint of a physical process.

If one sets $m=0$, then the above transformations (\ref{LorentzlikeSch}) become the Lorentz-like transformation (\ref{LorentzlikeMink}) we saw in the flat-space case.  Recall those transformations did not change the geometry for our modified Newman-Janis trick.  

In the Schwarzschild case, the geometry changes to Kerr.  We see that difference between the transformations (\ref{LorentzlikeMink}) and (\ref{LorentzlikeSch}) is that the mass parameter, $m$, is involved in the tetrad leg transformations.  Hence we can see that the geometrical transformation depends on the mass parameter.  
 
The point is that both the standard Newman-Janis trick and our modified Newman-Janis trick is not very clean as a method (both involve tetrads which are not null and arbitrary tetrad leg transformations).  The difference is that we can talk about physics (as we shall see in the next section) and well-defined metrics in the latter compared to the former.

But the most important thing is that both of these methods are equivalent from the viewpoint of obtaining the Kerr metric from the Schwarzschild metric through a complex coordinate transformation.

\section{Explanation}

Given our new modified Newman-Janis trick, we aim to provide a partial geometric and physical explanation for the trick.  To accomplish this task, requires us to build a table to explicate and amplify upon certain features, thereby setting up a framework to do our investigations on as well as carry out future work.  In particular our objective is to deconstruct the geometric changes in the modified Newman-Janis trick and analyze each geometric change in an independent manner.

\subsubsection*{Isolating geometrical change}

We start by considering the thee different versions of the flat metric in our modified Newman-Janis trick on flat space.  The first one is denoted $\eta_{1}$ and is expressed as

\begin{align} 
l^{a}&={\partial}_{r},\nonumber\\
n^{a}&={\partial}_{u}-{\frac{1}{2}} {\partial}_{r}\nonumber,\\
m^{a}&={\frac{1}{\sqrt{2}{r}}}({\partial}_{\theta}+\frac{i}{\sin \theta}{\partial}_{\phi})\nonumber,\\
\bar{m}^{a}&={\frac{1}{\sqrt{2}{r}}}({\partial}_{\theta}-\frac{i}{\sin \theta}{\partial}_{\phi})\nonumber.\\
\end{align} 

Notice that this is the first tetrad (\ref{ModifiedMinkTetrad1}) of the modified trick .  The second tetrad of the trick (\ref{ModifiedMinkTetrad2a}) is denoted by $\eta_{2}$ and is given by

\begin{align}
l^{a}&={\partial}_{r},\nonumber\\
n^{a}&={\partial}_{u}-{\frac{1}{2}} {\partial}_{r}\nonumber,\\
m^{a}&={\frac{1}{\sqrt{2}({r-i\:a\:\cos\theta})}}({\partial}_{\theta}+i\:a\:\sin\theta({\partial}_{u}-{\partial}_{r})+\frac{i}{\sin \theta}{\partial}_{\phi})\nonumber,\\
\bar{m}^{a}&= {\frac{1}{\sqrt{2}({r-i\:a\:\cos\theta})}}({\partial}_{\theta}+i\:a\:\sin\theta({\partial}_{u}-{\partial}_{r})-\frac{i}{\sin \theta}{\partial}_{\phi})\nonumber.\\
\end{align}

The final tetrad (\ref{ModifiedMinkTetrad4}) of the modified trick is denoted by $\eta_{3}$ and is expressed in a different but equivalent form given by

\begin{align}
l^{a}&={\partial}_{r},\nonumber\\
n^{a}&={\partial}_{u}-{\frac{1}{2}} {\partial}_{r}\nonumber,\\
m^{a}&={\frac{1}{\sqrt{2}({r-i\:a\:\cos\theta})}}({\partial}_{\theta}+i\:a\:\sin\theta({\partial}_{u}-{\partial}_{r})+\frac{i}{\sin \theta}{\partial}_{\phi})\nonumber,\\
\bar{m}^{a}&= {\frac{1}{\sqrt{2}({r+i\:a\:\cos\theta})}}({\partial}_{\theta}-i\:a\:\sin\theta({\partial}_{u}-{\partial}_{r})-\frac{i}{\sin \theta}{\partial}_{\phi})\nonumber.\\
\end{align}

These three versions of the flat metrics will be the basis for the construction of three different families of Kerr-Schild metrics, which can be expressed as 

\begin{equation}
g_{i}^{ab} = \eta_{i}^{ab}+2\:P\:l^{a}\:l^{b},
\end{equation}

where $i$ would differentiate between the different flat metrics considered above.  In addition to that, it will differentiate between the different Kerr-Schild families.  

First start by letting $l^{a}={\partial}_{r}$ and $l^{b}={\partial}_{r}$.  Then consider four different expressions for $P$ where

\begin{align}
\begin{split}
P_{1}&=0 \\
P_{2}&=\frac{m}{r} \\
P_{3}&= \frac{m}{r-i\:a\:\text{cos}\theta} \\
P_{4}&= \frac{mr}{r^{2}+a^{2}\:\text{cos}^{2}\theta}.
\end{split}
\end{align}

Metrics will belong to the same family if they share the same flat space metric, even though they may have different values for $P$. In the following table below, we construct three different families of Kerr-Schild metrics which correspond to the three different rows.

	\begin{table}[h!]
		\begin{center}
			\caption{Kerr-Schild family of metrics}
			\label{tab:table1}
			\begin{tabular}{cccc}
				\toprule
				$P_{1}=0$ (flat space) & $P_{2}$ & $P_{3}$ & $P_{4}$\\
				\midrule
				$\eta_{1}^{ab}$ & $\eta_{1}^{ab} + 2P_{2}\:l^{a}\:l^{b} (Sch.)$  & $\eta_{1}^{ab} + 2P_{3}\:l^{a}\:l^{b}$ & $\eta_{1}^{ab} + 2P_{4}\:l^{a}\:l^{b}$\\
				$\eta_{2}^{ab}$ & $\eta_{2}^{ab} + 2P_{2}\:l^{a}\:l^{b}$ & $\eta_{2}^{ab} + 2P_{3}\:l^{a}\:l^{b}$ & $\eta_{2}^{ab} + 2P_{4}\:l^{a}\:l^{b}$\\
				$\eta_{3}^{ab}$ & $\eta_{3}^{ab} + 2P_{2}\:l^{a}\:l^{b}$ & $\eta_{3}^{ab} + 2P_{3}\:l^{a}\:l^{b}$ & $\eta_{3}^{ab} + 2P_{4}\:l^{a}\:l^{b} (Kerr)$\\
				\bottomrule
			\end{tabular}
		\end{center}
	\end{table}
	
\emph{It is important to note that in all the metrics in the table $l^{a}={\partial}_{r}$ and $l^{b}={\partial}_{r}$.}  

It is also very important to note that the metrics in the first row and the second row were built from tetrads which were not null tetrads.  This is due to the fact the flat space metrics $\eta_{1}$ and $\eta_{2}$ were built from tetrads in which $\overline{m}^{a}$ is not the actual complex conjugate of $m^{a}$.

As can be seen, the different rows represent the different Kerr-Schild family.  So all the metrics in the first row belong to the same family and so on.  Metrics in the family only vary in what the $P$ function is.

All metrics are have well-defined geometric quantities from the viewpoint of standard complex analysis.  

Another interesting characteristic to note is that the difference vertically between the metrics is that they differ in the expression for flat space.  Flat space is given either by $\eta_{1}$, $\eta_{2}$, or $\eta_{3}$. Therefore $P$ stays constant as one moves vertically through the table.  These patterns of the table will be important when we consider the analysis of our modified Newman-Janis trick.

It was mentioned before that the three flat space metrics of our modified Newman-Janis trick on flat space is given by the first column.  What about the tetrads in our modified Newman-Janis trick for Schwarzschild to Kerr?

It turns out that our initial complex Schwarzschild tetrad (\ref{ModifiedTetrad 1}) is given by the metric in the first row and second column, i.e. $\eta_{1}^{ab} +2P_{2}\:l^{a}\:l^{b}$.  In our modified trick, we perform a holomorphic coordinate transformation that gives us a tetrad (\ref{ModifiedNJTetrad 3}) which is a version of a complex Schwarzschild metric.  In the table this is expressed in the second row and third column, i.e. $\eta_{2} + 2P_{3}\:l^{a}\:l^{b}$.  Finally in our modified Newman-Janis trick we do a tetrad leg transformation that gives us the tetrad (\ref{ModifiedNJTetrad 4}) for the complexified Kerr metric which is equivalently written as  

\begin{align}
l^{a}&={\partial}_{r},\nonumber\\
n^{a}&={\partial}_{u}-{\frac{1}{2}}\:\Bigl(1-\frac{2mr}{r^{2}+a^{2}\:\text{cos}^{2}\:\theta}\Bigl) {\partial}_{r}\nonumber,\\
m^{a}&={\frac{1}{\sqrt{2}({r-i\:a\:\cos\theta})}}({\partial}_{\theta}+i\:a\:\sin\theta({\partial}_{u}-{\partial}_{r})+\frac{i}{\sin \theta}{\partial}_{\phi})\nonumber,\\
\bar{m}^{a}&= {\frac{1}{\sqrt{2}({r+i\:a\:\cos\theta})}}({\partial}_{\theta}-i\:a\:\sin\theta({\partial}_{u}-{\partial}_{r})-\frac{i}{\sin \theta}{\partial}_{\phi})\nonumber.\\
\end{align}

In our table, this complex Kerr metric would be located on in the third row and fourth column, i.e. $\eta_{3} + 2P_{4}\:l^{a}\:l^{b}$.

To express the other metrics, it would be more fulfilling to show the the tetrad leg procedures which allows one to move horizontally in the  table, if one were to start with one of the flat space metrics.

We start by taking the relevant null tetrad for one of the flat space metrics and performing the tetrad leg transformation

\begin{align} \label{tetradlegtable}
\begin{split}
l^{a} \; &\longrightarrow \; \hat{l}^{a} = l^{a}, \\
n^{a} \; &\longrightarrow \; \hat{n}^{a} = n^{a} + P\:l^{a},\\
m^{a} \; &\longrightarrow \; \hat{m}^{a} = m^{a}, \\
\overline{m}^{a} \; &\longrightarrow \; \hat{\overline{m}}^{a} =\overline{m}^{a}.
\end{split}
\end{align}

to move horizontally.  The value of $P$ would depend on which column you would want to move to.  An example of this would be that suppose we start with $\eta_{3}$.  Then we can move to the Kerr metric which is a horizontal movement by performing the tetrad leg transformation (\ref{tetradlegtable}) with

\begin{equation}
 P=P_{4}=\frac{mr}{r^{2}+a^{2}\:\text{cos}^{2}\theta}.
\end{equation}

Hence, movement within the same Kerr-Schild family can be facilitated through (\ref{tetradlegtable}).  It is important to note that each of the metrics in the same family are different in terms of geometry and the only feature different among them is the $P$ variable.

Therefore the geometrical change as one moves horizontally through the table is \emph{solely} due to the change in the $n^{a}$ vector of the null tetrad.  Specifically this change is due to the change to the value of $P$.  

This forms the first part of constructing the framework for analysing the geometrical change in our modified Newman-Janis trick.

What happens when one wants to move from a metric on the first row to a metric directly below on the second row.  An example of this would be to start from metric $\eta_{1}^{ab} + 2P_{2}\:l^{a}\:l^{b}$ and move below to $\eta_{2}^{ab} + 2P_{2}\:l^{a}\:l^{b}$.  

This would involve starting with the relevant tetrads for the metric in the first row and then performing the tetrad leg transformations 

\begin{align} \label{tetradlegtable2}
\begin{split}
l^{a} \; &\longrightarrow \; \hat{l}^{a} = l^{a}, \\
n^{a} \; &\longrightarrow \; \hat{n}^{a} = n^{a} \\
m^{a} \; &\longrightarrow \; \hat{m}^{a} = \Bigl(\frac{r}{r-i\:a\:\text{cos}\:\theta}\Bigl)m^{a} + \frac{i\:a\:\text{sin}\:\theta(n^{a} - \frac{{1}}{2}(1+2\:P\:)l^{a})}{\sqrt{2}\:(r-i\:a\:\text{cos}\:\theta)},\\\
\overline{m}^{a} \; &\longrightarrow \; \hat{\overline{m}}^{a} =
\Bigl(\frac{r}{r-i\:a\:\text{cos}\:\theta}\Bigl)\overline{m}^{a} + \frac{i\:a\:\text{sin}\:\theta(n^{a} - \frac{{1}}{2}(1+2\:P\:)l^{a})}{\sqrt{2}\:(r-i\:a\:\text{cos}\:\theta)}. 
\end{split}
\end{align}

The function $P$ would depend on the metric in question since both the start and target metric have the same $P$.

Notice that the geometrical change as one moves from the first row to the second row, resides only with the vectors $m^{a}$ and $\overline{m}^{a}$.  More striking is that the change in these vectors depends on the value of $P$. But the value of $P$ stays constant within each column.  If the underlying metric was flat where $P=0$, then this transformation would not change the geometry.  Otherwise, there would be a change in geometry going vertically down.  

The next geometric change we want to isolate is given by the analysis of what happens when we move from the second row to the metric directly below on the third row.  Note that this operation is equivalent to fixing up the vector $\overline{m}^{a}$ to be the actual complex conjugate of $m^{a}$.  So if one starts with a tetrad in the second row, then to move to the metric directly below on the third row, one has to perform the tetrad leg transformation 

\begin{align} 
\begin{split}
l^{a} \; &\longrightarrow \; \hat{l}^{a} = l^{a}, \\
n^{a} \; &\longrightarrow \; \hat{n}^{a} = n^{a} \\
m^{a} \; &\longrightarrow \; \hat{m}^{a} = m^{a} ,\\
\overline{m}^{a} \; &\longrightarrow \; \hat{\overline{m}}^{a} =
\Bigl(\frac{r-i\:a\:\text{cos}\:\theta}{r+i\:a\:\text{cos}\:\theta}\Bigl)\overline{m}^{a} - \frac{2\:i\:a\:\text{sin}\:\theta(n^{a} - \frac{{1}}{2}(1+2\:P\:)l^{a})}{\sqrt{2}\:(r+i\:a\:\text{cos}\:\theta)}. 
\end{split}
\end{align} 

Notice that the geometrical change resides only in the $\overline{m}^{a}$ vector and this change depends on the variable $P$.  Once again if the underlying metric was flat, then one obtains a flat metric, with $P=0$.

\subsubsection*{Analyzing the modified Newman-Janis trick}

We've now set up a framework to analyse and interpret our modified Newman-Janis trick.  We can now isolate and see the geometrical changes in terms of function $P$.

Our modified Newman-Janis trick starts with the complex Schwarzschild metric (\ref{ModifiedTetrad 1}) which is located on the first row and second column of the table.  One then performs a holomorphic coordinate transformation 

\begin{align} 
u &\mapsto u'=u-{i}\:{a}\:{\cos\theta},\\
r &\mapsto r'=r+{i}\:{a}\:{\cos\theta},\\
\theta &\mapsto {\theta}'=\theta,\\
\phi &\mapsto {\phi}'=\phi,
\end{align}

which results in a complex Schwarzschild metric (\ref{ModifiedNJTetrad 3}).  The position of the metric on table, is that it is located the second row and third column.  

This implies two things have happened because of the complex coordinate transformation.  The first is that the value of $P$ has changed when we move horizontally to the right.  The metric there is well defined but not Ricci flat.  The second is that one moved vertically downward which also depended on the new value of $P$.  Though we have isolated the geometrical changes due to a complex coordinate transformation, the explanatory statements are not needed since the resulting metric is a complex Schwarzschild metric.

The main aspect of this framework is to analyze what happens in the second part of our modified Newman-Janis trick when perform the tetrad leg transformation.  This results in moving from our complex Schwarzschild metric to the complex Kerr metric which is located in the third row and fourth column.  

This last step of the Newman-Janis trick, involves changing the value of $P$ and then moving from the second row to the third row.  This geometric change also depends on the value of $P$.  

\emph{Another way to look at this, is that complex Schwarzschild and complex Kerr are related by a change in $P$ and making sure that $\overline{m}^{a}$ is the actual conjugate of $m^{a}$}.  This is the connection between Schwarzschild and Kerr in the modified Newman-Janis trick.  The complex coordinate transformation is simply a matter of setting up Schwarzschild in an appropriate way.

We are also able to quantify how the geometry changes in different isolated parts and it can be seen that this is attributed to the variable $P$.

The most important aspect of this framework is left for future work.  This involves rewriting all of the metrics in the table in the form $\eta_{1}^{ab} + 2\:Q\:k^{a}\:k^{a}$.  In other words, putting all the metrics, including the ones in the second and third row, in the form where the flat space metric in the Kerr-Schild form is given by $\eta_{1}$.  

Hence one would be able to observe the exact change happening at each step of the way with the variable $Q$ and null vector $k$, as one performs the modified Newman-Janis trick.  The flat space metric would be untouched and hence this will provide a cleaner geometric interpretation for the trick.

So the final question is what can be said of the physics. 

\subsubsection*{Physical interpretation}

It is interesting to observe that when we went from the second row with the complex Schwarzschild metric to the third row with complex Kerr, the only change going downward involves changes in the $\overline{m}^{a}$ (as well as a change in the variable $P$).  

Therefore, it seems reasonable to give a hypothesis that the source of physical angular momentum resides from going vertically down from the second row to the third row, i.e. to making sure that we fix up $\overline{m}^{a}$ to be the actual complex conjugate of $m^{a}$.  One can test these ideas on the adjacent metrics on the table and provide an analogous case.  If this turns out to be the case, then we can show a physical property can be actualized by a specific null tetrad condition.

What is the physical significance of $P$?  Well we can see that $P$ depends on the total mass of the spacetime.  Hence one can deduce that the source of geometrical change in table (7.1) is due to the mass.   

The most interesting physical question is: does the change $P$ correspond to a change in $m$?  The deeper question would be to ask if there is a change in mass as one does the Newman-Janis trick?  Is the physics of the situation due to some dynamics of mass?  We will explore the tip of the iceberg of this question.

Suppose we take the $P$ value of the complex Kerr metric given by $P_{4}$.  To move to the left to $P_{3}$, one can make the following action on the mass term in $P_{4}$, 

\begin{equation}
m \; \rightarrow \; \Bigl(1 + \frac{i\:a\:\text{cos}\theta}{r} \Bigl)\:m. 
\end{equation} 

Hence we see that to go from $P_{4}$ to $P_{3}$, involves adding an imaginary contribution of mass to the total mass.  

To move from the value of $P_{3}$ to $P_{2}$, requires performing the following action on the mass term in $P_{3}$:  

\begin{equation}
m \; \rightarrow \; \Bigl(1 - \frac{i\:a\:\text{cos}\theta}{r} \Bigl)m. 
\end{equation} 

This amounts to subtracting an imaginary contribution of the mass from the total mass.  Combining this together, we see that to from the complex Kerr value of $P_{4}$ to the initial complex Schwarzschild metric and its value of $P_{2}$ requires one to perform the transformation

\begin{equation}
m \; \rightarrow \; \Bigl(1 + \frac{a^{2}\:\text{cos}^{2}\theta}{r^{2}} \Bigl)m. 
\end{equation}

Hence this involves adding a real valued contribution of mass to the total amount.  

If one were to reverse the roles and go from $P_{2}$ to $P_{4}$, it is fair to say that this would involve subtracting a real-valued component of the total mass.  Furthermore this process happens through the complexified vacuum spacetimes of our modified Newman-Janis trick.

What makes this investigation more attractive is that the Schwarzschild metric (i.e. $P_{2}$) has all of its mass as the irreducible mass.  By contrast, the Kerr spacetime has the same total mass as the Schwarzschild value in the Newman-Janis trick but its mass has a composition made of the irreducible mass as well as the reducible mass.  Hence the Kerr spacetime has a decrease in irreducible mass compared to the Schwarzschild spacetime.

The hypothesis is that the irreducible mass is decreasing smoothly as one moves horizontally through the table.  The hypothesis is that the change in $P$ is related to the change in irreducible mass, even as it travels through the complexified spaces.

\subsubsection*{Future directions}

We have outlined in the previous subsections an insightful geometric understanding of the Newman-Janis trick.  It still begs to ask the question: Why does the trick work and how does the initial coordinate parameter $a$ become a physical parameter at the end of the trick?

In this subsection, we aim to provide a few observations that may provide directions on answering the above question.

Let us observe that on Table $(7.1)$, the Schwarzschild metric, which is on the first row and second column, can be explicitly written as

\begin{equation}\label{SchExplain}
ds^{2}= \Bigl(1-\frac{2\:m}{r}\Bigl)\:{du}^{2}+2\:du\:dr-{r}^2({d\theta}^{2}+{{\sin}^{2}\theta}\:{d\phi}^2).
\end{equation}

Notice that when $m=0$, the Schwarzschild metric becomes $\eta_{1}$ which is on the first row and expressed as 

\begin{equation}
ds^{2}= {du}^{2}+2\:du\:dr-{r}^2({d\theta}^{2}+{{\sin}^{2}\theta}\:{d\phi}^2).
\end{equation}

The Kerr metric, which is on the third row and fourth column, can be expressed as 

	\begin{multline}
	ds^{2}= \Bigl(1-\frac{2\:m\:r}{r^{2}+a^{2}{\cos}^{2}\theta}\Bigl)\:{du}^{2}+2\:du\:dr+\frac{4\:m\:r\:a\sin^{2}\theta}{r^{2}+a^{2}{\cos}^{2}\theta}du\:d\phi-2\:a\:\sin^{2}\theta\:d\phi\:dr \\
	-(({r^{2}+a^{2}{\cos}^{2}\theta})\:a^{2}\sin^{2}\theta + 2\:m\:r\:a^{2}\sin^{2}\theta\\
	+({r^{2}+a^{2}{\cos}^{2}\theta})^{2})\frac{\sin^{2}\theta}{({r^{2}+a^{2}{\cos}^{2}\theta})}d\phi^{2} 
	-({r^{2}+a^{2}{\cos}^{2}\theta})\:{d\theta}^{2},
	\end{multline}

and if one sets $a=0$, then the resulting metric is the Schwazschild spacetime (\ref{SchExplain}).  The more interesting case is when the expression $m=0$, is substituted into the Kerr metric and the resulting metric is expressed as  

	\begin{multline}\label{MinkKerra}
	ds^{2}= {du}^{2}+2\:du\:dr - 2\:a\sin^{2}\theta \: d\phi \: dr-(r^{2}+a^{2}{\cos}^{2}\theta) d{\theta}^2 \\
	-(a^{2} + r^{2}){\sin}^{2}{\theta}\:{d\phi}^2,
	\end{multline}  
	
which is $\eta_{3}$ in Table $(7.1)$.  

On the one hand, this is as expected but on the other hand this is absolutely striking.  In the Kerr metric the angular momentum per unit mass $a$ is expressed as 

\begin{equation}
a=\frac{J}{m},
\end{equation}

where $J$ is the angular momentum and $m$ is the usual mass parameter.  

Therefore, when one performs the operation to set $m=0$, in the Kerr metric and obtains the metric (\ref{MinkKerra}), it can be deduced that only certain $m$ parameters are set to $0$.  The parameter $m$ residing in the $a$ term, is not set to $0$ (though one can argue that one should include the extra condition that $J$ should be set to $0$ but then we lose the $a$ coordinate which is visible in the flat metric $\eta_{3}$).  

To summarize, from the Kerr metric, we can set certain $m$ parameters to zero and we obtain the flat metric $\eta_{3}$.  Through this process, the physical parameter $m$ becomes a mathematical parameter.  This is exactly what is happening in the Newman-Janis trick but in the opposite direction!  We start with a mathematical parameter $a=J/m$, which then turns to a physical parameter.

To expand upon this observation, let us look at Table $7.1$ and deduce the locations of the event horizons for each of the metrics involved.  In the Schwarzschild metric, we have the location of the event horzin as $r_{H}=2m$. When $m$ is set to $0$, one can in some sense say the event horizon for $\eta_{1}$ is $r_{H}=0$.  

In the Kerr metric, the location of the event horizon is given by $r_{H}=m+\sqrt{m^{2} -a^{2}}$.  It is very important to note that when certain $m$ terms (not the ones inside the $a$ term) are set to $0$, we see that $r_{H}=i\:a$ as the location of the event horizon for $\eta_{3}$.  This means that the flat space has a complex Schwarzschild radius associated to it and this comes from certain physical parameters turning to mathematical parameters.

Using this structure, we can deduce a number of different properties. For the metric $\eta_{1}$, we have that $r_{H}=0$, which can be rewritten as $r_{H}=i\:a\:\text{cos}\:\frac{\pi}{2}$.  Similarly for the flat space $\eta_{3}$, we can rewrite $r_{H}=i\:a$,  as $r_{H}=i\:a\:\text{cos}\:0$.  

Therefore one can make a ``guess" that the location of the event horizon for $\eta_{2}$ will be $r_{H}=i\:a\:\text{cos}\:\theta$.  

It is interesting to note that the coefficient of $d\theta^{2}$ in the different flat metrics vanishes when the above considerations are substituted in their respective metrics.  More precisely, the coefficient of $d\theta^{2}$ vanishes when $r=i\:a\:\text{cos}\:\frac{\pi}{2}$ is substituted in $\eta_{1}$, $r=i\:a\:\text{cos}\:\theta$ is substituted in $\eta_{2}$, and $r=i\:a\:\text{cos}\:0$ is substituted into $\eta_{3}$.  Hence, our guess that the location of the event horizon for $\eta_{2}$ will be $r=i\:a\:\text{cos}\:\theta$, seems to be on the right track.

We can go back to the notation that $r_{H}=0$, is the location of the event horizon for $\eta_{1}$ and also rewrite $r_{H}=i\:a\:\text{cos}\:\theta$, as $r_{H}-i\:a\:\text{cos}\:\theta = 0$ as the location of the event horizon for $\eta_{2}$.  Hence, we can see a transformation of the form, $r_{H} \rightarrow r_{H}-i\:a\:\text{cos}\:\theta$.  

This is exactly the coordinate transformation of the Newman-Janis trick.  In some sense, by using the complex Schwarzschild radius for $\eta_{3}$, we are able to deduce properties of the Newman-Janis trick.

From this consideration, we can also ``guess" what the event horizon will be for our complexified Schwarzschild metric which is on the second row and third column of Table $(7.1)$.  It has to reduce to $r_{H}=i\:a\:\text{cos}\theta$, when certain $m$ terms are set to $0$ (not the ones in the $a$ term).  The reason for this, is due that fact that the complexified Schwarzschild metric becomes $\eta_{2}$, when those same terms are set to $0$.

The guess would be that the event horizon for this complexified Schwarzschild metric is located at $r=m+\sqrt{m^{2}-a^{2}\:\text{cos}^{2}\:\theta}$. If we set $m=0$ in that expression, it gives us the desired result of  $r=i\:a\:\text{cos}\theta$, hence once again our guess seems to show promising signs.  

Notice that all of these constructions, seems to align with our interpretation of $P$ where the irreducible mass decreases.  The area of the event horizon and the irreducible mass are directly proportional, hence a reduction in $P$ as we move along the Newman-Janis trick should be equivalent to a reduction in the event horizon which is what we observe.  The event horizon location starts at $r=2m$, then reduces to $r=m+\sqrt{m^{2}-a^{2}\:\text{cos}^{2}\:\theta}$, and finally the Kerr event horizon location is at $r=m+\sqrt{m^{2} - a^{2}}$.

Future work will involve mathematical rigorous construction of our guesses and in some sense ``deriving" the Newman-Janis trick from the complex Schwarzschild radius $r_{H}=i\:a$ of $\eta_{3}$.  Hence one would be able to show that the Newman-Janis trick is nothing more than a sophisticated version of the problem of why the Kerr metric reduces to a flat metric when certain $m$ terms are set to zero, while others become mathematical parameters?   

More precisely, the latter has the problem of why does the physical parameter $m$ turn to a mathematical parameter when this procedure is done?  The exact reverse phrasing of this problem is at the heart of the Newman-Janis trick. 

It may be possible that the complex Schwarzschild radius of $\eta_{3}$, represents some physically meaningful hidden complex-valued structure.

\section{Resemblance between Schwarzschild and Kerr}

In this section, we're going to show some calculations which highlight the similarities between the Schwarzschild metric and the Kerr metric in oblate spheroidal coordinates.   For this section, we shall work in the Lorentzian signature $(-+++)$.

The reason for these calculations is that they may provide clues to why the Newman-Janis trick is successful to relate these two spacetimes.  In particular the Newman-Janis trick involves introducing an $a$ parameter into the complex coordinate transformation, which afterwards turns out to become a physical quantity for the Kerr metric.  From the perspective of oblate spheroidal coordinates, the parameter $a$ can be seen in a different manner.

We first start by writing Minkowski space, $ds^{2} = -dt^{2} + dx^{2} + dy^{2} + dz^{2}$, in oblate spheroidal coordinates

\begin{equation}
ds_{M}^{2} = -dt^{2}  + \frac{r^{2}+a^{2}\:\text{cos}^{2}\:\theta}{r^{2} + a^{2}}\: dr^{2} + (r^{2} + a^{2}\:\text{cos}^{2}\:\theta)\:d\theta^{2} + (r^{2}+a^{2})\:\text{sin}^{2}\:\theta d\phi^{2}, 
\end{equation}

through the coordinate transformation

\begin{align}
\begin{split}
x &= \sqrt{r^{2} + a^{2}}\:\text{sin}\:\theta \: \text{cos}\:\phi, \\
y &= \sqrt{r^{2} + a^{2}}\:\text{sin}\:\theta \: \text{sin}\:\phi, \\
z &= r\:\text{cos}\:\theta.
\end{split}
\end{align}

From this construction, one can write the Kerr-Schild ``Cartesian" version of the Kerr geometry in these oblate spheroidal coordinates leading to

\begin{equation} \label{Kerroblate}
ds_{\text{Kerr}}^{2} = ds_{M}^{2} \: +\: \frac{2mr}{r^{2} + a^{2}\:\text{cos}^{2}\:\theta} \Bigl(dt + \frac{r^{2} + a^{2}\:\text{cos}^{2}\:\theta}{r^{2} + a^{2}}\:dr - a\:\text{sin}^{2}\:\theta d\phi  \Bigl)^{2}.
\end{equation}

An important thing to note is that the $d\theta$ term vanishes in the null covector.  From there, if we write the Schwarzschild geometry in the same oblate spheroidal coordinates, it results in

\begin{align} \label{Schoblate}
\begin{split}
ds_{Schwarzschild}^{2} = ds_{M}^{2} \:+ \: \frac{2m}{\sqrt{r^{2} + a^{2}\:\text{sin}^{2}\:\theta}} \Bigl(dt \\
 + \frac{r}{\sqrt{r^{2} + a^{2}\:\text{sin}^{2}\:\theta}}\:dr + \frac{a^{2}\:\text{sin}\:\theta\:\text{cos}\:\theta}{\sqrt{r^{2} + a^{2}\:\text{sin}^{2}\:\theta}} d\theta \Bigl)^{2} 
 \end{split}
\end{align}

In comparison with the Kerr metric (\ref{Kerroblate}), it is interesting to note that in (\ref{Schoblate}) the d$\phi$ term vanishes in the null covector.  

Expanding our observation on the similarities and differences between the two metrics, it can be said that both

\begin{equation}
 \Bigl(dt + \frac{r^{2} + a^{2}\:\text{cos}^{2}\:\theta}{r^{2} + a^{2}}\:dr - a\:\text{sin}^{2}\:\theta d\phi  \Bigl),
\end{equation}

and 

\begin{align}
\begin{split}
\Bigl(dt + \frac{r}{\sqrt{r^{2} + a^{2}\:\text{sin}^{2}\:\theta}}\:dr + \frac{a^{2}\:\text{sin}\:\theta\:\text{cos}\:\theta}{\sqrt{r^{2} + a^{2}\:\text{sin}^{2}\:\theta}} d\theta \Bigl) \\
= \Bigl(dt + d\:\sqrt{r^{2} + a^{2}\:\text{sin}^{2}\:\theta}\Bigl) \\
=d\:\Bigl(t + d\:\sqrt{r^{2} + a^{2}\:\text{sin}^{2}\:\theta}  \Bigl)
\end{split}
\end{align}

are null one-forms with respect to the Minkowski metric.

If one were to keep the coordinate system fixed in oblate spheroidal coordinates, then somehow the transition from Schwarzschild metric to the Kerr metric amounts to replacing 

\begin{align}
\begin{split}
\frac{2m}{\sqrt{r^{2} + a^{2}\:\text{sin}^{2}\:\theta}} \Bigl(dt + \frac{r}{\sqrt{r^{2} + a^{2}\:\text{sin}^{2}\:\theta}}\:dr + \frac{a^{2}\:\text{sin}\:\theta\:\text{cos}\:\theta}{\sqrt{r^{2} + a^{2}\:\text{sin}^{2}\:\theta}} d\theta \Bigl)^{2} \\
= \frac{2m}{\sqrt{r^{2} + a^{2}\:\text{sin}^{2}\:\theta}}\: \Bigl(dt + d\:\sqrt{r^{2} + a^{2}\:\text{sin}^{2}\:\theta}\Bigl)^{2} \\
= \frac{2m}{\sqrt{r^{2} + a^{2}\:\text{sin}^{2}\:\theta}}\: \Bigl[ d\:\Bigl(t + \sqrt{r^{2} + a^{2}\:\text{sin}^{2}\:\theta}  \Bigl)\Bigl]^{2}
\end{split}
\end{align}

by

\begin{equation}
\frac{2mr}{r^{2} + a^{2}\:\text{cos}^{2}\:\theta} \Bigl(dt + \frac{r^{2} + a^{2}\:\text{cos}^{2}\:\theta}{r^{2} + a^{2}}\:dr - a\:\text{sin}^{2}\:\theta d\phi  \Bigl)^{2}.
\end{equation}

Notice the similarities between the two expressions, as well as the differences.

Finally for the case where we have oblate spheroidal rational polynomial coordinates $(\chi = \text{cos}\:\theta)$, Minkowski space $ds^{2} = -dt^{2} + dx^{2} + dy^{2} + dz^{2}$ can be represented as 

\begin{equation}
ds_{M}^{2} = -dt^{2} + \frac{r^{2} + a^{2}\:\chi^{2}}{r^{2}+a^{2}}\:dr^{2} + \frac{r^{2} + a^{2}\:\chi^{2}}{1-\chi^{2}}\:d\chi^{2} + (r^{2}+a^{2})\:(1-\chi^{2})\:d\phi^{2},
\end{equation}

by the coordinate transformation

\begin{align}
\begin{split}
x &= \sqrt{r^{2}+a^{2}}\:\sqrt{1-\chi^{2}}\:\text{cos}\:\phi, \\
y &= \sqrt{r^{2}+a^{2}}\:\sqrt{1-\chi^{2}}\:\text{sin}\:\phi, \\
z &= r\:\chi.
\end{split}
\end{align}

The Kerr metric in these coordinates, is expressed 

\begin{equation} \label{Kerrpolynomial}
ds_{\text{Kerr}}^{2} = ds_{M}^{2} \: +\: \frac{2mr}{r^{2} + a^{2}\:\chi^{2}} \Bigl(dt + \frac{r^{2} + a^{2}\:\chi^{2}}{r^{2} + a^{2}}\:dr - a\:(1-\chi^{2})\: d\phi  \Bigl)^{2},
\end{equation}

and the Schwarzschild spacetime is expressed as

\begin{align}
\begin{split}
ds_{Schwarzschild}^{2} = ds_{M}^{2} \:+ \: \frac{2m}{\sqrt{r^{2} + a^{2}\:(1-\chi^{2})}} \Bigl(dt + \frac{r}{\sqrt{r^{2} + a^{2}\:(1-\chi^{2})}}\:dr  \\
+ \frac{a^{2}\:\chi}{\sqrt{r^{2} + a^{2}(1-\chi^{2})}} d\theta \Bigl)^{2}  \\
= ds_{M}^{2}\:+ \: \frac{2m}{\sqrt{r^{2} + a^{2}\:(1-\chi^{2})}} \:\Bigl[ d\:\Bigl(t + \sqrt{r^{2} + a^{2}(1-\chi^{2})}  \Bigl)\Bigl]^{2}.
\end{split}
\end{align}

In this coordinate system, the two metrics share a lot of similarities and some differences.  Also the parameter $a$ in this coordinate system does not have the feature that when $a \rightarrow 0$, the Kerr metric goes to Schwarzschild.


\chapter{Summary and Conclusions}

This thesis has dealt primarily with two facets of the intersection of General Relativity with complex variables.  The first involved complex spacetimes, and the second was regarding the Newman-Janis trick.  The overall key conclusions and possible future directions are stated below.

In our discussions of complex spacetimes, there was the approach of modifying spacetime metrics into Euclidean metrics and thereby, turning them into complex Hermitian form.  The other approach would be to consider how to modify complex manifold theory for Lorentzian manifolds.  

These approaches came about, due to a key theorem which stated that Lorentzian signature metrics cannot admit an almost Hermitian structure.  The necessary modifications by Flaherty \cite{flaherty1976hermitian}, was the introduction of a modified almost Hermitian structure which was complex-valued.  Among the key results was that this modified almost Hermitian structure was integrable for spacetimes such as Schwarzschild and Kerr.

The second part of this thesis consisted of reviewing the Newman-Janis trick and a number of variations on that theme.  What was particularly interesting is that the trick seems to work naturally on a special subclass of Kerr-Schild metrics.  But so far, the necessity of a metric belonging to this Kerr-Schild class for the trick to be successful, has not been proven.

The explanations by Kerr-Talbot and Newman suffered from explaining the ambiguity involved in conjugating certain coordinate terms.  We highlighted that this non-holomorphic transformation and the derivative of the modulus of the complex number $r$ plays the central role in the lack of explanation of the trick.

We also noted that the original trick had a statement which was of crucial importance, namely that the vectors $m^{a}$ and $\overline{m}^{a}$ be complex conjugates throughout the procedure.

When this subtle statement was not enforced, it allowed us to identify a hidden tetrad in the Newman-Janis trick.  The hidden tetrad was used to explain the equivalence between the Newman-Janis trick and Giampieri's method, thereby removing the question of what the ansatz in Giampieri's method is.  

However, it is interesting to observe that the hidden tetrad does not satisfy the conditions of a null tetrad, where $m^{a}$ and $\overline{m}^{a}$ should be complex conjugates of each other.

We rewrote the Newman-Janis trick where non-holomorphic transformations and the ambiguities involved were not present.  In this new procedure, we exploited tetrads which mathematically did not satisfy the conditions of being a null tetrad, since the vectors  $m^{a}$ and $\overline{m}^{a}$ were not complex conjugates of each other, but yet still produced meaningful answers.  This version of the trick allows us to start considering the Newman-Janis trick in a more fruitful way, with the possibility of understanding the physics of the procedure.

Future work in the direction of this project would involve explorations of extending this modified Newman-Janis trick, to the generalized case of the Kerr-Schild metrics. 

An interesting exploration would be to consider that since null tetrads and the mass term \cite{witten1981new} can be built from the theory of 2-component spinors, it may be advantageous to employ this modified Newman-Janis trick in the spinor language.  A hope of this exercise, may be that it helps us observe how the mass parameter behaves under the trick, and thereby help us explain the physics of the situation.  

Directions such as the ones mentioned above, would help us move into the wider objective of elucidating the relationships between General Relativity and complex-valued structures.  Ultimately, this may help us in understanding how General Relativity interacts with the complex-valued structure of quantum theory at a fundamental level, and thereby lead us to answer the question:  What is space, time and the quantum?


\bibliographystyle{unsrt}
\bibliography{bibfile} 

\begin{thebibliography}{10}

\bibitem{flaherty1976hermitian}
Edward~J. Flaherty.
\newblock {\em Hermitian and K{\"a}hlerian geometry in relativity}.
\newblock Springer-Verlag, 1976.

\bibitem{newman1965note}
Ezra~T Newman and AI~Janis.
\newblock Note on the {Kerr Spinning-Particle Metric}.
\newblock {\em Journal of Mathematical Physics}, 6(6):915--917, 1965.

\bibitem{isaacson2007einstein}
Walter Isaacson.
\newblock {\em {Einstein: His life and universe}}.
\newblock Simon and Schuster, 2007.

\bibitem{einstein1915field}
John Norton.
\newblock {How Einstein found his field equations: 1912-1915}.
\newblock {\em Historical studies in the physical sciences}, pages 253--316,
  1984.

\bibitem{wald2010general}
Robert~M Wald.
\newblock {\em {General Relativity}}.
\newblock University of Chicago Press, 2010.

\bibitem{carroll2004spacetime}
Sean~M Carroll.
\newblock {\em {Spacetime and Geometry. An Introduction to General
  Relativity}}.
\newblock Addison-Wesley, 2004.

\bibitem{beem1996global}
John~K Beem, Paul Ehrlich, and Kevin Easley.
\newblock {\em {Global Lorentzian geometry}}.
\newblock CRC Press, 1996.

\bibitem{chrusciel2010mathematical}
Piotr Chru{\'s}ciel, Gregory Galloway, and Daniel Pollack.
\newblock {Mathematical General Relativity: a sampler}.
\newblock {\em Bulletin of the American Mathematical Society}, 47(4):567--638,
  2010.

\bibitem{schwarzschild1916gravitationsfeld}
Karl Schwarzschild.
\newblock {\"U}ber das gravitationsfeld eines massenpunktes nach der
  einsteinschen theorie.
\newblock {\em Sitzungsberichte der K{\"o}niglich Preu{\ss}ischen Akademie der
  Wissenschaften (Berlin), 1916}, 1:189--196, 1916.

\bibitem{kerr1963gravitational}
Roy~P Kerr.
\newblock Gravitational field of a spinning mass as an example of algebraically
  special metrics.
\newblock {\em Physical Review Letters}, 11(5):237, 1963.

\bibitem{wiltshire2009kerr}
David~L Wiltshire, Matt Visser, and Susan~M Scott.
\newblock {\em {The Kerr Spacetime}}.
\newblock Cambridge University Press, 2009.

\bibitem{chandrasekhar1998mathematical}
Subrahmanyan Chandrasekhar.
\newblock {\em The mathematical theory of black holes}.
\newblock Oxford University Press, 1998.

\bibitem{o2003introduction}
Peter~J O'Donnell.
\newblock {\em Introduction to 2-spinors in General Relativity}.
\newblock World Scientific, 2003.

\bibitem{penrose1988spinors}
Roger Penrose and Wolfgang Rindler.
\newblock {\em Spinors and space-time: Volume 1}.
\newblock Cambridge University Press, 1988.

\bibitem{geroch1968spinor}
Robert Geroch.
\newblock {Spinor structure of space-times in General Relativity}.
\newblock {\em Journal of Mathematical Physics}, 9(11):1739--1744, 1968.

\bibitem{penrose2006road}
Roger Penrose.
\newblock {\em {The Road to Reality: A Complete Guide to the Laws of the
  Universe}}.
\newblock Vintage, 2007.

\bibitem{einstein1945generalization}
Albert Einstein.
\newblock {A Generalization of the Relativistic Theory of Gravitation}.
\newblock {\em Annals of Mathematics}, pages 578--584, 1945.

\bibitem{huybrechts2006complex}
Daniel Huybrechts.
\newblock {\em Complex geometry: an introduction}.
\newblock Springer-Verlag, 2006.

\bibitem{griffiths2014principles}
Phillip Griffiths and Joseph Harris.
\newblock {\em Principles of {Algebraic Geometry}}.
\newblock John Wiley \& Sons, 2014.

\bibitem{kodaira2012complex}
Kunihiko Kodaira.
\newblock {\em Complex manifolds and deformation of complex structures}.
\newblock Springer-Verlag, 2012.

\bibitem{hwangcomplex}
AD~Hwang.
\newblock Complex manifolds and hermitian differential geometry (math 1360
  lecture notes, toronto, web draft 1997).

\bibitem{brody2001geometric}
Dorje~C Brody and Lane~P Hughston.
\newblock {Geometric Quantum Mechanics}.
\newblock {\em Journal of geometry and physics}, 38(1):19--53, 2001.

\bibitem{bengtsson2006geometry}
Ingemar Bengtsson and Karol Zyczkowski.
\newblock {\em {Geometry of Quantum states: an Introduction to Quantum
  Entanglement}}.
\newblock Cambridge University Press, 2006.

\bibitem{lee2006riemannian}
John~M Lee.
\newblock {\em {Riemannian manifolds: an Introduction to Curvature}}.
\newblock Springer-Verlag, 2006.

\bibitem{nijenhuis}
Albert Nijenhuis.
\newblock $x_{n-1}$ - forming sets of eigenvectors.
\newblock {\em Proceedings of the Koninklijke Nederlandse Akademie van
  Wetenschappen}, 54:200--212, 1951.

\bibitem{bryant2014s}
Robert~L Bryant.
\newblock {S.-S. Chern's study of Almost-Complex Structures on the Six-sphere}.
\newblock {\em arXiv preprint arXiv:1405.3405}, 2014.

\bibitem{morrow1971complex}
James~A Morrow and Kunihiko Kodaira.
\newblock {\em Complex manifolds}.
\newblock American Mathematical Soc., 1971.

\bibitem{gibbons1993euclidean}
Gary~W Gibbons and Stephen~W Hawking.
\newblock {\em {Euclidean Quantum Gravity}}.
\newblock World Scientific, 1993.

\bibitem{ortin2004gravity}
Tom{\'a}s Ort{\'\i}n.
\newblock {\em {Gravity and Strings}}.
\newblock Cambridge University Press, 2004.

\bibitem{duggal1986cr}
KL~Duggal.
\newblock {CR-structures and Lorentzian geometry}.
\newblock {\em Acta Applicandae Mathematica}, 7(3):211--223, 1986.

\bibitem{robinson2002holomorphic}
DC~Robinson.
\newblock {Holomorphic 4-metrics and Lorentzian structures}.
\newblock {\em General Relativity and Gravitation}, 34(8):1173--1191, 2002.

\bibitem{zee2010quantum}
Anthony Zee.
\newblock {\em Quantum field theory in a nutshell}.
\newblock Princeton University press, 2010.

\bibitem{mattvisserwick}
Matt Visser.
\newblock How to wick rotate generic curved spacetime.
\newblock {\em {Gravity Research Foundation (unpublished)}}, 1991.

\bibitem{stephani2003exact}
Hans Stephani, Dietrich Kramer, Malcolm MacCallum, Cornelius Hoenselaers, and
  Eduard Herlt.
\newblock {\em Exact Solutions of Einstein's field equations}.
\newblock Cambridge University Press, 2003.

\bibitem{adamo2009null}
Timothy~M Adamo, Ezra~T Newman, and Carlos Kozameh.
\newblock {Null Geodesic Congruences, Asymptotically-Flat Spacetimes and Their
  Physical Interpretation}.
\newblock {\em Living Rev. Relativity}, 15:47, 2009.

\bibitem{adamo2014kerr}
Tim Adamo and ET~Newman.
\newblock {The Kerr-Newman metric: A Review}.
\newblock {\em Scholarpedia}, 9(31791), 2014.

\bibitem{erbin2015demianski}
Harold Erbin.
\newblock {{Demia{\'n}ski}--{Janis}--{Newman} algorithm}.
\newblock {\em Unpublished pre-print}, 2015.

\bibitem{whisker2008braneworld}
Richard Whisker.
\newblock Braneworld black holes.
\newblock {\em PhD thesis (University of Durham, 2006), arXiv
  preprint:0810.1534}, 2008.

\bibitem{newman1965metric}
Ezra~T Newman, E~Couch, K~Chinnapared, A~Exton, A~Prakash, and R~Torrence.
\newblock {Metric of a rotating, charged mass}.
\newblock {\em Journal of Mathematical Physics}, 6(6):918--919, 1965.

\bibitem{demianski1972new}
M~Demia{\'n}ski.
\newblock {New Kerr-like space-time}.
\newblock {\em Physics Letters A}, 42(2):157--159, 1972.

\bibitem{demianski1966combined}
M~Demianski and Ezra~T Newman.
\newblock {Combined Kerr-NUT solution of the Einstein field equations}.
\newblock Technical report, Univ., Warsaw. Univ. of Pittsburgh, 1966.

\bibitem{kerr2007discovering}
Roy Kerr.
\newblock {Discovering the Kerr and Kerr-Schild metrics}.
\newblock {\em The Kerr Spacetime, eds. D.L. Wiltshire, M. Visser, and S.M.
  Scott, (Cambridge University Press)}, pages 38--71 [arXiv:0706.1109], 2009.

\bibitem{talbot1969newman}
CJ~Talbot.
\newblock {Newman-Penrose approach to twisting degenerate metrics}.
\newblock {\em Communications in Mathematical Physics}, 13(1):45--61, 1969.

\bibitem{keane2014extension}
Aidan~J Keane.
\newblock {An extension of the Newman--Janis algorithm}.
\newblock {\em Classical and Quantum Gravity}, 31(15):155003, 2014.

\bibitem{lessner2008complex}
G~Lessner.
\newblock The “complex trick” in five-dimensional relativity.
\newblock {\em General Relativity and Gravitation}, 40(10):2177--2184, 2008.

\bibitem{erbin2014five}
Harold Erbin and Lucien Heurtier.
\newblock {Five-dimensional Janis-Newman algorithm}.
\newblock {\em arXiv preprint arXiv:1411.2030}, 2014.

\bibitem{herrera1982complexification}
L~Herrera and J~Jim{\'e}nez.
\newblock {The Complexification of a Nonrotating Sphere: an extension of the
  Newman--Janis algorithm}.
\newblock {\em Journal of Mathematical Physics}, 23(12):2339--2345, 1982.

\bibitem{drake1997application}
SP~Drake and R~Turolla.
\newblock {The application of the Newman-Janis algorithm in obtaining interior
  solutions of the Kerr metric}.
\newblock {\em Classical and Quantum Gravity}, 14(7):1883, 1997.

\bibitem{ibohal2005rotating}
Ng~Ibohal.
\newblock Rotating metrics admitting non-perfect fluids.
\newblock {\em General Relativity and Gravitation}, 37(1):19--51, 2005.

\bibitem{viaggiu2006interior}
Stefano Viaggiu.
\newblock {Interior Kerr solutions with the Newman--Janis algorithm starting
  with static physically reasonable space--times}.
\newblock {\em International Journal of Modern Physics D}, 15(09):1441--1453,
  2006.

\bibitem{azreg2014static}
Mustapha Azreg-A{\"\i}nou.
\newblock From static to rotating to conformal static solutions: rotating
  imperfect fluid wormholes with (out) electric or magnetic field.
\newblock {\em The European Physical Journal C}, 74(5):1--11, 2014.

\bibitem{azreg2014generating}
Mustapha Azreg-A{\"\i}nou.
\newblock Generating rotating regular black hole solutions without
  complexification.
\newblock {\em Physical Review D}, 90(6):064041, 2014.

\bibitem{drake2000uniqueness}
SP~Drake and Peter Szekeres.
\newblock {Uniqueness of the Newman--Janis algorithm in generating the
  Kerr--Newman metric}.
\newblock {\em General relativity and Gravitation}, 32(3):445--457, 2000.

\bibitem{glass2004kottler}
EN~Glass and JP~Krisch.
\newblock {Kottler-Lambda-Kerr Spacetime}.
\newblock {\em arXiv preprint gr-qc/0405143}, 2004.

\bibitem{yazadjiev2000letter}
S~Yazadjiev.
\newblock {Letter: Newman--Janis method and rotating dilaton-axion black hole}.
\newblock {\em General Relativity and Gravitation}, 32(12):2345--2352, 2000.

\bibitem{kim1999spinning}
Hongsu Kim.
\newblock {Spinning BTZ black hole versus Kerr black hole: A closer look}.
\newblock {\em Physical Review D}, 59(6):064002, 1999.

\bibitem{dianyan1988exact}
Xu~Dianyan.
\newblock {Exact solutions of Einstein and Einstein-Maxwell equations in
  higher-dimensional spacetime}.
\newblock {\em Classical and Quantum Gravity}, 5(6):871, 1988.

\bibitem{lombardo2004newman}
Diego Julio~Cirilo Lombardo.
\newblock {The Newman--Janis algorithm, rotating solutions and
  Einstein--Born--Infeld black holes}.
\newblock {\em Classical and Quantum Gravity}, 21(6):1407, 2004.

\bibitem{pirogov2013towards}
Yu~F Pirogov.
\newblock Towards the rotating scalar-vacuum black holes.
\newblock {\em arXiv preprint arXiv:1306.4866}, 2013.

\bibitem{dadhich2013rotating}
Naresh Dadhich and Sushant~G Ghosh.
\newblock {Rotating black hole in Einstein and pure Lovelock gravity}.
\newblock {\em arXiv preprint arXiv:1307.6166}, 2013.

\bibitem{ghosh2013radiating}
Sushant~G Ghosh, Sunil~D Maharaj, and Uma Papnoi.
\newblock {Radiating Kerr--Newman black hole in $f(R)$ gravity}.
\newblock {\em The European Physical Journal C}, 73(6):1--11, 2013.

\bibitem{hansen2013applicability}
Devin Hansen and Nicol{\'a}s Yunes.
\newblock {Applicability of the Newman-Janis algorithm to black hole solutions
  of modified gravity theories}.
\newblock {\em Physical Review D}, 88(10):104020, 2013.

\bibitem{caravelli2010spinning}
Francesco Caravelli and Leonardo Modesto.
\newblock Spinning loop black holes.
\newblock {\em Classical and Quantum Gravity}, 27(24):245022, 2010.

\bibitem{gutierrez2014computer}
Carlos Gutierrez-Chavez, Francisco Frutos-Alfaro, and Ivan Cordero-Garcia.
\newblock {A Computer Program for the Newman-Janis Algorithm}.
\newblock {\em arXiv preprint arXiv:1405.3008}, 2014.

\bibitem{giampieri}
Giacomo Giampieri.
\newblock {Introducing Angular Momentum into a Black Hole using Complex
  Variables}.
\newblock {\em Gravity Research Foundation (unpublished)}, 1990.

\bibitem{newman1973complex}
ET~Newman.
\newblock {Complex coordinate transformations and the Schwarzschild-Kerr
  metrics}.
\newblock {\em Journal of Mathematical Physics}, 14(6):774--776, 1973.

\bibitem{newman1976heaven}
Ezra~T Newman.
\newblock Heaven and its properties.
\newblock {\em General Relativity and Gravitation}, 7(1):107--111, 1976.

\bibitem{schiffer1973kerr}
Menahem~M Schiffer, Ronald~J Adler, James Mark, and Charles Sheffield.
\newblock {Kerr geometry as complexified Schwarzschild geometry}.
\newblock {\em Journal of Mathematical Physics}, 14(1):52--56, 1973.

\bibitem{finkelstein1975general}
Robert~J Finkelstein.
\newblock The general relativistic fields of a charged rotating source.
\newblock {\em Journal of Mathematical Physics}, 16:1271--1277, 1975.

\bibitem{szekeres1998explanation}
P~Szekeres and SP~Drake.
\newblock {An explanation of the Newman-Janis algorithm}.
\newblock {\em arXiv preprint gr-qc/9807001}, 1998.

\bibitem{erbin2014janis}
Harold Erbin.
\newblock {Janis-Newman algorithm: simplifications and gauge field
  transformation}.
\newblock {\em General Relativity and Gravitation}, 47(3):19, 2015.

\bibitem{witten1981new}
Edward Witten.
\newblock A new proof of the positive energy theorem.
\newblock {\em Communications in Mathematical Physics}, 80(3):381--402, 1981.

\end{thebibliography}
\end{document}